\documentclass[notoc]{JHEP3}

\usepackage{epsfig}

\newcommand\fverb{\setbox\pippobox=\hbox\bgroup\verb}
\newcommand\fverbdo{\egroup\medskip\noindent%
                        \Fbox{\unhbox\pippobox}\ }
\newcommand\fverbit{\egroup\item[\fbox{\unhbox\pippobox}]}
\newbox\pippobox

\def\JHEP{\it JHEP}

\def\app#1#2#3{{\it Astropart. Phys. }{\bf #1} (#2) {#3}}

\def\mw{m_W}

\def\gm{\gamma}
\def\Re{{\rm Re}}
\def\wt{\widetilde{w}}
\def\prop#1{s-m_{#1}^2+i \Gamma_{#1}m_{#1}}
\def\slp#1{\widetilde{\ell}_{#1}}
\def\sn#1{\widetilde{\nu}_{#1}}
\def\neu#1{\chi_{#1}^0}
\def\cha#1{\chi_{#1}^\pm}
\def\msl#1{m_{\widetilde{\ell}_{#1}}}
\def\msli{m_{\widetilde{\ell}_{a}}}
\def\mslj{m_{\widetilde{\ell}_{b}}}
\def\mslk{m_{\widetilde{\ell}_{c}}}

\def\mslpj{m_{\widetilde{\ell}'_{b}}}
\def\msn{m_{\widetilde{\nu}_{\ell}}}
\def\mch{m_{H^\pm}}
\def\ma{m_A}

\def\ml{m_{\ell}}
\def\mlp{m_{\ell'}}
\def\mnq{m_{\chi_i^0}}
\def\mnp{m_{\chi_j^0}}
\def\mcp{m_{\chi_k^{\pm}}}
\def\mcr{m_{\chi_l^{\pm}}}
\def\mnmn{m_{\chi^0_i}m_{\chi^0_j}}
\def\mnpmn{m_{\chi^0_i}+m_{\chi^0_j}}

\def\mcmc{m_{\chi_k^{\pm}}m_{\chi_l^{\pm}}}
\def\slsl#1{\msl{a}^{#1}\msl{b}^{#1}}
\def\slpsl#1{\msl{a}^{#1}+\msl{b}^{#1}}
\def\slmsl#1{\msl{a}^{#1}-\msl{b}^{#1}}
\def\Cijqp#1{C^{abij}_{#1}}
\def\CCijqp#1{\widetilde{C}^{abij}_{#1}}
\def\Dijqp#1{D^{abij}_{#1}}
\def\DDijqp#1{\widetilde{D}^{abij}_{#1}}
\def\CAzq{C_A^{\chi^0_i\chi Z}}
\def\CAzpc{C_A^{\chi^0_j\chi Z *}}
\def\CS#1#2{C_S^{\chi^0_{#1}\slp{#2}^* \ell}}
\def\CP#1#2{C_P^{\chi^0_{#1}\slp{#2}^* \ell}}

\def\CAzc{C_A^{\ell\ell Z *}}
\def\CVzc{C_V^{\ell\ell Z *}}
\def\CZsl#1#2{C^{\slp{#1}^*\slp{#2}Z}}
\def\CWLp{C_L^{\chi_k^+\chi W^- *}}
\def\CWLr{C_L^{\chi_l^+\chi W^-}}
\def\CWRp{C_R^{\chi_k^+\chi W^- *}}
\def\CWRr{C_R^{\chi_l^+\chi W^-}}
\def\ffb{f\bar{f}}
\def\Fu{{\mathcal F}^u}
\def\Ft{{\mathcal F}^t}
\def\Tu{{\mathcal T}^u}
\def\Tt{{\mathcal T}^t}
\def\cosb{\cos\beta}

\def\cosapb{\cos(\alpha+\beta)}
\def\sinapb{\sin(\alpha+\beta)}
\def\cosa{\cos\alpha}
\def\sina{\sin\alpha}
\def\gm{\gamma}
\def\Re{{\rm Re}}
\def\wt{\widetilde{w}}
\def\prop#1{s-m_{#1}^2+i \Gamma_{#1}m_{#1}}
\def\slp#1{\widetilde{\ell}_{#1}}
\def\sn#1{\widetilde{\nu}_{#1}}
\def\neu#1{\chi_{#1}^0}
\def\cha#1{\chi_{#1}^\pm}
\def\msl#1{m_{\widetilde{\ell}_{#1}}}
\def\msli{m_{\widetilde{\ell}_{a}}}
\def\mslj{m_{\widetilde{\ell}_{b}}}
\def\mslk{m_{\widetilde{\ell}_{c}}}

\def\mslpj{m_{\widetilde{\ell}'_{b}}}
\def\msn{m_{\widetilde{\nu}_{\ell}}}
\def\mch{m_{H^\pm}}
\def\ma{m_A}

\def\ml{m_{\ell}}
\def\mlp{m_{\ell'}}
\def\mnq{m_{\chi_i^0}}
\def\mnp{m_{\chi_j^0}}
\def\mcp{m_{\chi_k^{\pm}}}
\def\mcr{m_{\chi_l^{\pm}}}
\def\mnmn{m_{\chi^0_i}m_{\chi^0_j}}
\def\mnpmn{m_{\chi^0_i}+m_{\chi^0_j}}

\def\mcmc{m_{\chi_k^{\pm}}m_{\chi_l^{\pm}}}
\def\slsl#1{\msl{a}^{#1}\msl{b}^{#1}}
\def\slpsl#1{\msl{a}^{#1}+\msl{b}^{#1}}
\def\slmsl#1{\msl{a}^{#1}-\msl{b}^{#1}}
\def\Cijqp#1{C^{abij}_{#1}}
\def\CCijqp#1{\widetilde{C}^{abij}_{#1}}
\def\Dijqp#1{D^{abij}_{#1}}
\def\DDijqp#1{\widetilde{D}^{abij}_{#1}}
\def\CAzq{C_A^{\chi^0_i\chi Z}}
\def\CAzpc{C_A^{\chi^0_j\chi Z *}}
\def\CS#1#2{C_S^{\chi^0_{#1}\slp{#2}^* \ell}}
\def\CP#1#2{C_P^{\chi^0_{#1}\slp{#2}^* \ell}}

\def\CAzc{C_A^{\ell\ell Z *}}
\def\CVzc{C_V^{\ell\ell Z *}}
\def\CZsl#1#2{C^{\slp{#1}^*\slp{#2}Z}}
\def\CWLp{C_L^{\chi_k^+\chi W^- *}}
\def\CWLr{C_L^{\chi_l^+\chi W^-}}
\def\CWRp{C_R^{\chi_k^+\chi W^- *}}
\def\CWRr{C_R^{\chi_l^+\chi W^-}}
\def\ffb{f\bar{f}}
\def\vsl{\widetilde{V}_{\ell}^\dagger}
\def\Fu{{\mathcal F}^u}
\def\Ft{{\mathcal F}^t}
\def\Tu{{\mathcal T}^u}
\def\Tt{{\mathcal T}^t}
\newcommand{\newc}{\newcommand}
\newc\eg{{\it {e.g.}}}  \newc\etal{{\it {et al.}}} \newc\ie{{\it i.e.}}
\newc\etc{{\it {etc}}}  

\newcommand\lsim{\mathrel{\rlap{\lower4pt\hbox{\hskip1pt$\sim$}}
    \raise1pt\hbox{$<$}}}
\newcommand\gsim{\mathrel{\rlap{\lower4pt\hbox{\hskip1pt$\sim$}}
    \raise1pt\hbox{$>$}}}

\newc{\mhalf}{m_{1/2}}      \newc{\mzero}{m_0}

\newc{\tanb}{\tan\beta}
\newc{\azero}{A_0}
\newc{\at}{A_t} \newc{\abot}{A_b} \newc{\atau}{A_\tau} 

\newc{\bmu}{B\mu}           \newc{\sgn}{{\rm sgn}}
\newc{\mone}{M_1}           \newc{\mtwo}{M_2}

\newc{\charone}{\chi_1^\pm} \newc{\mcharone}{m_{\chi_1^\pm}}

\newc{\hl}{h}               \newc{\mhl}{m_{\hl}}
\newc{\hh}{H}               \newc{\mhh}{m_{\hh}}
\newc{\ha}{A}               \newc{\mha}{m_{\ha}}
\newc{\hc}{H^{\pm}}         \newc{\mhc}{m_{\hc}}

\newc{\qzero}{Q_0}          \newc{\qstop}{Q_{\widetilde t}}

\newc{\amu}{a_{\mu}}        \newc{\amususy}{a_{\mu}^{\rm SUSY}}
\newc{\amuexpt}{a_{\mu}^{\rm expt}}        \newc{\amusm}{a_{\mu}^{\rm SM}}
\newc{\deltaamususy}{\Delta a_{\mu}^{\rm SUSY}}

\newc{\msbar}{\overline {\rm MS}} \newc{\drbar}{\overline {\rm DR}}

\newc{\mt}{m_t} \newc{\mb}{m_b} \newc{\mtau}{m_{\tau}}
\newc{\yt}{h_t} \newc{\yb}{h_b} \newc{\ytau}{h_{\tau}}
\newc{\mtpole}{m_t^{\rm pole}} \newc{\mbpole}{m_b^{\rm pole}} 
\newc{\mtaupole}{m_{\tau}^{\rm pole}} 

\newc{\mtmtsmmsbar}{m_t(m_t)^{\msbar}_{{\rm SM}}}
\newc{\mtmtsmdrbar}{m_t(m_t)^{\drbar}_{{\rm SM}}}
\newc{\mtmtmssmdrbar}{m_t(m_t)^{\drbar}_{{\rm SUSY}}}

\newc{\mbmbsmmsbar}{m_b(m_b)^{\msbar}_{{\rm SM}}}
\newc{\mbmzsmmsbar}{m_b(\mz)^{\msbar}_{{\rm SM}}}
\newc{\mbmzsmdrbar}{m_b(\mz)^{\drbar}_{{\rm SM}}}
\newc{\mbmzmssmdrbar}{m_b(\mz)^{\drbar}_{{\rm SUSY}}}
\newc{\mtaumzsmmsbar}{m_{\tau}(\mz)^{\msbar}_{{\rm SM}}}
\newc{\mtaumzsmdrbar}{m_{\tau}(\mz)^{\drbar}_{{\rm SM}}}
\newc{\mtaumzmssmdrbar}{m_{\tau}(\mz)^{\drbar}_{{\rm SUSY}}}

\newc{\mgut}{M_{\rm GUT}}

\newc{\mplanck}{M_{\rm P}}      \newc{\mpl}{M_{\rm Pl}}
\newc{\msusy}{M_{\rm SUSY}}      \newc{\ms}{M_{\rm S}}

\newc{\jxf}{J({\xf})}
\newc{\jxfexact}{J_{\rm exact}({\xf})}  \newc{\jxfexp}{J_{\rm exp}({\xf})}
\newc{\VEV}[1]{\langle #1 \rangle}

\newc{\xf}{x_f}
\newc\vrel{v_{\rm rel}}
\newcommand\mchi{m_{\chi}}              

\newc\sell{{\widetilde e}_L}      \newc\msell{m_{\sell}}
\newc\selr{{\widetilde e}_R}      \newc\mselr{m_{\selr}}

\newc\snue{{\widetilde \nu}_e}      \newc\msnue{m_{\snue}}
\newc\snutau{{\widetilde \nu}_\tau}      \newc\msnutau{m_{\snutau}}

\newc\supl{{\widetilde u}_L}      \newc\msupl{m_{\supl}}
\newc\supr{{\widetilde u}_R}      \newc\msupr{m_{\supr}}
\newc\sdl{{\widetilde d}_L}      \newc\msdl{m_{\sdl}}
\newc\sdr{{\widetilde d}_R}      \newc\msdr{m_{\sdr}}

\newcommand\stauone{{\widetilde \tau}_1}   
\newcommand\stautwo{{\widetilde \tau}_2}

\newc\hpm{H^\pm} \newc\hp{H^+} \newc\hm{H^-} 
\newc\sfermion{\tilde f}  \newc\msfermion{m_{\sfermion}}  

\newc\second{{\rm sec}} 

\newc\alphas{\alpha_s}

\newc\alphaem{\alpha_{em}}

\newc{\gstar}{g_\ast}           \newc{\gsstar}{g_{s\ast}}
\newc{\geff}{g_{\rm eff}}

\newcommand\mz{m_{Z}}

\newc{\sthw}{\sin\theta_W}              \newc{\cthw}{\cos\theta_W}
\newc{\bino}{\widetilde B}              \newc{\wino}{\widetilde W_3}
\newc{\higgsinob}{{\widetilde H}^0_b}   \newc{\higgsinot}{{\widetilde H}^0_t}

\newc{\abund}{\Omega h^2}
\newc{\abundchi}{\Omega_\chi h^2}
\newc{\abundcdm}{\Omega_{{\rm CDM}} h^2}
\newc{\omegam}{\Omega_{{\rm M}}}       \newc{\abundm}{\Omega_{{\rm M}} h^2}
\newc{\omegab}{\Omega_{{\rm b}}}        \newc{\abundb}{\Omega_{{\rm b}} h^2}
\newc{\omegacdm}{\Omega_{{\rm CDM}}}   \newc{\omegatot}{\Omega_{{\rm TOT}}}

\newc{\rhocrit}{\rho_{crit}}
\newc{\rhochi}{\rho_{\chi}}

\newcommand\gev{\,\mbox{GeV}}

\newc\br{\mbox{BR}}

\newc{\ra}{\rightarrow}
\newc{\beq}{\begin{equation}}
\newc{\eeq}{\end{equation}}
\newc{\bea}{\begin{eqnarray}}
\newc{\eea}{\end{eqnarray}}

\newcommand\vs{{\it {vs.}}}

\newcommand\mh{m_{h}}          
          
\newc\stoponetwo{{\widetilde t}_{1,2}}
\newc\sbotonetwo{{\widetilde b}_{1,2}}
\newc\stauonetwo{{\widetilde \tau}_{1,2}}

\long\def\begincomment#1\endcomment{%
        \begingroup\sf\baselineskip12pt#1\endgroup}

\newcommand\T{\mathcal{T}}       \newcommand\Y{\mathcal{Y}}
\newcommand\F{\mathcal{F}}

\newcommand\cw{\cos\theta_W}
\newcommand\tw{\tan\theta_W}
\newcommand{\swsq}{\sin^2 \theta_W}
\newcommand{\delmux}{\stackrel{\leftrightarrow}{\partial^{\,\mu}}}
\newcommand{\llb}{\ell\bar{\ell}}
\newcommand{\qqb}{q\bar{q}}

\title{Exact Cross Sections \\ for the Neutralino-Slepton Coannihilation}

\author{Takeshi Nihei\\
    Department of Physics, College of Science and Technology,
        Nihon University, \\
        1-8-14, Kanda-Surugadai, Chiyoda-ku, Tokyo, 101-8308, Japan \\
        E-mail: \email{nihei@phys.cst.nihon-u.ac.jp}}

\author{Leszek Roszkowski\\
        Department of Physics, Lancaster University,
        Lancaster LA1 4YB, England\\
        E-mail: \email{L.Roszkowski@lancaster.ac.uk}}

\author{Roberto Ruiz de Austri\\
        Physics Division, School of Technology, 
        Aristotle University of Thessaloniki, \\
        GR - 540 06 Thessaloniki, Greece \\
        E-mail: \email{rruiz@gen.auth.gr}}

\abstract{Coannihilation processes provide an important additional
        mechanism for reducing the density of stable relics in the
        Universe. In the case of the stable lightest neutralino of the
        MSSM, and in particular the Constrained MSSM (CMSSM), the
        coannihilation with sleptons plays a major role in opening up
        otherwise cosmologically excluded ranges of supersymmetric
        parameters. In this paper, we derive a full set of exact,
        analytic expressions for the coannihilation of the lightest
        neutralino with the sleptons into all two--body tree--level
        final states in the framework of minimal supersymmetry. We
        make no simplifying assumptions about the neutralino nor about
        sfermion masses and mixings other than the absence of explicit
        CP--violating terms and inter--family mixings. The expressions
        should be particularly useful in computing the neutralino WIMP
        relic abundance without the approximation of partial
        wave expansion. We illustrate the effect of our analytic
        results with numerical examples and demonstrate a sizeable
        difference with approximate expressions available in the
        literature.  }


\begin{document}



\section{Introduction}\label{intro:sec}
The relic density of stable weakly--interacting massive particles (WIMPs)
is determined primarily by how efficiently their number 
density  in the early Universe can be reduced. In the case of
the most popular WIMP candidate: the lightest neutralino of minimal
supersymmetry (SUSY), assumed to be the lightest SUSY particle (LSP),
there are two generic mechanisms~\cite{kt90,jkg96}. First, the neutralino can
pair--annihilate into ordinary particles. Second, in some cases they can
coannihilate with some some other species if these are nearly
mass--degenerate with the LSPs. 

The standard mechanism of neutralino pair--annihilation has been
considered in much detail in many papers~\cite{jkg96}.
In particular, complete sets of neutralino annihilation cross sections
were provided in~\cite{dn93} (see also~\cite{jkg96}) in the case
of partial wave approximation. Exact expressions applicable both near
resonances and new final--state thresholds were recently published
in~\cite{nrr2}. 

The mechanism of coannihilation was
originally pointed out by Griest and Seckel~\cite{gs91}. It applies
when there exists some other species $\chi^\prime$ which is not much
heavier than the WIMP and may therefore be still present in the
thermal plasma at the time of WIMP decoupling. Coannihilation becomes
important if its annihilation with the WIMP (and/or itself) is equally, or 
more, efficient than the pair--annihilation of the stable WIMPs.

These circumstances in particular are naturally realized in the case
of the higgsino--like lightest neutralino LSP in the framework of the
Minimal Supersymmetric Standard Model (MSSM).  In this case the
next--to--lightest neutralino and the lightest chargino are almost
mass--degenerate with the LSP~\cite{mizuta92,eg97}. In fact, the
coannihilation in this case is so efficient that it has a devastating
effect on the relic density of the higgsino--like neutralino, which is
the type of LSP strongly disfavored by naturalness~\cite{chiasdm} and
mass--unification~\cite{rr93,an93,kkrw94}.  It even plays some
role~\cite{eg97} in the strongly prefered case of the bino--like
LSP~\cite{chiasdm}.

Since, in the framework of general softly--broken low--energy SUSY, scalar
superpartner masses are a priori unrelated (or at best loosely
related) to the neutralino sector, in principle one might assume that
any of the scalar superpartners could be light enough to 
participate in coannihilation
with the neutralino LSP -- the case that would be technically rather
challenging and in any case not particularly well--motivated.

A more realistic scenario is the one in which one of the scalar partners
of the top--quark or the $\tau$--lepton is rather light and nearly
degenerate in mass with the neutralino LSP. This is because the
off--diagonal elements in their $2\times2$ mass matrices can under some
circumstances greatly reduce one of the eigenmasses relative to the
other.  The case of neutralino--stop coannihilation 
was considered in~\cite{bdd00} in the
framework of the MSSM and recently shown in~\cite{eos01} to be
also applicable in the case of the Constrained MSSM (CMSSM) but only
for rather large values of the trilinear soft SUSY--breaking term
$\azero$.

The importance of the neutralino coannihilation with the lighter of
the two staus in the framework of the CMSSM was pointed out
in~\cite{efo98}. In this model, when the common gaugino mass parameter
$\mhalf$ is much larger than the common scalar mass parameter
$\mzero$, it is the lightest stau that is the LSP~\cite{kkrw94}. The
effect of the neutralino--stau coannihilation is to open up a narrow
corridor~\cite{efo98} just above the boundary of equal
neutralino--stau masses into an otherwise cosmologically forbidden
region in the $(\mhalf,\mzero)$--plane.  (Without coannihilation, the
requirement of the relic abundance of the neutralino to be consistent
with that allowed by the age of the Universe ($\abundchi\lsim{\cal
O}(1)$) often provides a stringent upper bound on the parameters
$\mhalf$ and $\mzero$~\cite{rr93,kkrw94}.) The effect has since been
included in a number of recent analyses, \eg\
in~\cite{coann:recent,bbb02}, and in a publically available package 
for computing the relic density micrOMEGAs~\cite{micromegas}.

One should mention that there are two other ways of evading this
otherwise generic cosmological bound on $\mhalf$ and $\mzero$. One is
realized for $\mzero\gg\mhalf$.  In this region one invariably finds
it difficult to satisfy the conditions of radiative electroweak
symmetry breaking (EWSB); in other words the square of the
Higgs/higgsino mass parameter $\mu$ comes out to be negative. In a
very narrow corridor along the region of no--EWSB, $\mu$ grows rapidly
from zero but it is still less than $\mhalf$~\cite{kkrw94,fmw00}. As a
result, the LSP has a sizeable higgsino component (although it is
still mostly a bino, like in the rest of the $(\mhalf,\mzero)$--plane)
and its relic density is typically small. In fact, because of the
growing LSP mass and its gaugino fraction, the relic density increases
rapidly along a very steep slope from very small values,
characteristic for lighther neutralinos with a larger higgsino
admixture, to larger (and often cosmologically excluded) values
characteristic of heavier and bino--dominated neutralino. As a result,
the cosmologically favored range $0.1<\abundchi<0.2$ is only realized
there for a rather narrow range of $\mhalf$~\cite{fmw00,rrn1}.

The other important escape route from the cosmological bound on
$\mhalf$ and $\mzero$ occurs when $\tanb$, which is the usual ratio of the
vacuum expectation values of the neutral Higgs scalars, is large,
$\gsim 50$.  This is because the physical masses of the heavy Higgs
scalar $\hh$ and pseudoscalar $\ha$ decrease with increasing
$\tanb$~\cite{dn93}.  When the neutralino mass becomes large enough,
close to half of the heavy Higgs boson mass, the LSP relic abundance
becomes efficiently reduced through a relatively wide resonance
involving mostly the pseudoscalar exchange. The effect is amplified by
the coupling of $A$ to down--type fermions which grows like $\tanb$.
(The heavy scalar Higgs coupling also grows in a similar fashion but
its dominant contribution is only $p$--wave and therefore suppressed
by square of the WIMP relative velocity.)
In the framework of the CMSSM, the effect is to open up wide fractions
of the otherwise cosmologically excluded ranges of the
$(\mhalf,\mzero)$--plane along the wide
$A$--resonance~\cite{efo98,rrn1,bk01,ls01,ddk01}.  
The precise position of the resonance shows a sizeable
dependence on some input parameters, most notably on the ratio
$\mt/\mb$, $\azero$, \etc, and remains a subject of some ongoing
debate. The full two--loop Higgs effective potential would have to be
computed and implemented in the analysis to reduce the sensitivity to,
\eg, the scale dependence.

In the CMSSM a set of reasonably well--motivated
unification assumptions leads to only four parameters: a common
gaugino mass $\mhalf$, a common scalar mass $\mzero$, a trilinear
coupling $\azero$, as well as $\tanb$. One is also free to choose the
sign of the $\mu$--parameter, while its magnitude is determined by the
mechanism of electroweak radiative symmetry breaking (EWSB). The CMSSM
can therefore be considered as a well--motivated SUSY model with the
smallest number of independent parameters. (In particular, the
so--called minimal supergravity (mSUGRA) model can be viewed as a
specific realization of the framework.)  The CMSSM has become a
benchmark model for the LHC and other SUSY searches.

As mentioned above, the neutralino--stau coannihilation effect is
particularly important in the framework of the CMSSM, but it can
affect the neutralino relic density also in the more general
MSSM. This will be the framework in which we will work here for the
sake of generality.

In this paper, we will present a full set of exact, analytic
expressions for the tree--level cross section of the neutralino
coannihilation with sleptons into all two--body final states in the
general MSSM. In our analysis we will make no simplifying assumptions
about the neutralino, nor will we assume the degeneracy of the left--
and right--sfermion masses. We will not consider here the possibility
of CP and flavor violation in the slepton sector although we will
assume a general form of the left--right slepton mixing within each
generation.  We will include all tree--level final states and all
intermediate states. We will also keep finite widths in $s$--channel
resonances.  A set of expressions for the neutralino--slepton
coannihilaton was given in~\cite{efo98,efos00} but only in the
approximation of the partial wave expansion. Furthermore, these
formulae did not include the effects of the tau Yukawa, of the
$\stauone-\stautwo$ mixing and in some channels of the mass of the
$\tau$, \etc, which make them less reliable at large
$\tanb\gsim20$~\cite{efgos01}.

The results presented here are exact, include all the terms and are
valid both small and large values of $\tanb$, and both near and
away from resonances and thresholds for new final states. This
paper is meant to be a follow-up to~\cite{nrr2} where we have
calculated all the analytic cross sections of all tree--level
processes for the neutralino pair--annihilation into all two--body
final states. We follow the same conventions and notations as
in~\cite{nrr2}.

The plan of the paper is as follows. In Sect.~\ref{relicdensity:sec}
we briefly review the formalism for computing the relic density in the
presence of both annihilation and coannihilation.  In
Sect.~\ref{mssm:sec} we introduce the relevant ingredients of the MSSM
and list all the neutralino pair--annihilation channels. Explicit
expressions for the coannihilation cross secion are given in
Sect.~\ref{exact:sec}. In Sect.~\ref{numanalysis:sec} we present some
numerical examples and in Sect.~\ref{summary:sec} we summarize our
work. Appendix~A contains a complete list of relevant
couplings and in Appendix~B we provide
expressions for several auxiliary functions used in the text.

%
\section{Calculation of the Relic Density with Coannihilation
}\label{relicdensity:sec}
As summarized, \eg, in~\cite{nrr2}, the 
time evolution and subsequent
freeze--out of a stable relic in an expanding Universe are described by the
Boltzmann equation
\begin{eqnarray}
\frac{d n}{dt} & = & - 3 H n
 - \langle\sigma v_{\rm M{\o}l}\rangle
\left[n^2 - (n^{\rm eq})^2\right], 
\label{Boltzmann-eq:eq}
\end{eqnarray}
where $n$ stands for the species' number density, 
$n^{\rm eq}$ is the number density that it would
have had if it had remained in thermal equilibrium,
$H(T)$ is the Hubble expansion rate,
$\sigma$ denotes the cross section of the species
pair--annihilation into all allowed final states,
$v_{\rm M\o l}$ is the so--called M{\o}ller velocity~\cite{gg91} which is
effectively the relative velocity
of the two initial--state non--relativistic particles in the CM frame, 
and
$\langle\sigma v_{\rm M\o l}\rangle$ represents the thermal average of
$\sigma v_{\rm M\o l}$.
In the early Universe, the species were initially in thermal
equilibrium, $n=n^{\rm eq}$. The number density decreased
with the expanding Universe and
the relic froze out of the thermal equilibrium when its typical
interaction rate became less than the Hubble parameter.

In a more general framework with coannihilation, the stable particle
$\chi$ is the lightest of $N$ species $\chi_i$, each with mass $m_i$,
number density $n_i$, equilibrium number density $n_i^{\rm
eq}$ and the number of internal degrees of freedom $g_i$. 
Griest and Seckel~\cite{gs91} showed that
the total number density of all the species taking part in
the coannihilation process
\begin{equation}
n=\sum_i^N n_i
\end{equation}
obeys the Boltzmann equation as given in
eq.~(\ref{Boltzmann-eq:eq}). The quantity 
$\langle\sigma v_{\rm M\o l}\rangle$ now stands for
\begin{eqnarray}
\label{sigmavtot:eq}
\langle\sigma v_{\rm M\o l}\rangle
& = & \sum_{ij} \langle\sigma_{ij} v_{ij}\rangle
\frac{n_i^{\rm eq}}{n^{\rm eq}}\frac{n_j^{\rm eq}}{n^{\rm eq}},
\end{eqnarray}
where
\begin{eqnarray}
\sigma_{ij} & = & \sigma(\chi_i \chi_j \rightarrow {\rm all})
\end{eqnarray}
and $v_{ij}$ are the relative (M\o ller) velocities of the
coannihilating particles. Edsj{\" o} and Gondolo~\cite{eg97}
recast eq.~(\ref{sigmavtot:eq}) into a convenient form involving
only one--dimensional integrals
\begin{eqnarray}
\langle\sigma v_{\rm M\o l}\rangle
& = & \frac{
\int_{4 m_\chi^2}^\infty ds \, s^{3/2}
K_1\left(\frac{\sqrt{s}}{T}\right) 
\sum_{ij}^{N} 
\beta_f^2(s,m_i,m_j) \frac{g_i g_j}{g_\chi^2} \sigma_{ij}(s) 
}
{8 m_\chi^4 T \left[ \sum_i^N \frac{g_i}{g_\chi} 
\frac{m_i^2}{m_\chi^2} K_2^2(m_i/T) \right]^2}, 
\label{thermal-average:eq}
\end{eqnarray}
where  $s=(p_i+p_j)^2$ is the usual Mandelstam variable, $K_i$ denotes the
modified Bessel function of order $i$, 
and the kinematic factor $\beta_f(s,m_i,m_j)$ is given by
\begin{eqnarray}
\beta_f(s,m_i,m_j)=
\left[1-\frac{(m_i+m_j)^2}{s}\right]^{1/2}
\left[1-\frac{(m_i-m_j)^2}{s}\right]^{1/2}.
\label{kdef:eq}
\end{eqnarray}

Note that eq.~(\ref{kdef:eq}) reduces to a familiar expression derived by
Gondolo and Gelmini~\cite{gg91} in the case of single particle
annihilation. (Compare, \eg, eq.~(2.3) in~\cite{nrr2}.) Note also
that, because of the assumed $R$--parity, after the freeze--out only the
LSP $\chi$ will survive and its number density will at the end be $n_\chi=n$.

Eqs.~(\ref{Boltzmann-eq:eq})
and~(\ref{thermal-average:eq}) can be rather accurately solved by 
iteratively solving the equation
\begin{eqnarray}
x_f^{-1} & = & \ln \left[ \frac{m_\chi}{2 \pi^3} \frac{g_{\rm eff}}{2} 
\sqrt{\frac{45}{2g_* G_N}}
\langle\sigma v_{\rm M\o l}\rangle({x_f})\, x_f^{1/2} \right],
\label{freeze-out-temperature:eq}
\end{eqnarray}
where $g_{\rm eff}$ is the effective
number of degrees of freedom of the coannihilating
particles~\cite{gs91}, $g_*$ represents the effective
number of degrees of freedom at freeze--out
($\sqrt{g_*}\simeq 9$), $G_N$ is the gravitational
constant, $x=T/m_\chi$ and the freeze--out point
$x_f\equiv T_f/\mchi$ is roughly $1/25$ to $1/20$. 
(See, \eg,~\cite{nrr1}.) 

The neutralino LSP relic abundance is
$\abundchi=\rho_\chi/{\rho_{crit}}$, where the critical density is
$\rho_{crit} = 1.05\times 10^{-5}\,
(h^2)\gev/cm^3$, and $\rho_\chi$ today is to a good 
approximation given by
\begin{eqnarray}
\rho_\chi &=& \frac{1.66}{M_{\rm Pl}} \left(\frac{T_\chi}{T_\gamma}\right)^3
T_\gamma^3 \sqrt{g_*} \left[\int_0^{x_f}dx \langle\sigma v_{\rm M\o
l}\rangle(x)\right]^{-1}, 
\label{relic-density:eq}
\end{eqnarray}
where $M_{\rm Pl}=1/{\sqrt{G_N}}$ denotes the Planck mass,
$T_\chi$ and $T_\gamma$ are the present temperatures of the neutralino
and the photon, respectively. The suppression factor
$(T_\chi/T_\gamma)^3$ $\approx$ $1/20$ follows from entropy
conservation in a comoving volume~\cite{reheatfactor}.

%
\section{WIMP Coannihilation in the MSSM}\label{mssm:sec}
We will be working in the framework of the general
MSSM.\ 
(For a review, see, \eg,~\cite{susyreview}. We
follow the conventions of~\cite{gh}.) We will not define it here
but instead refer the reader to our previous publication~\cite{nrr2}
where also all the relevant quantities are introduced. All the remaining
couplings that we will need here are 
summarized in Appendix~A.

In this Section we summarize all the coannihilation channels of the
neutralino LSP with the sleptons into tree--level two--body final
states.  They are listed in Table~\ref{tableone}.
We only neglect slepton coannihilation with the lightest chargino
and next--to--lightest neutralino which might be of some importance in
the higgsino--dominant case. However, in this region the
neutralino--chargino coannihilation is very effective anyway in
reducing the relic density below any interesting level.

We further neglect (s--)lepton flavor (generation) mixing, but
include the left--right mixing for the sleptons. Thus $a=1,2$, where
the index $a$ denotes the slepton mass state within each family. 
$\ell$ and $\ell^{\prime}$ 
represent charged leptons of different generations.

\TABLE[h!] {\label{tableone} 
\begin{tabular}[h!]{|p{1.6in}|p{1.in}|p{1.in}|p{1.in}|p{0.5in}|} \hline
& \multicolumn{4}{c|}{ Exchanged particles } \\
\cline{2-5}
\multicolumn{1}{|c|} { Process } &
\raisebox{2ex}[5mm]{}
{$\ \  s$--channel} &{$\ \ \ t$--channel} &
                     {$\ \ \ u$--channel} & \ \ \ {PI} \\ 
\hline \hline
\hspace{0.1in}\raisebox{0ex}[5mm]{} $\slp{a} \slp{b}^* \rightarrow WW$ &
\hspace{.1in} $h,H,\gamma,Z $ &
\hspace{.35in} $ $ &
\hspace{.35in} $\sn{\ell} $ &
\hspace{.1in} 4P \\
\hline
\hspace{0.1in}\raisebox{0ex}[5mm]{} $\slp{a} \slp{b}^* \rightarrow ZZ$ &
\hspace{.1in} $h,H $ &
\hspace{0.35in} $\slp{c} $ &
\hspace{0.35in} $\slp{c} $ &
\hspace{.1in} 4P \\ 
\hline
\hspace{0.1in}\raisebox{0ex}[5mm]{} $\slp{a} \slp{b}^* \rightarrow Z\gamma$ &
\hspace{.1in} $ $ &
\hspace{0.35in} $\slp{c} $ &
\hspace{0.35in} $\slp{c} $ &
\hspace{.1in} 4P \\ 
\hline
\hspace{0.1in}\raisebox{0ex}[5mm]{} $\slp{a} \slp{b}^* \rightarrow \gamma\gamma$ &
\hspace{.1in} $ $ &
\hspace{0.35in} $\slp{c} $ &
\hspace{0.35in} $\slp{c} $ &
\hspace{.1in} 4P \\ 
\hline \hline 
\hspace{0.1in}\raisebox{0ex}[5mm]{}  $\slp{a} \slp{b}^* 
                                           \rightarrow W^\pm H^\mp$ &
\hspace{.1in} $h,H,A $ &
\hspace{0.35in} $ $ &
\hspace{0.35in} $\sn{\ell} $ &
\hspace{.1in} $ $ \\ 
\hline 
\hspace{0.1in}\raisebox{0ex}[5mm]{}  $\slp{a} \slp{b}^* \rightarrow Zh, ZH$ &
\hspace{.1in} $A,Z $ &
\hspace{0.35in} $\slp{c} $ &
\hspace{0.35in} $\slp{c} $ &
\hspace{.1in} $ $ \\ 
\hline 
\hspace{0.1in}\raisebox{0ex}[5mm]{}  $\slp{a} \slp{b}^* \rightarrow ZA$ &
\hspace{.1in} $h,H $ &
\hspace{0.35in} $\slp{c} $ &
\hspace{0.35in} $\slp{c} $ &
\hspace{.1in} $ $ \\ 
\hline
\hspace{0.1in}\raisebox{0ex}[5mm]{}  $\slp{a} \slp{b}^* \rightarrow \gamma h, \gamma H$ &
\hspace{.1in} $ $ &
\hspace{0.35in} $\slp{c} $ &
\hspace{0.35in} $\slp{c} $ &
\hspace{.1in} $ $ \\ 
\hline
\hspace{0.1in}\raisebox{0ex}[5mm]{}  $\slp{a} \slp{b}^* \rightarrow \gamma A$ &
\hspace{.1in} $ $ &
\hspace{0.35in} $\slp{c} $ &
\hspace{0.35in} $\slp{c} $ &
\hspace{.1in} $ $ \\ 
\hline \hline
\hspace{0.1in}\raisebox{0ex}[5mm]{} $\slp{a}\slp{b}^*\rightarrow hh, HH, hH$&
\hspace{.1in} $h,H $ &
\hspace{0.35in} $\slp{c} $ &
\hspace{0.35in} $\slp{c} $ &
\hspace{.1in} 4P \\
\hline 
\hspace{0.1in}\raisebox{0ex}[5mm]{}  $\slp{a} \slp{b}^* \rightarrow hA, HA$ &
\hspace{.1in} $A,Z $ &
\hspace{0.35in} $\slp{c} $ &
\hspace{0.35in} $\slp{c} $ &
\hspace{.1in} $ $ \\ 
\hline
\hspace{0.1in}\raisebox{0ex}[5mm]{}  $\slp{a} \slp{b}^* \rightarrow AA$ &
\hspace{.1in} $h,H $ &
\hspace{0.35in} $\slp{c} $ &
\hspace{0.35in} $\slp{c} $ &
\hspace{.1in} 4P \\ 
\hline
\hspace{0.1in}\raisebox{0ex}[5mm]{} $\slp{a} \slp{b}^* \rightarrow H^{+}H^{-}$ &
\hspace{.1in} $h,H,\gamma,Z $ &
\hspace{0.35in} $ $ &
\hspace{0.35in} $\sn{\ell} $ &
\hspace{.1in} 4P \\
\hline 
\hline
\hspace{0.1in}\raisebox{0ex}[5mm]{} 
        $\slp{a} \slp{b}^* \rightarrow \ell{\bar \ell}$ &
\hspace{.1in} $h,H,A,\gamma,Z $ &
\hspace{0.35in} $\neu{i} $ &
\hspace{0.35in} $ $  &
\hspace{.1in} $ $ \\ 
\hline 
\hspace{0.1in}\raisebox{0ex}[5mm]{}
        $\slp{a} \slp{b}^* \rightarrow q{\bar q}$ &
\hspace{.1in} $h,H,A,\gamma,Z $ &
\hspace{0.35in} $ $ &
\hspace{0.35in} $ $  &
\hspace{.1in} $ $ \\ 
\hline 
\hspace{0.1in}\raisebox{0ex}[5mm]{} 
      $\slp{a} \slp{b}^* \rightarrow \nu_{\ell} \bar{\nu}_{\ell}$ &
\hspace{.1in} $ Z $ &
\hspace{0.35in} $\cha{k} $ &
\hspace{0.35in} $ $  &
\hspace{.1in} $ $ \\ 
\hline 
\hspace{0.1in}\raisebox{0ex}[5mm]{}  $\slp{a} \slp{b}^{\prime *} 
                  \rightarrow \ell{\bar \ell}^{\prime}$ &
\hspace{.1in} $ $ &
\hspace{0.35in} $\neu{i} $ &
\hspace{0.35in} $ $  &
\hspace{.1in} $ $ \\ 
\hline 
\hspace{0.1in}\raisebox{0ex}[5mm]{} $\slp{a} \slp{b}^{\prime *} \rightarrow 
                                   \nu_\ell{\bar \nu}_{\ell^{\prime}}$ &
\hspace{.1in} $ $ &
\hspace{0.35in} $\cha{k} $ &
\hspace{0.35in} $ $  &
\hspace{.1in} $ $ \\ 
\hline \hline 
\hspace{0.1in}\raisebox{0ex}[5mm]{} $\slp{a} \slp{b} \rightarrow \ell\ell$ &
\hspace{.1in} $ $ &
\hspace{0.35in} $\neu{i} $ &
\hspace{0.35in} $\neu{i} $ &
\hspace{.1in} $ $ \\ 
\hline \hline 
\hspace{0.1in}\raisebox{0ex}[5mm]{} 
         $\slp{a} \slp{b}^{\prime} \rightarrow \ell\ell^{\prime}$ &
\hspace{.1in} $ $ &
\hspace{0.35in} $\neu{i} $ &
\hspace{0.35in} $ $ &
\hspace{.1in} $ $ \\ 
\hline \hline \hline
\hspace{0.1in}\raisebox{0ex}[5mm]{} $\slp{a} \chi \rightarrow Z\ell$ &
\hspace{.1in} $\ell $ &
\hspace{0.35in} $\slp{c} $ &
\hspace{0.35in} $\neu{i} $ &
\hspace{.1in} $ $ \\ 
\hline 
\hspace{0.1in}\raisebox{0ex}[5mm]{} $\slp{a} \chi \rightarrow \gamma \ell$ &
\hspace{.1in} $\ell $ &
\hspace{0.35in} $\slp{c} $ &
\hspace{0.35in} $ $ &
\hspace{.1in} $ $ \\ 
\hline
\hspace{0.1in}\raisebox{0ex}[5mm]{} $\slp{a} \chi \rightarrow W^{-}\nu_\ell$ &
\hspace{.1in} $\ell $ &
\hspace{0.35in} $\sn{\ell} $ &
\hspace{0.35in} $\cha{k} $ &
\hspace{.1in} $ $ \\ 
\hline \hline
\hspace{0.1in}\raisebox{0ex}[5mm]{} $\slp{a} \chi \rightarrow h \ell, H\ell$ &
\hspace{.1in} $\ell $ &
\hspace{0.35in} $\slp{c} $ &
\hspace{0.35in} $\neu{i} $ &
\hspace{.1in} $ $ \\ 
\hline 
\hspace{0.1in}\raisebox{0ex}[5mm]{} $\slp{a} \chi \rightarrow A \ell$ &
\hspace{.1in} $\ell $ &
\hspace{0.35in} $\slp{c} $ &
\hspace{0.35in} $\neu{i} $ &
\hspace{.1in} $ $ \\ 
\hline 
\hspace{0.1in}\raisebox{0ex}[5mm]{} $\slp{a} \chi \rightarrow H^{-}\nu_\ell$ &
\hspace{.1in} $\ell $ &
\hspace{0.35in} $\sn{\ell} $ &
\hspace{0.35in} $\cha{k} $ &
\hspace{.1in} $ $ \\ 
\hline 
\end{tabular}
\caption{A complete set of processes relevant for the 
neutralino--slepton coannihilation into 
tree--level two--body final states in the MSSM. 
`PI' denotes four--point (4P) interactions. 
The notation is as follows: 
$\slp{a}$ denotes $\widetilde{e}_{a}$, $\widetilde{\mu}_{a}$
and $\widetilde{\tau}_{a}$, where $a=1,2$ (and likewise $b$ and $c$) 
denotes a slepton mass state index for each generation. 
The symbols $\ell$ and $\ell^{\prime}$ represent charged leptons 
of different generations, and likewise for the sleptons. 
For the neutralino index $i=1,\ldots,4$ and for the chargino one $k=1,2$. 
Note that for each reaction with
a non--zero net electric charge there is a corresponding one with the 
the opposite net charge which is not included in the Table. For example,
in addition to $\slp{a} \slp{b} \rightarrow \ell\ell$ there is also
$\slp{a}^* \slp{b}^* \rightarrow {\bar\ell}{\bar\ell}$. 
}
}

%
\section{Exact Expressions}\label{exact:sec}
We now move on to present a full set of exact, analytic expressions
for the total cross sections 
for the neutralino--slepton coannihilation processes. We have
included all contributing diagrams as well as all interference terms
and kept finite widths of all $s$--channel resonances. 
In addition to neglecting CP violating phases in SUSY parameters we
have also assumed no mixing among different generations of leptons and
sleptons, although we have kept the left--right slepton mixing within
each generation. The results presented here are meant to supplement
and extend the ones presented in our previous paper~\cite{nrr2} for
the case of the neutralino--pair annihilation. 

In the presence of coannihilation the formalism used in~\cite{nrr2}
must be generalized to include different initial 
states. First, following~\cite{eg97,efos00,swo88} we introduce a
Lorentz--invariant function $w_{ij}(s)$
\beq
w_{ij}(s)= \frac{1}{4} \int d\, {\rm LIPS}\, |{\cal A} (ij
\rightarrow {\rm all})|^2
\label{wdef:eq}
\eeq
where $|{\cal A} (ij \rightarrow {\rm all})|^2$
denotes the absolute square of the reduced matrix
element for the annihilation of two initial particles $ij$ into all allowed
final states, averaged over
initial spins and summed over final spins. The function $w_{ij}(s)$ is
related to the coannihilation cross section $\sigma_{ij}(s)$ in
eq.~(\ref{thermal-average:eq}) via~\cite{eg97,efo98}
\beq
w_{ij}(s)=  \frac{s}{2} \beta_f(s,m_i,m_j)\, \sigma_{ij}(s).
\label{wtosigma:eq}
\eeq
The expression~(\ref{wtosigma:eq}) readily reduces to a more familiar
form in the case of single--particle annihilation. (Compare, \eg,
eq.~(4.2) of~\cite{nrr2}.)

Since $w_{ij}(s)$ receives contributions from all the kinematically allowed
annihilation process $ij\rightarrow f_1 f_2$, it can be written as
\begin{eqnarray}
w_{ij}(s)&=&\frac{1}{32\,\pi}\sum_{f_1 f_2} \bigg[
c \, \theta\left(s-(m_{f_1}+m_{f_2})^2 \right)\,
\beta_f(s,m_{f_1},m_{f_2})\,\widetilde{w}_{ij\rightarrow f_1f_2}(s)\bigg],
\label{wtowtilde:eq}
\end{eqnarray}
where
the summation extends over all possible two--body final states
$f_1f_2$, $m_{f_1}$ and $m_{f_2}$ denote their respective masses, and
\begin{eqnarray}
\label{color:eq}
c &=& \left\{ \begin{array}{ll}
c_{f}  & \mbox{if $f_{1 (2)}=f(\bar f)$} \\
 1     & \mbox{otherwise,}  \,
\end{array}
      \right.
\end{eqnarray}
where $c_{f}$ is the color factor of SM fermions
($c_{f}=3$ for quarks and $c_{f}=1$ for leptons).

Since $w_{ij}$ are Lorentz--invariant functions, we choose, for convenience, 
the CM frame in which the function 
$\widetilde{w}_{ij\rightarrow f_1 f_2}(s)$ can be expressed as
\begin{eqnarray}
\widetilde{w}_{ij\rightarrow f_1 f_2}(s)= 
\frac{1}{2}\int_{-1}^{+1} \!d \cos \theta_{CM} 
\,|{\cal A}(ij \rightarrow f_1 f_2)|^2, \label{wtildedef:eq}
\end{eqnarray}
where $\theta_{CM}$ denotes the scattering angle in the CM frame. In other
words, we write
$|{\cal A}(ij \rightarrow f_1 f_2)|^2$ as a function of
$s$ and $\cos \theta_{CM}$, which greatly simplifies the computation.

We will follow Table~\ref{tableone} in presenting explicit expressions
for $\widetilde{w}_{ij\rightarrow f_1 f_2}(s)$ for all the two--body final
states. All the couplings are defined in Appendix~A. All other
auxiliary functions, are listed in Appendix~B. Some symbols are
obvious (\eg, $m_W$ is the mass of the $W$--boson) and will not be
defined here.

\vspace{0.3in}
We begin by presenting the results for $\slp{a}\slp{b}^*\rightarrow$
vector boson pairs: $WW$, $ZZ$, $Z\gamma$ and $\gamma\gamma$.

\subsection{\boldmath ${\slp{a}\slp{b}^*\rightarrow WW}$}
\label{slslww:sec}
This process involves 
the $s$--channel CP--even Higgs boson ($h$ and $H$) exchange, 
the four--point interaction, 
the $s$--channel $Z$--boson and photon exchange, 
and the $u$--channel sneutrino ($\sn{\ell}$) exchange 
\begin{eqnarray}
 \wt_{\slp{a}\slp{b}^*\rightarrow WW}=
  \wt_{WW}^{(h,H,P)} + \wt_{WW}^{(Z,\gm)} + \wt_{WW}^{(\sn{})} 
+ \wt_{WW}^{(h,H,P-\sn{})} +  \wt_{WW}^{(Z,\gm-\sn{})}:
\end{eqnarray}
%
$\bullet$
 \underline{Higgs ($h,H$) exchange ($+$ Point interaction):}
\begin{eqnarray}
\wt_{WW}^{(h,H,P)} & = & 
\left|    
\sum_{r=h,H} \frac{C^{WWr}C^{\slp{b}^*\slp{a}r}}{\prop{r}}
-C^{\slp{b}\slp{a}WW}
\right|^2
\frac{s^2-4 \mw^2 s+12\mw^4}{4\mw^4};
\end{eqnarray}
%
$\bullet$
 \underline{$Z,\gm$ exchange:}
\begin{eqnarray}
\wt_{WW}^{(Z,\gm)} & = & 
\left| \frac{C^{WWZ}C^{\slp{b}^*\slp{a}Z}}{\prop{Z}}
+ \frac{e^2\delta_{ab}}{s} \right|^2
\{ s-(\slpsl{})^2 \} \nonumber \\
 & & \times \{ s-(\slmsl{})^2 \} 
        \cdot \frac{s^3+16 \mw^2 s^2-68\mw^4 s -48\mw^6}{12s\mw^4};
\end{eqnarray}
%
$\bullet$
 \underline{sneutrino ($\sn{\ell}$) exchange:}
\begin{eqnarray}
\wt_{WW}^{(\sn{})} & = & 
\frac{1}{\mw^4}
\left| C^{\slp{b}^*\sn{\ell}W^-} C^{\sn{\ell}^*\slp{a}W^+} \right|^2 \nonumber \\
 & & \times \Big[
\T_4 - 2(\slpsl{2}+2\mw^2) \T_3 \nonumber \\
 & & +\Big\{ \slpsl{4} +4\slsl{2} + 2\mw^2(\slpsl{2}+3\mw^2) \Big\}\T_2
 \nonumber \\
 & & -2 \Big\{ (\slpsl{2})(\slsl{2}-\mw^4) \nonumber \\
 & & + \mw^2(\slpsl{4} -4\slsl{2} +2\mw^4) \Big\}\T_1 \nonumber \\
 & & +(\msli^2-\mw^2)^2(\mslj^2-\mw^2)^2\T_0 \Big],
\end{eqnarray}
where the auxiliary functions $\T_i$ ($i=0,1,2,3,4$) are
defined in Appendix~B. The arguments for these functions 
in the above should be understood as $\T_i$ $=$ 
$\T_i(s$, $\msli^2$, $\mslj^2$, $\mw^2$, $\mw^2$, $\msn^2$, $\msn^2)$. 
Throughout the text, the second and the third arguments denote
the masses for the two initial particles. The fourth and the fifth
ones correspond to the masses for the two final particles. The last
two arguments are the masses for the exchanged particles in the 
relevant $t$-- and/or $u$--channel diagrams. Assuming this convention,
we will omit the arguments for $\T_i$ in the following. 
\vspace{3mm} \\
%
$\bullet$
 \underline{Higgs ($h,H$) ($+$ Point) -- sneutrino ($\sn{\ell}$) interference:}
\begin{eqnarray}
\wt_{WW}^{(h,H,P-\sn{})} & = & 
\frac{1}{2\mw^4} \Re 
\left[ \left( \sum_{r=h,H} 
\frac{C^{WWr}C^{\slp{b}^*\slp{a}r}}{\prop{r}}-C^{\slp{b}^*\slp{a}WW}
\right)^* C^{\slp{b}^*\sn{\ell}W^-} C^{\sn{\ell}^*\slp{a}W^+} \right]
\nonumber \\
 & & \times \Big[ 
s^2 + s (\slpsl{2}-2\msn^2-4\mw^2) 
    + 2\mw^2(\slpsl{2}-2\msn^2+2\mw^2) \nonumber \\
 & & -2 [ s(\msli^2-\msn^2-\mw^2)(\mslj^2-\msn^2-\mw^2) \nonumber \\
 & & +2\mw^2 \{ \slpsl{4}-\slsl{2} 
     -(\msn^2+\mw^2)(\slpsl{2})+(\msn^2-\mw^2)^2 \} ]\F \Big], \nonumber \\
 & & 
\end{eqnarray}
where the auxiliary function $\F$ is
defined in Appendix~B. The arguments for these functions 
in the above should be understood as 
$\F$ $=$ $\F(s$, $\msli^2$, $\mslj^2$, $\mw^2$, $\mw^2$, $\msn^2)$. 
Throughout the text, the second and the third arguments of $\F$
denote the masses for the two initial particles. The fourth and the fifth
ones correspond to the masses for the two final particles. The last
argument is the mass for the exchanged particles in the 
relevant $t$-- and/or $u$--channel diagrams. With this understanding,
we will omit the arguments for $\F$ in the following. 
\vspace{3mm} \\
%
$\bullet$
\underline{$Z,\gm$ -- sneutrino ($\sn{\ell}$) interference:}
\begin{eqnarray}
\wt_{WW}^{(Z,\gm-\sn{})} & = & 
\frac{1}{6\mw^4} \Re 
\left[ \left(  
\frac{C^{WWZ}C^{\slp{b}^*\slp{a}Z}}{\prop{Z}}
+ \frac{e^2\delta_{ab}}{s}
\right)^* C^{\slp{b}^*\sn{\ell}W^-} C^{\sn{\ell}^*\slp{a}W^+} \right]
\nonumber \\
 & & \times \left[ I_1^{WW} + 6 I_2^{WW} \F \right],
\end{eqnarray}
where
\begin{eqnarray}
I_1^{WW} & = &  
-s^3 + 2s^2 (\slpsl{2}-9\mw^2) \nonumber \\
 & & + s\{ -\msli^4-\mslj^4-10\slsl{2}+12(\msn^2+2\mw^2)(\slpsl{2})
                              \nonumber \\
 & & +4(-3\msn^4+6\mw^2\msn^2+7\mw^4) \} \nonumber \\
 & & +2\mw^2 \{ -9(\slpsl{4})+30\slsl{2} \nonumber \\
 & & -4(3\msn^2+\mw^2)(\slpsl{2}) +12(\msn^2-\mw^2)^2 \} \nonumber \\
 & & -\frac{8}{s}\mw^4(\slmsl{2})^2, \\
I_2^{WW} & = & 
-s^2(\msli^2-\msn^2-\mw^2)(\mslj^2-\msn^2-\mw^2) \nonumber \\
 & & + s \{ \slsl{2}(\slpsl{2})-(\msn^2+3\mw^2)(\slpsl{4}) \nonumber \\
 & & -2(2\msn^2+\mw^2)\slsl{2} + (3\msn^4+5\mw^4)(\slpsl{2}) \nonumber \\
 & & -2(\msn^6-4\mw^2\msn^4+\mw^4\msn^2+2\mw^6) \} \nonumber \\
 & & +2\mw^2 \{ (\slpsl{2})(\slpsl{4}-3\slsl{2}) \nonumber \\
 & & -(\msn^2+\mw^2)(\slpsl{4}) + 4\slsl{2}(2\msn^2+\mw^2) \nonumber \\
 & & -(\slpsl{2})(\msn^2-\mw^2)(3\msn^2+\mw^2) + 2(\msn^2-\mw^2)^3 \}.
\nonumber \\
 & & 
\end{eqnarray}
%

\subsection{\boldmath ${\slp{a}\slp{b}^*\rightarrow ZZ}$}
\label{slslzz:sec}
This process proceeds via 
the $s$--channel CP--even Higgs boson ($h$ and $H$) exchange, 
the four--point interaction, 
and the $t$-- and $u$--channel slepton ($\slp{a}$, $a=1,2$) exchange 
\begin{eqnarray}
 \wt_{\slp{a}\slp{b}^*\rightarrow ZZ}=
 \wt_{ZZ}^{(h,H,P)} + \wt_{ZZ}^{(\slp{})} + \wt_{ZZ}^{(h,H,P-\slp{})}:
\end{eqnarray}
%
$\bullet$
 \underline{Higgs ($h,H$) exchange ($+$ Point interaction):}
\begin{eqnarray}
\wt_{ZZ}^{(h,H,P)} & = & 
\left|    
\sum_{r=h,H} \frac{C^{ZZr}C^{\slp{b}^*\slp{a}r}}{\prop{r}}
-C^{\slp{b}^*\slp{a}ZZ}
\right|^2
\frac{s^2-4 \mz^2 s+12\mz^4}{8\mz^4};
\end{eqnarray}
%
$\bullet$
 \underline{slepton\ ($\slp{c}$) exchange:}
\begin{eqnarray}
\wt_{ZZ}^{(\slp{})} & = & 
\frac{1}{\mz^4}\sum_{c,d=1}^{2}
C^{\slp{b}^*\slp{c}Z} C^{\slp{c}^*\slp{a}Z} 
C^{\slp{b}^*\slp{d}Z *} C^{\slp{d}^*\slp{a}Z *} \nonumber \\
 & & \times \left[
       \T_4 - 2(\slpsl{2}+2\mz^2) \T_3 \right. \nonumber \\
 & & +[ \slpsl{4} +4\slsl{2} + 2\mz^2(\slpsl{2}+3\mz^2) ]\T_2 \nonumber \\
 & & -2[ (\slpsl{2})(\slsl{2}-\mz^4) 
         + \mz^2(\slpsl{4} -4\slsl{2} +2\mz^4) ]\T_1 \nonumber \\
 & & +(\msli^2-\mz^2)^2(\mslj^2-\mz^2)^2\T_0 \nonumber \\
  & & -\Y_4
      +[ s(\slpsl{2}-2\mz^2)
               -2(\msli^2-\mz^2)(\mslj^2-\mz^2)  ]\Y_2 \nonumber \\
 & & -[ s^2(\msli^2-\mz^2)(\mslj^2-\mz^2) \nonumber \\
 & & 
  +s\{ -\slsl{2}(\slpsl{2})+3\mz^2(\slpsl{4})-3\mz^4(\slpsl{2})+2\mz^6 \}
                   \nonumber \\
 & & \left.
 + (\msli^2-\mz^2)^2(\mslj^2-\mz^2)^2
]\Y_0 
\right];
\end{eqnarray}
%
$\bullet$
 \underline{Higgs ($h,H$) ($+$ Point) -- slepton\ ($\slp{c}$) interference:}
\begin{eqnarray}
\wt_{ZZ}^{(h,H,P-\slp{})} & = & 
\frac{1}{2\mz^4}\sum_{c=1}^2 \Re 
\left[ \left( \sum_{r=h,H}
\frac{C^{ZZr}C^{\slp{b}^*\slp{a}r}}{\prop{r}}-C^{\slp{b}^*\slp{a}ZZ}
\right)^* C^{\slp{b}^*\slp{c}Z} C^{\slp{c}^*\slp{a}Z} \right]
\nonumber \\
 & & \times \left[ 
s^2 + s (\slpsl{2}-2\mslk^2-4\mz^2) 
+ 2\mz^2(\slpsl{2}-2\mslk^2+2\mz^2) \right. \nonumber \\
 & & 
-2 [ s(\msli^2-\mslk^2-\mz^2)(\mslj^2-\mslk^2-\mz^2) \nonumber \\
 & & 
+2\mz^2 \{ \slpsl{4}-\slsl{2}
+\mslk^2(\mslk^2-\msli^2-\mslj^2-2\mz^2) \nonumber \\
 & & 
\left. -\mz^2 (\slpsl{2}-\mz^2) \} ]\F \right]. 
\end{eqnarray}
%

\subsection{\boldmath ${\slp{a}\slp{b}^*\rightarrow Z\gamma}$}
\label{slslzgm:sec}
This process proceeds via 
the four--point interaction, 
and the $t$-- and $u$--channel slepton ($\slp{a}$, $a=1,2$) exchange 
\begin{eqnarray}
 \wt_{\slp{a}\slp{b}^*\rightarrow Z\gm} = \wt_{Z\gm}^{(P)} 
             + \wt_{Z\gm}^{(\slp{})} + \wt_{Z\gm}^{(P-\slp{})}:
\end{eqnarray}
%
$\bullet$
 \underline{Point interaction:}
\begin{eqnarray}
\wt_{Z\gm}^{(P)} & = & 
3 \left| C^{\slp{b}^*\slp{a}Z\gm} \right|^2;
\end{eqnarray}
%
$\bullet$
 \underline{slepton\ ($\slp{c}$) exchange:}
\begin{eqnarray}
\wt_{Z\gm}^{(\slp{})} & = & 
\frac{2}{\mz^2}\sum_{c,d=1}^{2}
C^{\slp{c}^*\slp{a}Z} C^{\slp{b}^*\slp{c}\gm} 
C^{\slp{d}^*\slp{a}Z *} C^{\slp{b}^*\slp{d}\gm *} \nonumber \\
 & & \times \left[
     - \Tt_3 + (2\msli^2-\mslj^2+2\mz^2) \Tt_2 \right. \nonumber \\
 & & \left. -\{(\msli^2-\mz^2)^2 - 2\mslj^2(\msli^2+\mz^2) \}\Tt_1 
      -\mslj^2(\msli^2-\mz^2)^2 \Tt_0 \right] \nonumber \\
& & + 
\frac{2}{\mz^2}\sum_{c,d=1}^{2}
C^{\slp{c}^*\slp{a}\gm} C^{\slp{b}^*\slp{c}Z} 
C^{\slp{d}^*\slp{a}\gm *} C^{\slp{b}^*\slp{d}Z *} \nonumber \\
 & & \times \left[
     - \Tu_3 + (2\mslj^2-\msli^2+2\mz^2) \Tu_2 \right. \nonumber \\
 & & \left. -\{(\mslj^2-\mz^2)^2 - 2\msli^2(\mslj^2+\mz^2) \}\Tu_1 
      -\msli^2(\mslj^2-\mz^2)^2 \Tu_0 \right] \nonumber \\
 & & + \frac{1}{\mz^2} \Re \sum_{c,d=1}^{2}
C^{\slp{c}^*\slp{a}Z} C^{\slp{b}^*\slp{c}\gm} 
C^{\slp{d}^*\slp{a}\gm *} C^{\slp{b}^*\slp{d}Z *}
(s-2\msli^2-2\mslj^2+\mz^2) \nonumber \\
 & & \times 
      \left[ -2\Y_2 
             +\frac{1}{s}(s+\mz^2)(\slmsl{2}) \Y_1 \right. \nonumber \\
 & & +[ s(\slpsl{2}-2\mz^2)-(\slpsl{4})+3\mz^2(\slpsl{2}) \nonumber \\
 & & \left. -\frac{1}{s}\mz^2(\slmsl{2})^2 
          - \frac{1}{2s^2}\mz^4(\slmsl{2})^2 ]\Y_0 
\right];
\end{eqnarray}
%
$\bullet$
 \underline{Point -- slepton\ ($\slp{c}$) interference:}
\begin{eqnarray}
\wt_{Z\gm}^{(P-\slp{})} & = & 
\frac{1}{\mz^2} \Re  \sum_{c=1}^2 
C^{\slp{b}^*\slp{a}Z\gm *} 
C^{\slp{c}^*\slp{a}Z} C^{\slp{b}^*\slp{c}\gm} \nonumber \\
 & & \times \Big[ 
-3\msli^2+\mslj^2+2\mslk^2-2\mz^2 - \frac{1}{s}\mz^2(\slmsl{2})
        \nonumber \\
 & & + [ s(\mslk^2-\msli^2+\mz^2) +(2\msli^2-\mz^2)(\msli^2-\mz^2) 
\nonumber \\
 & &         + \mslk^2(2\mslk^2-4\msli^2-3\mz^2)-2\mz^2\mslj^2
      ] \Ft \Big] \nonumber \\
 & & +\frac{1}{\mz^2} \Re  \sum_{c=1}^2 
C^{\slp{b}^*\slp{a}Z\gm *} 
C^{\slp{c}^*\slp{a}\gm} C^{\slp{b}^*\slp{c}Z} \nonumber \\
 & & \times \Big[ 
-3\mslj^2+\msli^2+2\mslk^2-2\mz^2 + \frac{1}{s}\mz^2(\slmsl{2})
        \nonumber \\
 & & + [ s(\mslk^2-\mslj^2+\mz^2) +(2\mslj^2-\mz^2)(\mslj^2-\mz^2) 
\nonumber \\
 & &         + \mslk^2(2\mslk^2-4\mslj^2-3\mz^2)-2\mz^2\msli^2
      ] \Fu \Big]. 
\end{eqnarray}

\subsection{\boldmath ${\slp{a}\slp{b}^*\rightarrow \gm\gm}$}
\label{slslgmgm:sec}
This process proceeds via 
the four--point interaction, 
and the $t$-- and $u$--channel slepton ($\slp{a}$, $a=1,2$) exchange 
\begin{eqnarray}
 \wt_{\slp{a}\slp{b}^*\rightarrow \gm\gm} = \wt_{\gm\gm}^{(P)} 
             + \wt_{\gm\gm}^{(\slp{})} + \wt_{\gm\gm}^{(P-\slp{})}:
\end{eqnarray}
%
$\bullet$
 \underline{Point interaction:}
\begin{eqnarray}
\wt_{\gm\gm}^{(P)} & = & 8 e^4 \delta_{ab};
\end{eqnarray}
%
$\bullet$
 \underline{slepton\ ($\slp{a}$) exchange:}
\begin{eqnarray}
\wt_{\gm\gm}^{(\slp{})} & = & 
e^4 \delta_{ab} \left[  
 4(\T_2 + 2 \msli^2 \T_1 + \msli^4 \T_0) - (s-4\msli^2)^2 \Y_0 
\right];
\end{eqnarray}
%
$\bullet$
 \underline{Point -- slepton\ ($\slp{c}$) interference:}
\begin{eqnarray}
\wt_{\gm\gm}^{(P-\slp{})} & = & 
2 e^4 \delta_{ab} 
\left[ - 4 + (s-8\msli^2)\F \right]. 
\end{eqnarray}

\vspace{0.3in}
Next, there are processes of the type $\slp{a}\slp{b}^*\rightarrow$ 
vector boson -- Higgs boson: $W^{\pm}H^{\mp}$, $Zh$, $ZH$,
$ZA$, $\gamma h$, $\gamma H$ and $\gamma A$.


\subsection{\boldmath ${\slp{a}\slp{b}^*\rightarrow W^{\pm}H^{\mp}}$}
\label{slslwh:sec}
The process $\slp{a}\slp{b}^*\rightarrow W^+H^-$ involves 
the $s$--channel CP--even Higgs boson ($h$ and $H$)
            and CP--odd Higgs boson ($A$) exchange, 
and the $u$--channel sneutrino ($\sn{\ell}$) exchange 
\begin{eqnarray}
 \wt_{\slp{a}\slp{b}^*\rightarrow W^+H^-} = \wt_{W^+H^-}^{(h,H,A)} 
             + \wt_{W^+H^-}^{(\sn{})} + \wt_{W^+H^-}^{(h,H,A-\sn{})}:
\end{eqnarray}
%
$\bullet$
 \underline{Higgs ($h,H,A$) exchange:}
\begin{eqnarray}
\wt_{W^+H^-}^{(h,H,A)} & = & 
\left|    
\sum_{r=h,H,A} \frac{C^{W^-H^+r}C^{\slp{b}^*\slp{a}r}}{\prop{r}}
\right|^2
\frac{s^2-2s(\mch^2+\mw^2)+(\mch^2-\mw^2)^2}{\mw^2}; \nonumber \\
 & & 
\end{eqnarray}
%
$\bullet$
 \underline{sneutrino\ ($\sn{\ell}$) exchange:}
\begin{eqnarray}
\wt_{W^+H^-}^{(\sn{})} & = & 
\frac{1}{\mw^2}
\left| C^{\sn{\ell}^*\slp{a}W^+} C^{\slp{b}^*\sn{\ell}H^-} \right|^2 \nonumber \\
 & & \times
\left[ \Tu_2 -2 (\mslj^2+\mw^2) \Tu_1 + (\mslj^2-\mw^2)^2 \Tu_0 \right];
\end{eqnarray}
%
$\bullet$
 \underline{Higgs ($h,H,A$) -- sneutrino\ ($\sn{\ell}$) interference:}
\begin{eqnarray}
\wt_{W^+H^-}^{(h,H,A-\sn{})} & = & 
- \frac{2}{\mw^2}\Re 
\left[ \left( \sum_{r=h,H,A}
\frac{C^{W^-H^+r}C^{\slp{b}^*\slp{a}r}}{\prop{r}}\right)^* 
C^{\sn{\ell}^*\slp{a}W^+} C^{\slp{b}^*\sn{\ell}H^-} \right]
\nonumber \\
 & & \times \left[ s - \mch^2 + \mw^2 
        + \{ s(\msn^2+\mw^2-\mslj^2) \right. \nonumber \\
 & & + \mw^2 (\mslj^2-2\msli^2+\msn^2-\mw^2) \nonumber \\
 & & \left. + \mch^2 (\mslj^2-\msn^2+\mw^2) \} \Fu \right]. 
\end{eqnarray}
The contribution from $\slp{a}\slp{b}^*\rightarrow W^-H^+$ can be 
obtained by interchanging the indices $a$ and $b$, but this is
obviously equal to the one above:
\begin{eqnarray}
 \wt_{\slp{a}\slp{b}^*\rightarrow W^-H^+} =
 \wt_{\slp{b}\slp{a}^*\rightarrow W^+H^-}. 
\end{eqnarray}
%

\subsection{\boldmath ${\slp{a}\slp{b}^*\rightarrow Zh, ZH}$}
\label{slslzh:sec}
The process ${\slp{a}\slp{b}^*\rightarrow Zh}$ proceeds via 
the $s$--channel CP--odd Higgs boson ($A$) exchange, 
the $s$--channel $Z$--boson exchange, 
and the $t$-- and $u$--channel slepton ($\slp{a}$, $a=1,2$) exchange 
\begin{eqnarray}
 \wt_{\slp{a}\slp{b}^*\rightarrow Zh}=
  \wt_{Zh}^{(A)} + \wt_{Zh}^{(Z)} + \wt_{Zh}^{(\slp{})} 
+ \wt_{Zh}^{(A-Z)} + \wt_{Zh}^{(A-\slp{})} + \wt_{Zh}^{(Z-\slp{})}:
\end{eqnarray}
%
$\bullet$
 \underline{$A$ exchange:}
\begin{eqnarray}
\wt_{Zh}^{(A)} & = & 
\left| \frac{C^{ZhA}C^{\slp{b}^*\slp{a}A}}{\prop{A}} \right|^2
\frac{s^2-2s(\mh^2+\mz^2)+(\mh^2-\mz^2)^2}{\mz^2};
\end{eqnarray}
%
$\bullet$
 \underline{$Z$ exchange:}
\begin{eqnarray}
\wt_{Zh}^{(Z)} & = & 
\frac{1}{12\mz^6}
\left| \frac{C^{ZZh}C^{\slp{b}^*\slp{a}Z}}{\prop{Z}} \right|^2
\nonumber \\
 & & \times \Bigg[ s^2 \Big\{ 3(\slmsl{2})^2+\mz^4 \Big\} \nonumber \\
 & & -2s \Big\{ 3(\slmsl{2})^2(\mh^2+2\mz^2) 
             + \mz^4(\slpsl{2}+\mh^2-5\mz^2) \Big\} \nonumber \\
 & & +\mz^4 \Big\{ (\mh^2-\mz^2)^2 +4(\slpsl{2})(\mh^2-5\mz^2) \Big\}
 \nonumber \\
 & & +(\slmsl{2})^2(3\mh^4+6\mz^2\mh^2+19\mz^4) \nonumber \\
 & & - \frac{2}{s}\mz^2 \Big\{ (\slmsl{2})^2(3\mh^4-2\mz^2\mh^2+\mz^4)
      +\mz^2(\mh^2-\mz^2)^2(\slpsl{2}) \Big\}
     \nonumber \\
 & &     + \frac{4}{s^2}\mz^4(\mh^2-\mz^2)^2(\slmsl{2})^2 \Bigg];
\end{eqnarray}
%
$\bullet$
 \underline{slepton\ ($\slp{c}$) exchange:}
\begin{eqnarray}
\wt_{Zh}^{(\slp{})} & = & 
\frac{1}{\mz^2}\sum_{c,d=1}^{2}
C^{\slp{c}^*\slp{a}Z} C^{\slp{b}^*\slp{c}h} 
C^{\slp{d}^*\slp{a}Z *} C^{\slp{b}^*\slp{d}h *} \nonumber \\
 & & \times \left[ \Tt_2 - 2(\msli^2+\mz^2) \Tt_1 
                  +(\msli^2-\mz^2)^2 \Tt_0 \right] \nonumber \\
& & + 
\frac{1}{\mz^2}\sum_{c,d=1}^{2}
C^{\slp{c}^*\slp{a}h} C^{\slp{b}^*\slp{c}Z} 
C^{\slp{d}^*\slp{a}h *} C^{\slp{b}^*\slp{d}Z *} \nonumber \\
 & & \times \left[ \Tu_2 - 2(\mslj^2+\mz^2) \Tu_1 
                  +(\mslj^2-\mz^2)^2 \Tu_0 \right] \nonumber \\
 & & + 
\frac{1}{\mz^2} \Re \sum_{c,d=1}^{2}
C^{\slp{c}^*\slp{a}Z} C^{\slp{b}^*\slp{c}h} 
C^{\slp{d}^*\slp{a}h *} C^{\slp{b}^*\slp{d}Z *} \nonumber \\
 & & \times \left[ -2\Y_2 
+\frac{1}{s}(s-\mh^2+\mz^2)(\slmsl{2}) \Y_1 \right. \nonumber \\
 & & + \Big\{ s(\slpsl{2}-2\mz^2)-(\slpsl{4}) -2\mz^2\mh^2 \nonumber \\
 & &    +(3\mz^2-\mh^2)(\slpsl{2}) 
        +\frac{1}{s}(\mh^2-\mz^2)(\slmsl{2})^2 \nonumber \\
 & & \left. - \frac{1}{2s^2}(\mh^2-\mz^2)^2(\slmsl{2})^2 \Big\} \Y_0 \right];
\end{eqnarray}
%
$\bullet$
 \underline{$A$ -- $Z$ interference:}
\begin{eqnarray}
\wt_{Zh}^{(A-Z)} & = & 
- \frac{1}{s\mz^4}\Re 
\left[ \left( \frac{C^{ZhA}C^{\slp{b}^*\slp{a}A}}{\prop{A}} \right)^* 
\left( \frac{C^{ZZh}C^{\slp{b}^*\slp{a}Z}}{\prop{Z}} \right) \right]
\nonumber \\
 & & \times (\slmsl{2})(s-\mz^2)
\left[ s^2 - 2s(\mh^2+\mz^2)+(\mh^2-\mz^2)^2 \right]; \nonumber \\
 & & 
\end{eqnarray}
%
$\bullet$
 \underline{Higgs ($A$) -- slepton\ ($\slp{c}$) interference:}
\begin{eqnarray}
\wt_{Zh}^{(A-\slp{})} & = & 
- \frac{2}{\mz^2}\Re \sum_{c=1}^{2}
\left( \frac{C^{ZhA}C^{\slp{b}^*\slp{a}A}}{\prop{A}} \right)^* 
\nonumber \\
 & & \times \left[ 
C^{\slp{c}^*\slp{a}Z} C^{\slp{b}^*\slp{c}h}
\left\{ -(s - \mh^2 + \mz^2 ) + [ s(\msli^2-\mslk^2-\mz^2)
\right. \right. \nonumber \\
 & & \left. -\msli^2(\mh^2+\mz^2) 
+ (\mslk^2-\mz^2)(\mh^2-\mz^2) + 2\mz^2\mslj^2 ] \Ft 
                                                \right\} \nonumber \\
 & & -
C^{\slp{c}^*\slp{a}h} C^{\slp{b}^*\slp{c}Z}
\left\{ -(s - \mh^2 + \mz^2 ) + [ s(\mslj^2-\mslk^2-\mz^2)
\right. \nonumber \\
 & & \left. \left. -\mslj^2(\mh^2+\mz^2) 
+ (\mslk^2-\mz^2)(\mh^2-\mz^2) + 2\mz^2\msli^2 ] \Fu
                                          \right\} \right]; \nonumber \\
 & &
\end{eqnarray}
%
$\bullet$
 \underline{$Z$ -- slepton\ ($\slp{c}$) interference:}
\begin{eqnarray}
\wt_{Zh}^{(Z-\slp{})} & = & 
\frac{1}{\mz^4}\Re \sum_{c=1}^{2}
\left( \frac{C^{ZZh}C^{\slp{b}^*\slp{a}Z}}{\prop{Z}} \right)^* 
\nonumber \\
 & & \times \Bigg[ 
C^{\slp{c}^*\slp{a}Z} C^{\slp{b}^*\slp{c}h}
\Big\{ -(s - \mh^2)(\slmsl{2}) \nonumber \\
 & & +2\mz^2(\msli^2-\mslk^2+\mz^2) 
     -\frac{1}{s}\mz^2(\mh^2-\mz^2)(\slmsl{2}) \nonumber \\
 & & + \Big[ s(\slmsl{2}+\mz^2)(\msli^2-\mslk^2-\mz^2) \nonumber \\
 & & +\mz^2 \{ -2\mslk^4+\mslk^2(3\msli^2+\mslj^2+3\mz^2)-3\msli^4
                                                   \nonumber \\
 & & +3\slsl{2}-2\mslj^4+\mz^2(4\msli^2+\mslj^2)-\mz^4 \} \nonumber \\
 & &     -\mh^2(\slmsl{2}+\mz^2)(\msli^2-\mslk^2+\mz^2) \Big] \Ft \Big\} 
     \nonumber \\
 & & + C^{\slp{c}^*\slp{a}h} C^{\slp{b}^*\slp{c}Z}
\Big\{ (s - \mh^2)(\slmsl{2}) \nonumber \\
 & & +2\mz^2(\mslj^2-\mslk^2+\mz^2) 
     +\frac{1}{s}\mz^2(\mh^2-\mz^2)(\slmsl{2}) \nonumber \\
 & & + \Big[ -s(\slmsl{2}-\mz^2)(\mslj^2-\mslk^2-\mz^2) \nonumber \\
 & & +\mz^2 \{ -2\mslk^4+\mslk^2(3\mslj^2+\msli^2+3\mz^2)-3\mslj^4
                                                   \nonumber \\
 & & +3\slsl{2}-2\msli^4+\mz^2(4\mslj^2+\msli^2)-\mz^4 \} \nonumber \\
 & &     +\mh^2(\slmsl{2}-\mz^2)(\mslj^2-\mslk^2+\mz^2) \Big] \Fu \Big\} 
 \Bigg]. 
\end{eqnarray}
The expression for $\wt_{\slp{a}\slp{b}^*\rightarrow ZH}$ can be 
obtained by a simple substitution $h\rightarrow H$.
%

\subsection{\boldmath ${\slp{a}\slp{b}^*\rightarrow ZA}$}
\label{slslza:sec}
This process proceeds via 
the $s$--channel CP--even Higgs boson ($h$ and $H$) exchange, 
and the $t$-- and $u$--channel slepton ($\slp{a}$, $a=1,2$) exchange 
\begin{eqnarray}
 \wt_{\slp{a}\slp{b}^*\rightarrow ZA}=
  \wt_{ZA}^{(h,H)} + \wt_{ZA}^{(\slp{})} + \wt_{ZA}^{(h,H-\slp{})}:
\end{eqnarray}
%
$\bullet$
 \underline{Higgs ($h,H$) exchange:}
\begin{eqnarray}
\wt_{ZA}^{(h,H)} & = & 
\left|\sum_{r=h,H}  \frac{C^{ZAr}C^{\slp{b}^*\slp{a}r}}{\prop{r}} \right|^2
\frac{s^2-2s(\ma^2+\mz^2)+(\ma^2-\mz^2)^2}{\mz^2};
\end{eqnarray}
The expression for $\wt_{ZA}^{(\slp{})}$
can be obtained from 
$\wt_{Zh}^{(\slp{})}$ by a simple substitution $h\rightarrow A$. \\
$\bullet$
 \underline{Higgs ($h,H$) -- slepton\ ($\slp{c}$) interference:}
\begin{eqnarray}
\wt_{ZA}^{(h,H-\slp{})} & = & 
 \frac{2}{\mz^2}\Re \sum_{c=1}^{2}
\left(\sum_{r=h,H} \frac{C^{ZAr}C^{\slp{b}^*\slp{a}r}}{\prop{r}} \right)^* 
\nonumber \\
 & & \times \left[ 
C^{\slp{c}^*\slp{a}Z} C^{\slp{b}^*\slp{c}A}
\left\{ -(s - \ma^2 + \mz^2 ) + [ s(\msli^2-\mslk^2-\mz^2)
\right. \right. \nonumber \\
 & & \left. -\msli^2(\ma^2+\mz^2) 
+ (\mslk^2-\mz^2)(\ma^2-\mz^2) + 2\mz^2\mslj^2 ] \Ft 
                                                \right\} \nonumber \\
 & & -
C^{\slp{c}^*\slp{a}A} C^{\slp{b}^*\slp{c}Z}
\left\{ -(s - \ma^2 + \mz^2 ) + [ s(\mslj^2-\mslk^2-\mz^2)
\right. \nonumber \\
 & & \left. \left. -\mslj^2(\ma^2+\mz^2) 
+ (\mslk^2-\mz^2)(\ma^2-\mz^2) + 2\mz^2\msli^2 ] \Fu
                                          \right\} \right].
\end{eqnarray}


\subsection{\boldmath ${\slp{a}\slp{b}^*\rightarrow \gm h, \gm H}$}
\label{slslgmh:sec}
The process ${\slp{a}\slp{b}^*\rightarrow \gm h}$ proceeds only via 
the $t$-- and $u$--channel slepton ($\slp{a}$, $a=1,2$) exchange 
\begin{eqnarray}
 \wt_{\slp{a}\slp{b}^*\rightarrow \gm h}= \wt_{\gm h}^{(\slp{})}:
\end{eqnarray}
%
$\bullet$
 \underline{slepton\ ($\slp{c}$) exchange:}
\begin{eqnarray}
\wt_{\gm h}^{(\slp{})} & = & 
-2e^2 \left| C^{\slp{b}^*\slp{a}h} \right|^2 
\left[ (\Tt_1+\msli^2\Tt_0)+(\Tu_1+\mslj^2\Tu_0) \right. \nonumber \\
 & & \left. + (s+\mh^2-2\msli^2-2\mslj^2)\Y_0 \right].
\end{eqnarray}
The expression for $\wt_{\slp{a}\slp{b}^*\rightarrow \gm H}$ can be 
obtained by a simple substitution $h\rightarrow H$.

\subsection{\boldmath ${\slp{a}\slp{b}^*\rightarrow \gm A}$}
\label{slslgma:sec}
This process proceeds only via 
the $t$-- and $u$--channel slepton ($\slp{a}$, $a=1,2$) exchange 
\begin{eqnarray}
 \wt_{\slp{a}\slp{b}^*\rightarrow \gm A}= \wt_{\gm A}^{(\slp{})}:
\end{eqnarray}
$\wt_{\gm A}^{(\slp{})}$ can be obtained from 
$\wt_{\gm h}^{(\slp{})}$
by a simple substitution $h\rightarrow A$. 

\vspace{0.3in}
Next we proceed to present the results for
${\slp{a}\slp{b}^*\rightarrow}$ \ Higgs--Higgs pairs: 
$hh$, $HH$, $hH$, $hA$, $HA$, $AA$ and $H^+H^-$. 

\subsection{\boldmath ${\slp{a}\slp{b}^*\rightarrow hh, HH, hH}$}
\label{slslhh:sec}
The process $\slp{a}\slp{b}^*\rightarrow hH$ proceeds via 
the $s$--channel CP--even Higgs boson ($h$ and $H$) exchange, 
the four--point interaction, 
and the $t$-- and $u$--channel slepton ($\slp{a}$, $a=1,2$) exchange 
\begin{eqnarray}
 \wt_{\slp{a}\slp{b}^*\rightarrow hH}=
 \wt_{hH}^{(h,H,P)} + \wt_{hH}^{(\slp{})} + \wt_{hH}^{(h,H,P-\slp{})}:
\end{eqnarray}
%
$\bullet$
 \underline{Higgs ($h,H$) exchange ($+$ Point interaction):}
\begin{eqnarray}
\wt_{hH}^{(h,H,P)} & = & 
\left| 
\sum_{r=h,H} \frac{C^{hHr}C^{\slp{b}^*\slp{a}r}}{\prop{r}}
-C^{\slp{b}^*\slp{a}hH}
\right|^2;
\end{eqnarray}
%
$\bullet$
 \underline{slepton\ ($\slp{c}$) exchange:}
\begin{eqnarray}
\wt_{hH}^{(\slp{})} & = & 
\sum_{c,d=1}^2 
C^{\slp{c}^*\slp{a}h} C^{\slp{b}^*\slp{c}H} 
C^{\slp{d}^*\slp{a}h *} C^{\slp{b}^*\slp{d}H *} \Tt_0 
     + \sum_{c,d=1}^2 C^{\slp{c}^*\slp{a}H} C^{\slp{b}^*\slp{c}h} 
C^{\slp{d}^*\slp{a}H *} C^{\slp{b}^*\slp{d}h *} \Tu_0 \nonumber \\
 & & - 2 \Re \sum_{c,d=1}^2 
C^{\slp{c}^*\slp{a}h} C^{\slp{b}^*\slp{c}H} 
C^{\slp{d}^*\slp{a}H *} C^{\slp{b}^*\slp{d}h *} \Y_0;
\end{eqnarray}
%
$\bullet$
 \underline{Higgs ($h,H$) ($+$ Point) -- slepton ($\slp{c}$) interference:}
\begin{eqnarray}
\wt_{hH}^{(h,H,P-\slp{})} & = & 
2 \Re \sum_{c=1}^2 
   \left( 
\sum_{r=h,H} \frac{C^{hHr}C^{\slp{b}^*\slp{a}r}}{\prop{r}}
-C^{\slp{b}^*\slp{a}hH}
\right)^* \nonumber \\
 & & 
   \times \Big[ 
C^{\slp{c}^*\slp{a}h} C^{\slp{b}^*\slp{c}H} \Ft
+ C^{\slp{c}^*\slp{a}H} C^{\slp{b}^*\slp{c}h} \Fu \Big]. 
\end{eqnarray}

The expressions for $hh$ final state are obtained from
the above by replacing $C^{hHr}$, $C^{\slp{b}^*\slp{c}H}$, 
$C^{\slp{c}^*\slp{a}H}$, $C^{hH\slp{b}^*\slp{a}}$ with
$C^{hhr}$, $C^{\slp{b}^*\slp{c}h}$, 
$C^{\slp{c}^*\slp{a}h}$, $C^{hh\slp{b}^*\slp{a}}$, respectively,
and multiplying $\widetilde{w}$ by a factor of $1/2$ for identical 
particles in the final state. The contributions for $HH$ final 
state are obtained in an analogous way.

\subsection{\boldmath ${\slp{a}\slp{b}^*\rightarrow hA, HA}$}
\label{slslha:sec}
The process ${\slp{a}\slp{b}^*\rightarrow hA}$ proceeds via 
the $s$--channel CP--odd Higgs boson ($A$) exchange, 
the $s$--channel $Z$--boson exchange, 
and the $t$-- and $u$--channel slepton ($\slp{a}$, $a=1,2$) exchange 
\begin{eqnarray}
 \wt_{\slp{a}\slp{b}^*\rightarrow hA}=
  \wt_{hA}^{(A)} + \wt_{hA}^{(Z)} + \wt_{hA}^{(\slp{})} 
+ \wt_{hA}^{(A-Z)} + \wt_{hA}^{(A-\slp{})} + \wt_{hA}^{(Z-\slp{})}:
\end{eqnarray}
%
$\bullet$
 \underline{$A$ exchange:}
\begin{eqnarray}
\wt_{hA}^{(A)} & = & 
\left| \frac{C^{hAA}C^{\slp{b}^*\slp{a}A}}{\prop{A}} \right|^2;
\end{eqnarray}
%
$\bullet$
 \underline{$Z$ exchange:}
\begin{eqnarray}
\wt_{hA}^{(Z)} & = & 
\frac{1}{3s^2\mz^4}
\left| \frac{C^{hAZ}C^{\slp{b}^*\slp{a}Z}}{\prop{Z}} \right|^2
\nonumber \\
 & & \times \left[ s^4\mz^4 
- 2 s^3 \mz^4 (\slpsl{2}+\ma^2+\mh^2) \right. \nonumber \\
 & & +s^2 \left\{ (\slmsl{2})^2[3(\ma^2-\mh^2)^2+\mz^4] \right. \nonumber \\
 & & \left. 
+ 4\mz^4(\ma^2+\mh^2)(\slpsl{2})+\mz^4(\ma^2-\mh^2)^2 \right\} \nonumber \\
 & & -2s \mz^2 \left\{ 
(\slmsl{2})^2 [3(\ma^2-\mh^2)^2+\mz^2(\ma^2+\mh^2)] \right. \nonumber \\
 & & \left. + \mz^2 (\ma^2-\mh^2)^2 (\slpsl{2}) \right\} \nonumber \\
 & & \left. + 4\mz^4(\ma^2-\mh^2)^2(\slmsl{2})^2 \right];
\end{eqnarray}
%
$\bullet$
 \underline{slepton\ ($\slp{c}$) exchange:}
\begin{eqnarray}
\wt_{hA}^{(\slp{})} & = & 
\sum_{c,d=1}^2 
C^{\slp{c}^*\slp{a}h} C^{\slp{b}^*\slp{c}A} 
C^{\slp{d}^*\slp{a}h *} C^{\slp{b}^*\slp{d}A *} \Tt_0 
     + \sum_{c,d=1}^2 C^{\slp{c}^*\slp{a}A} C^{\slp{b}^*\slp{c}h} 
C^{\slp{d}^*\slp{a}A *} C^{\slp{b}^*\slp{d}h *} \Tu_0 \nonumber \\
 & & - 2 \Re \sum_{c,d=1}^2 
C^{\slp{c}^*\slp{a}h} C^{\slp{b}^*\slp{c}A} 
C^{\slp{d}^*\slp{a}A *} C^{\slp{b}^*\slp{d}h *} \Y_0; 
\end{eqnarray}
%
$\bullet$
 \underline{$A$ -- $Z$ interference:}
\begin{eqnarray}
\wt_{hA}^{(A-Z)} & = & 
2 \Re 
\left[ \left( \frac{C^{hAA}C^{\slp{b}^*\slp{a}A}}{\prop{A}} \right)^* 
\left( \frac{C^{hAZ}C^{\slp{b}^*\slp{a}Z}}{\prop{Z}} \right) \right]
\nonumber \\
 & & \times
\frac{(\ma^2-\mh^2)(\slmsl{2})(s-\mz^2)}{s\mz^2};
\end{eqnarray}
%
$\bullet$
 \underline{Higgs ($A$) -- slepton\ ($\slp{c}$) interference:}
\begin{eqnarray}
\wt_{hA}^{(A-\slp{})} & = & 
2 \Re \sum_{c=1}^2 
  \left( \frac{C^{hAA}C^{\slp{b}^*\slp{a}A}}{\prop{A}} \right)^* 
   \Big[
C^{\slp{c}^*\slp{a}h} C^{\slp{b}^*\slp{c}A} \Ft
+ C^{\slp{c}^*\slp{a}A} C^{\slp{b}^*\slp{c}h} \Fu \Big];
\nonumber \\
 & & 
\end{eqnarray}
%
$\bullet$
 \underline{$Z$ -- slepton\ ($\slp{c}$) interference:}
\begin{eqnarray}
\wt_{hA}^{(Z-\slp{})} & = & 
-\frac{2}{\mz^2} \sum_{c=1}^2 \Re 
\left[ \left( \frac{C^{hAZ}C^{\slp{b}^*\slp{a}Z}}{\prop{Z}} \right)^* 
C^{\slp{c}^*\slp{a}h} C^{\slp{b}^*\slp{c}A} \right] \nonumber \\
 & & \times \left[ 2\mz^2 
- \{ (\ma^2-\mh^2)(\slmsl{2})+\mz^2(\slpsl{2}) \right. \nonumber \\
 & & \left. + \mz^2(-s-2\mslk^2+\ma^2+\mh^2) \} \Ft \right] \nonumber \\
 & & + \frac{2}{\mz^2} \sum_{c=1}^2 \Re 
\left[ \left( \frac{C^{hAZ}C^{\slp{b}^*\slp{a}Z}}{\prop{Z}} \right)^* 
C^{\slp{c}^*\slp{a}A} C^{\slp{b}^*\slp{c}h} \right] \nonumber \\
 & & \times \left[ 2\mz^2 
- \{ -(\ma^2-\mh^2)(\slmsl{2})+\mz^2(\slpsl{2}) \right. \nonumber \\
 & & \left. + \mz^2(-s-2\mslk^2+\ma^2+\mh^2) \} \Fu \right].
\end{eqnarray}
The expression for $\wt_{\slp{a}\slp{b}^*\rightarrow  HA}$ can be 
obtained by a simple substitution $h\rightarrow H$.
%

\subsection{\boldmath ${\slp{a}\slp{b}^*\rightarrow AA}$}
\label{slslaa:sec}
This process proceeds via 
the $s$--channel CP--even Higgs boson ($h$ and $H$) exchange, 
the four--point interaction, 
and the $t$-- and $u$--channel slepton ($\slp{a}$, $a=1,2$) exchange 
\begin{eqnarray}
 \wt_{\slp{a}\slp{b}^*\rightarrow AA}=
 \wt_{AA}^{(h,H,P)} + \wt_{AA}^{(\slp{})} + \wt_{AA}^{(h,H,P-\slp{})}:
\end{eqnarray}
%
$\bullet$
 \underline{Higgs ($h,H$) exchange ($+$ Point interaction):}
\begin{eqnarray}
\wt_{AA}^{(h,H,P)} & = & 
\frac{1}{2}
\left| 
\sum_{r=h,H} \frac{C^{AAr}C^{\slp{b}^*\slp{a}r}}{\prop{r}}
-C^{\slp{b}^*\slp{a}AA}
\right|^2;
\end{eqnarray}
%
$\bullet$
 \underline{slepton\ ($\slp{c}$) exchange:}
\begin{eqnarray}
\wt_{AA}^{(\slp{})} & = & 
\sum_{c,d=1}^2 
C^{\slp{c}^*\slp{a}A} C^{\slp{b}^*\slp{c}A} 
C^{\slp{d}^*\slp{a}A *} C^{\slp{b}^*\slp{d}A *} (\T_0-\Y_0);
\end{eqnarray}
%
$\bullet$
 \underline{Higgs ($h,H$) ($+$ Point) -- slepton ($\slp{c}$) interference:}
\begin{eqnarray}
\wt_{AA}^{(h,H,P-\slp{})} & = & 
2 \sum_{c=1}^2 \Re 
\left[ \left( 
\sum_{r=h,H} \frac{C^{AAr}C^{\slp{b}^*\slp{a}r}}{\prop{r}}
-C^{\slp{b}^*\slp{a}AA} \right)^* 
C^{\slp{c}^*\slp{a}A} C^{\slp{b}^*\slp{c}A} \right]\F.
 \nonumber \\
 & & 
\end{eqnarray}
%

\subsection{\boldmath ${\slp{a}\slp{b}^*\rightarrow H^+H^-}$}
\label{slslhphm:sec}
This process proceeds via 
the $s$--channel CP--even Higgs boson ($h$ and $H$) exchange, 
the four--point interaction, 
the $s$--channel $Z$--boson and photon exchange, 
and the $u$--channel sneutrino ($\sn{\ell}$) exchange 
\begin{eqnarray}
 \wt_{\slp{a}\slp{b}^*\rightarrow H^+H^-} =  
        \wt_{H^+H^-}^{(h,H,P)} + \wt_{H^+H^-}^{(Z,\gm)}
        + \wt_{H^+H^-}^{(\sn{})} + \wt_{H^+H^-}^{(h,H,P-\sn{})}
        + \wt_{H^+H^-}^{(Z,\gm-\sn{})} :
\end{eqnarray}
%
$\bullet$
 \underline{Higgs ($h,H$) exchange ($+$ Point interaction):}
\begin{eqnarray}
\wt_{H^+H^-}^{(h,H,P)} & = & 
\left| 
\sum_{r=h,H} \frac{C^{H^+H^-r}C^{\slp{b}^*\slp{a}r}}{\prop{r}}
-C^{\slp{b}^*\slp{a}H^+H^-}
\right|^2;
\end{eqnarray}
%
$\bullet$
 \underline{$Z,\gm$ exchange:}
\begin{eqnarray}
\wt_{H^+H^-}^{(Z,\gm)} & = & 
\frac{1}{3s}
\left| 
\frac{C^{H^+H^-Z}C^{\slp{b}^*\slp{a}Z}}{\prop{Z}}
+ \frac{e^2\delta_{ab}}{s} \right|^2 \nonumber \\
 & & \times (s-4\mch^2)[s-(\slpsl{})^2][s-(\slmsl{})^2];
\end{eqnarray}
%
$\bullet$
 \underline{sneutrino\ ($\sn{\ell}$) exchange:}
\begin{eqnarray}
\wt_{H^+H^-}^{(\sn{})} & = & 
\left| C^{\sn{\ell}^*\slp{a}H^+} C^{\slp{b}^*\sn{\ell}H^-} \right|^2 \T_0;
\end{eqnarray}
%
$\bullet$
 \underline{Higgs ($h,H$) ($+$ Point) -- sneutrino ($\sn{\ell}$) interference:}
\begin{eqnarray}
\wt_{H^+H^-}^{(h,H,P-\sn{})} & = & 
2 \Re \left[ \left( 
\sum_{r=h,H} \frac{C^{H^+H^-r}C^{\slp{b}^*\slp{a}r}}{\prop{r}}
-C^{\slp{b}^*\slp{a}H^+H^-} \right)^* 
C^{\sn{\ell}^*\slp{a}H^+} C^{\slp{b}^*\sn{\ell}H^-} \right]\F;
 \nonumber \\
 & & 
\end{eqnarray}
%
$\bullet$
 \underline{$Z,\gm$ -- sneutrino ($\sn{\ell}$) interference:}
\begin{eqnarray}
\wt_{H^+H^-}^{(Z,\gm-\sn{})} & = & 
2 \Re 
\left[ \left( 
\frac{C^{H^+H^-Z}C^{\slp{b}^*\slp{a}Z}}{\prop{Z}}
+ \frac{e^2\delta_{ab}}{s} \right)^* 
C^{\sn{\ell}^*\slp{a}H^+} C^{\slp{b}^*\sn{\ell}H^-} 
\right] \nonumber \\
 & & 
   \times \left[
 2 + (s+2\msn^2-2\mch^2-\msli^2-\mslj^2) \F \right]. 
\end{eqnarray}

%
\vspace{0.3in}
Finally, we present the results for
${\slp{a}\slp{b}^*\rightarrow \ffb}$ 
where $f$ denotes any of the SM fermions. 

\subsection{\boldmath ${\slp{a}\slp{b}^*\rightarrow \ell\bar{\ell}}$}
\label{slslff:sec}
This process involves 
the $s$--channel CP--even Higgs boson ($h$ and $H$)
            and CP--odd Higgs boson ($A$) exchange, 
the $s$--channel $Z$--boson and photon exchange, 
and the $t$--channel neutralino ($\neu{i}$, $i=1,2,3,4$) exchange 

\begin{eqnarray}
 \wt_{\slp{a}\slp{b}^*\rightarrow \llb} 
 & = &
\wt_{\llb}^{(h,H,A)} + \wt_{\llb}^{(Z,\gm)}
 + \wt_{\llb}^{(\neu{})} + \wt_{\llb}^{(A-Z)}
 + \wt_{\llb}^{(h,H,A-\neu{})} + \wt_{\llb}^{(Z,\gm-\neu{})} :
\end{eqnarray}
%
$\bullet$
 \underline{Higgs ($h,H,A$) exchange:}
\begin{eqnarray}
\wt_{\llb}^{(h,H,A)} & = & 
2 \left| \sum_{r=h,H} \frac{C^{\ell\ell r}_S C^{\slp{b}^*\slp{a}r}}{\prop{r}}
  \right|^2 (s-4\ml^2) 
 + 2 \left| \frac{C^{\ell\ell A}_PC^{\slp{b}^*\slp{a}A}}{\prop{A}}
  \right|^2 s; \nonumber \\
 & & 
\end{eqnarray}
%
$\bullet$
 \underline{$Z,\gm$ exchange:}
\begin{eqnarray}
\wt_{\llb}^{(Z,\gm)} & = & 
\frac{4}{3}
\left| \frac{C^{\ell\ell Z}_V C^{\slp{b}^*\slp{a}Z}}{\prop{Z}}
       - \frac{e e_{\ell} \delta_{ab}}{s} \right|^2 \nonumber \\
 & & \times (s+2\ml^2) \Big\{ s-(\slpsl{})^2 \Big\} 
             \Bigg\{ 1-\frac{(\slmsl{})^2}{s} \Bigg\} \nonumber \\
 & & 
 + \frac{4}{3s\mz^4}
   \left| \frac{C^{\ell\ell Z}_A C^{\slp{b}^*\slp{a}Z}}{\prop{Z}} \right|^2 
\nonumber \\
 & & \times
\Bigg[ s^3 \mz^4 -2s^2 \{ \mz^4(\slpsl{2}+2\ml^2)-3\ml^2(\slmsl{2})^2 \}
                         \nonumber \\
 & & + s\mz^2 \{ (\mz^2-12\ml^2)(\slmsl{2})^2 
     + 8 \ml^2\mz^2(\slpsl{2}) \} \nonumber \\
 & & + 2\ml^2\mz^4(\slmsl{2})^2 \Bigg],
\end{eqnarray}
where $e_{\ell}=-e<0$ is the electric charge for the charged
lepton $\ell$ in the final state. For other fermion pair production,
this should be replaced with the electric charge of the final particle.
\vspace{3mm} \\
%
$\bullet$
 \underline{neutralino\ ($\neu{i}$) exchange:}
\begin{eqnarray}
\wt_{\llb}^{(\neu{})} & = & 
\sum_{i,j=1}^4 \Bigg[ 
\Cijqp{LLLL}\Big\{
-\T_2 -(s-\msli^2-\mslj^2)\T_1-(\msli^2-\ml^2)(\mslj^2-\ml^2)\T_0 \Big\} 
   \nonumber \\
 & & 
  -2 \ml^2 \Big[ \Cijqp{LRLR}\mnmn \T_0 + \Cijqp{LLRR}\T_1 \Big]
     + \Cijqp{LRRL}\mnmn (s-2\ml^2)\T_0 \nonumber \\
 & & -\ml\mnq \Big[ \Cijqp{LRLL} \{ \T_1-(\msli^2-\ml^2)\T_0 \}
     + \Cijqp{RLLL} \{ \T_1-(\mslj^2-\ml^2)\T_0 \} \Big] \nonumber \\
 & & -\ml\mnp \Big[ \Cijqp{LLRL} \{ \T_1-(\msli^2-\ml^2)\T_0 \}
     + \Cijqp{LLLR} \{ \T_1-(\mslj^2-\ml^2)\T_0 \} \Big] \Bigg],
\nonumber \\
 & & 
\end{eqnarray}
where 
\begin{eqnarray}
\Cijqp{LLLL} & = & C^{\neu{i}\slp{a}^*\ell *}_L C^{\neu{i}\slp{b}^*\ell}_L
                   C^{\neu{j}\slp{b}^*\ell *}_L C^{\neu{j}\slp{a}^*\ell}_L
                    + (L \rightarrow R), \nonumber \\
\Cijqp{LRLR} & = & C^{\neu{i}\slp{a}^*\ell *}_L C^{\neu{i}\slp{b}^*\ell}_R
                   C^{\neu{j}\slp{b}^*\ell *}_L C^{\neu{j}\slp{a}^*\ell}_R
                    + (L \leftrightarrow R), \nonumber \\
\Cijqp{LLRR} & = & C^{\neu{i}\slp{a}^*\ell *}_L C^{\neu{i}\slp{b}^*\ell}_L
                   C^{\neu{j}\slp{b}^*\ell *}_R C^{\neu{j}\slp{a}^*\ell}_R
                    + (L \leftrightarrow R), \nonumber \\
\Cijqp{LRRL} & = & C^{\neu{i}\slp{a}^*\ell *}_L C^{\neu{i}\slp{b}^*\ell}_R
                   C^{\neu{j}\slp{b}^*\ell *}_R C^{\neu{j}\slp{a}^*\ell}_L
                    + (L \leftrightarrow R), \nonumber \\
\Cijqp{LRLL} & = & C^{\neu{i}\slp{a}^*\ell *}_L C^{\neu{i}\slp{b}^*\ell}_R
                   C^{\neu{j}\slp{b}^*\ell *}_L C^{\neu{j}\slp{a}^*\ell}_L
                    + (L \leftrightarrow R), \nonumber \\
\Cijqp{RLLL} & = & C^{\neu{i}\slp{a}^*\ell *}_R C^{\neu{i}\slp{b}^*\ell}_L
                   C^{\neu{j}\slp{b}^*\ell *}_L C^{\neu{j}\slp{a}^*\ell}_L
                    + (L \leftrightarrow R), \nonumber \\
\Cijqp{LLRL} & = & C^{\neu{i}\slp{a}^*\ell *}_L C^{\neu{i}\slp{b}^*\ell}_L
                   C^{\neu{j}\slp{b}^*\ell *}_R C^{\neu{j}\slp{a}^*\ell}_L
                    + (L \leftrightarrow R), \nonumber \\
\Cijqp{LLLR} & = & C^{\neu{i}\slp{a}^*\ell *}_L C^{\neu{i}\slp{b}^*\ell}_L
                   C^{\neu{j}\slp{b}^*\ell *}_L C^{\neu{j}\slp{a}^*\ell}_R
                    + (L \leftrightarrow R); \nonumber 
\end{eqnarray}
%
$\bullet$
 \underline{$A-Z$ interference:}
\begin{eqnarray}
\wt_{\llb}^{(A-Z)} & = & 
-8 \Re
\left[ 
\left( \frac{C^{\ell\ell A}_P C^{\slp{b}^*\slp{a}A}}{\prop{A}} \right)^*
\left( \frac{C^{\ell\ell Z}_A C^{\slp{b}^*\slp{a}Z}}{\prop{Z}} \right)
\right] \cdot \ml (\slmsl{2})(s-\mz^2); \nonumber \\
 & & 
\end{eqnarray}
%
$\bullet$
 \underline{Higgs ($h,H,A$) -- neutralino ($\neu{i}$) interference:}
\begin{eqnarray}
\wt_{\llb}^{(h,H,A-\neu{})} & = & 
4 \Re \sum_{i=1}^4 \sum_{r=h,H}
\left( \frac{C^{\ell\ell r}_S C^{\slp{b}^*\slp{a}r}}{\prop{r}} \right)^*
  \nonumber \\
 & & \times
\Bigg[ C_{+ii}^{ba}\ml \Big\{ -2 + (\slpsl{2}-2\mnq^2-2\ml^2)\F \Big\}
       + C_{-ii}^{ba}\mnq (s-4\ml^2)\F \Bigg] \nonumber \\
 & & 
+ 4 \Re \sum_{i=1}^4 
\left( \frac{C^{\ell\ell A}_P C^{\slp{b}^*\slp{a}A}}{\prop{A}} \right)^*
\Bigg[ D_{+ii}^{ba}\ml (\slmsl{2}) + D_{-ii}^{ba}\mnq s \Bigg] \F,
\nonumber \\
 & & 
\end{eqnarray}
where
\begin{eqnarray}
C_{\pm ij}^{ab} & = & 
     \CS{i}{a}\Big(\CS{j}{b}\Big)^* \pm \CP{i}{a}\Big(\CP{j}{b}\Big)^*,  \\
D_{\pm ij}^{ab} & = &
     \CS{i}{a}\Big(\CP{j}{b}\Big)^* \pm \CP{i}{a}\Big(\CS{j}{b}\Big)^* ;
\end{eqnarray}
%
$\bullet$
 \underline{$Z,\gm$ -- neutralino ($\neu{i}$) interference:}
\begin{eqnarray}
\wt_{\llb}^{(Z,\gm-\neu{})} & = & 
4 \Re \sum_{i=1}^4 
\left( \frac{C^{\ell\ell Z}_V C^{\slp{b}^*\slp{a}Z}}{\prop{Z}} 
       - \frac{e e_{\ell} \delta_{ab}}{s} \right)^*
  \nonumber \\
 & & \times
\Bigg[ C_{+ii}^{ba} \Big\{ -s + \slpsl{2} -2\mnq^2-2\ml^2 \nonumber \\
 & & + 2 [ -s \mnq^2 -(\msli^2-\mnq^2)(\mslj^2-\mnq^2) +\ml^4 ]\F \Big\}
        \nonumber \\
 & &        + C_{-ii}^{ba} \cdot 2\ml\mnq 
\Big\{ -2 + ( -s+\slpsl{2}-2\mnq^2+2\ml^2 )\F \Big\} \Bigg] \nonumber \\
 & & 
+ 4 \Re \sum_{i=1}^4 
\left( \frac{C^{\ell\ell Z}_A C^{\slp{b}^*\slp{a}Z}}{\prop{Z}} \right)^*
\frac{1}{\mz^2}
  \nonumber \\
 & & \times
\Bigg[ D_{+ii}^{ba} \Big\{ \mz^2(-s + \slpsl{2} -2\mnq^2+2\ml^2) \nonumber \\
 & & - 2 [ \mz^2 \{ s\mnq^2 +(\msli^2-\mnq^2)(\mslj^2-\mnq^2) \}
\nonumber \\ 
 & & + \ml^2 \{ (\slmsl{2})^2-\mz^2(\slpsl{2}+2\mnq^2-\ml^2) \} ]\F \Big\}
        \nonumber \\
 & &  - D_{-ii}^{ba} \cdot 2\ml\mnq (\slmsl{2})(s-\mz^2)\F \Bigg]. 
\end{eqnarray}

\subsection{\boldmath ${\slp{a}\slp{b}^*\rightarrow q\bar{q}}$}
This process involves 
the $s$--channel CP--even Higgs boson ($h$ and $H$)
            and CP--odd Higgs boson ($A$) exchange, 
and the $s$--channel $Z$--boson and photon exchange
\begin{eqnarray}
 \wt_{\slp{a}\slp{b}^*\rightarrow \qqb}
  & = &
\wt_{\qqb}^{(h,H,A)} + \wt_{\qqb}^{(Z,\gm)}
 + \wt_{\qqb}^{(A-Z)} :
\end{eqnarray}
Each contribution can be found from 
the respective ones for the $\ell\bar{\ell}$ final
state by replacing $\ell$ with $q$. 

\subsection{\boldmath ${\slp{a}\slp{b}^*\rightarrow 
                                    \nu_{\ell}\bar{\nu}_{\ell}}$}
This process involves 
the $s$--channel $Z$--boson, 
and the $t$--channel chargino ($\cha{k}$, $k=1,2$) exchange 
\begin{eqnarray}
 \wt_{\slp{a}\slp{b}^*\rightarrow \nu_{\ell}\bar{\nu}_{\ell}}
 & = &
\wt_{\nu_{\ell}\bar{\nu}_{\ell}}^{(Z)}
 + \wt_{\nu_{\ell}\bar{\nu}_{\ell}}^{(\cha{})}
 + \wt_{\nu_{\ell}\bar{\nu}_{\ell}}^{(Z-\cha{})} : 
\end{eqnarray}
Each contribution can be found from 
the respective ones for the $\ell\bar{\ell}$ final
state by replacing $\ell$ with $\nu_{\ell}$. 

\subsection{\boldmath ${\slp{a}\slp{b}^{\prime *}\rightarrow 
                                    \ell\bar{\ell}'}$}
This process involves 
the $t$--channel neutralino ($\neu{i}$, $i=1,2,3,4$) exchange
\begin{eqnarray}
 \wt_{\slp{a}\slp{b}^{\prime *}\rightarrow \ell\bar{\ell}'}
 & = & \wt_{\ell\bar{\ell}'}^{(\neu{})}: 
\end{eqnarray}
%
$\bullet$
 \underline{neutralino\ ($\neu{i}$) exchange:}
\begin{eqnarray}
\wt_{\ell\bar{\ell'}}^{(\neu{})} & = & 
\sum_{i,j=1}^4 \Bigg[ 
\CCijqp{LLLL}\Big[ -\Tt_2
  -(s-\msli^2-\mslpj^2)\Tt_1-(\msli^2-\ml^2)(\mslpj^2-\mlp^2)\Tt_0 \Big]
\nonumber \\
 & & -2\ml\mlp \Big[ \CCijqp{LRLR}\mnmn\Tt_0+\CCijqp{LLRR}\Tt_1 \Big]
+ \CCijqp{LRRL} \mnmn (s-\ml^2-\mlp^2)\Tt_0 \nonumber \\
 & & -\mnq \Big[ \CCijqp{LRLL}\mlp \{ \Tt_1-(\msli^2-\ml^2)\Tt_0 \}
     + \CCijqp{RLLL}\ml \{ \Tt_1-(\mslpj^2-\mlp^2)\Tt_0 \} \Big] \nonumber \\
 & & -\mnp \Big[ \CCijqp{LLRL}\mlp \{ \Tt_1-(\msli^2-\ml^2)\Tt_0 \}
     + \CCijqp{LLLR}\ml \{ \Tt_1-(\mslpj^2-\mlp^2)\Tt_0 \} \Big] \Bigg],
\nonumber \\
 & & 
\end{eqnarray}
where
\begin{eqnarray}
\CCijqp{LLLL} & = & 
         C^{\neu{i}\slp{a}^*\ell *}_L C^{\neu{i}\slp{b}^{'*}\ell'  }_L
         C^{\neu{j}\slp{a}^*\ell *}_L C^{\neu{j}\slp{b}^{'*}\ell'  }_L
                    + (L \rightarrow R), \nonumber \\
\CCijqp{LRLR} & = & 
         C^{\neu{i}\slp{a}^*\ell *}_L C^{\neu{i}\slp{b}^{'*}\ell'  }_R
         C^{\neu{j}\slp{a}^*\ell *}_L C^{\neu{j}\slp{b}^{'*}\ell'  }_R
                    + (L \leftrightarrow R), \nonumber \\
\CCijqp{LLRR} & = & 
         C^{\neu{i}\slp{a}^*\ell *}_L C^{\neu{i}\slp{b}^{'*}\ell'  }_L
         C^{\neu{j}\slp{a}^*\ell *}_R C^{\neu{j}\slp{b}^{'*}\ell'  }_R
                    + (L \leftrightarrow R), \nonumber \\
\CCijqp{LRRL} & = & 
         C^{\neu{i}\slp{a}^*\ell *}_L C^{\neu{i}\slp{b}^{'*}\ell'  }_R
         C^{\neu{j}\slp{a}^*\ell *}_R C^{\neu{j}\slp{b}^{'*}\ell'  }_L
                    + (L \leftrightarrow R), \nonumber \\
\CCijqp{LRLL} & = & 
         C^{\neu{i}\slp{a}^*\ell *}_L C^{\neu{i}\slp{b}^{'*}\ell'  }_R
         C^{\neu{j}\slp{a}^*\ell *}_L C^{\neu{j}\slp{b}^{'*}\ell'  }_L
                    + (L \leftrightarrow R), \nonumber \\
\CCijqp{RLLL} & = & 
         C^{\neu{i}\slp{a}^*\ell *}_R C^{\neu{i}\slp{b}^{'*}\ell'  }_L
         C^{\neu{j}\slp{a}^*\ell *}_L C^{\neu{j}\slp{b}^{'*}\ell'  }_L
                    + (L \leftrightarrow R),  \nonumber \\
\CCijqp{LLRL} & = & 
         C^{\neu{i}\slp{a}^*\ell *}_L C^{\neu{i}\slp{b}^{'*}\ell'  }_L
         C^{\neu{j}\slp{a}^*\ell *}_R C^{\neu{j}\slp{b}^{'*}\ell'  }_L
                    + (L \leftrightarrow R), \nonumber \\
\CCijqp{LLLR} & = & 
         C^{\neu{i}\slp{a}^*\ell *}_L C^{\neu{i}\slp{b}^{'*}\ell'  }_L
         C^{\neu{j}\slp{a}^*\ell *}_L C^{\neu{j}\slp{b}^{'*}\ell'  }_R
                    + (L \leftrightarrow R). \nonumber 
\end{eqnarray}

\subsection{\boldmath ${\slp{a}\slp{b}^{\prime *}\rightarrow 
                                    \nu_{\ell}\bar{\nu}_{\ell'}}$}
This process involves 
the $t$--channel chargino ($\cha{k}$, $k=1,2$) exchange 
\begin{eqnarray}
 \wt_{\slp{a}\slp{b}^{\prime *}\rightarrow \nu_{\ell}\bar{\nu}_{\ell'}}
 & = & \wt_{\nu_{\ell}\bar{\nu}_{\ell'}}^{(\cha{})} : 
\end{eqnarray}
The expression for $\wt_{\nu_{\ell}\bar{\nu}_{\ell'}}^{(\cha{})}$
can be found from $\wt_{\ell\bar{\ell}'}^{(\neu{})}$ by replacing
$\ell$ and $\neu{}$ with $\nu_{\ell}$ and $\cha{}$, respectively. 

\vspace{0.3in}\noindent
The second class of processes involves ${\slp{a}\slp{b}\rightarrow
\ell\ell}$ and ${\slp{a}\slp{b}^{\prime}\rightarrow
\ell\ell^{\prime}}$. For each there is a corresponding process involving
particles with the oposite electric charge.

\subsection{\boldmath ${\slp{a}\slp{b}\rightarrow \ell\ell}$}
\label{slslelel:sec}
This process proceeds only via 
the $t$-- and $u$--channel neutralino ($\neu{i}$, $i=1,2,3,4$) exchange
\begin{eqnarray}
 \wt_{\slp{a}\slp{b}\rightarrow \ell\ell}=\wt_{\ell\ell}^{(\neu{})}:
\end{eqnarray}
%
$\bullet$
 \underline{neutralino\ ($\neu{i}$) exchange:}
\begin{eqnarray}
\wt_{\slp{a}\slp{b}\rightarrow \ell\ell}^{(\neu{})} & = & 
\sum_{i,j=1}^4 \Bigg[ 
\Dijqp{LLLL} \mnmn (s-2\ml^2)\T_0 \nonumber \\
 & & -2\ml^2 \Big[ \Dijqp{LLRR}\mnmn\T_0+\Dijqp{LRRL}\T_1 \Big]
\nonumber \\
 & & + \Dijqp{LRLR}\Big[ -\T_2
  -(s-\msli^2-\mslj^2)\T_1-(\msli^2-\ml^2)(\mslj^2-\ml^2)\T_0 \Big]
\nonumber \\
 & & -\ml\mnq \Big[ \Dijqp{LLLR} \{ \T_1-(\msli^2-\ml^2)\T_0 \}
     + \Dijqp{LLRL} \{ \T_1-(\mslj^2-\ml^2)\T_0 \} \Big] \nonumber \\
 & & -\ml\mnp \Big[ \Dijqp{LRLL} \{ \T_1-(\msli^2-\ml^2)\T_0 \}
     + \Dijqp{RLLL} \{ \T_1-(\mslj^2-\ml^2)\T_0 \} \Big] \Bigg]
\nonumber \\
 & & 
 + \frac{1}{2}\sum_{i,j=1}^4 
\Bigg[ 
-2 \Dijqp{LLLL} \mnmn (s-2\ml^2)\Y_0 \nonumber \\
 & & +4\Dijqp{LLRR}\mnmn\ml^2\Y_0 
     -2\Dijqp{LRLR}\ml^2(\slpsl{2}-2\ml^2)\Y_0 \nonumber \\
 & & + 2 \Dijqp{LRRL}\Big[ -\Y_2 -(\slsl{2}-\ml^4)\Y_0 \Big] \nonumber \\
 & & -\ml\mnq \Big[ \Dijqp{LLLR} \{ \Y_1+(s+\slmsl{2}-4\ml^2)\Y_0 \}
      \nonumber \\
 & & 
     + \Dijqp{LLRL} \{ \Y_1+(s-\msli^2+\mslj^2-4\ml^2) \Y_0 \} 
                                                       \Big] \nonumber \\
 & & +\ml\mnp \Big[ \Dijqp{LRLL} \{ \Y_1-(s+\slmsl{2}-4\ml^2)\Y_0 \}
      \nonumber \\
 & & 
     + \Dijqp{RLLL} \{ \Y_1-(s-\msli^2+\mslj^2-4\ml^2) \Y_0 \} 
\Big] \Bigg], \nonumber \\
 & & 
\end{eqnarray}
where
\begin{eqnarray}
\Dijqp{LLLL} & = & 
           C^{\neu{i}\slp{a}^*\ell *}_L C^{\neu{i}\slp{b}^*\ell *}_L
           C^{\neu{j}\slp{b}^*\ell  }_L C^{\neu{j}\slp{a}^*\ell  }_L
                    + (L \rightarrow R), \nonumber \\
\Dijqp{LLRR} & = & 
           C^{\neu{i}\slp{a}^*\ell *}_L C^{\neu{i}\slp{b}^*\ell *}_L
           C^{\neu{j}\slp{b}^*\ell  }_R C^{\neu{j}\slp{a}^*\ell  }_R
                    + (L \leftrightarrow R), \nonumber \\
\Dijqp{LRRL} & = & 
           C^{\neu{i}\slp{a}^*\ell *}_L C^{\neu{i}\slp{b}^*\ell *}_R
           C^{\neu{j}\slp{b}^*\ell  }_R C^{\neu{j}\slp{a}^*\ell  }_L
                    + (L \leftrightarrow R), \nonumber \\
\Dijqp{LRLR} & = & 
           C^{\neu{i}\slp{a}^*\ell *}_L C^{\neu{i}\slp{b}^*\ell *}_R
           C^{\neu{j}\slp{b}^*\ell  }_L C^{\neu{j}\slp{a}^*\ell  }_R
                    + (L \leftrightarrow R), \nonumber \\
\Dijqp{LLLR} & = & 
           C^{\neu{i}\slp{a}^*\ell *}_L C^{\neu{i}\slp{b}^*\ell *}_L
           C^{\neu{j}\slp{b}^*\ell  }_L C^{\neu{j}\slp{a}^*\ell  }_R
                    + (L \leftrightarrow R), \nonumber \\
\Dijqp{LLRL} & = & 
           C^{\neu{i}\slp{a}^*\ell *}_L C^{\neu{i}\slp{b}^*\ell *}_L
           C^{\neu{j}\slp{b}^*\ell  }_R C^{\neu{j}\slp{a}^*\ell  }_L
                    + (L \leftrightarrow R), \nonumber \\
\Dijqp{LRLL} & = & 
           C^{\neu{i}\slp{a}^*\ell *}_L C^{\neu{i}\slp{b}^*\ell *}_R
           C^{\neu{j}\slp{b}^*\ell  }_L C^{\neu{j}\slp{a}^*\ell  }_L
                    + (L \leftrightarrow R), \nonumber \\
\Dijqp{RLLL} & = & 
           C^{\neu{i}\slp{a}^*\ell *}_R C^{\neu{i}\slp{b}^*\ell *}_L
           C^{\neu{j}\slp{b}^*\ell  }_L C^{\neu{j}\slp{a}^*\ell  }_L
                    + (L \leftrightarrow R). 
\end{eqnarray}

\subsection{\boldmath ${\slp{a}\slp{b}'\rightarrow \ell\ell'}$}
\label{slslpelelp:sec}
This process proceeds only via 
the $t$--channel neutralino ($\neu{i}$, $i=1,2,3,4$) exchange
\begin{eqnarray}
 \wt_{\slp{a}\slp{b}'\rightarrow \ell\ell'}=\wt_{\ell\ell'}^{(\neu{})}:
\end{eqnarray}
%
$\bullet$
 \underline{neutralino\ ($\neu{i}$) exchange:}
\begin{eqnarray}
\wt_{\ell\ell'}^{(\neu{})} & = & 
\sum_{i,j=1}^4 \Bigg[ 
\DDijqp{LLLL} \mnmn (s-\ml^2-\mlp^2)\Tt_0 \nonumber \\
 & & -2\ml\mlp \Big[ \DDijqp{LLRR}\mnmn\Tt_0+\DDijqp{LRRL}\Tt_1 \Big]
\nonumber \\
 & & + \DDijqp{LRLR}\Big[ -\Tt_2
  -(s-\msli^2-\mslpj^2)\Tt_1-(\msli^2-\ml^2)(\mslpj^2-\mlp^2)\Tt_0 \Big]
\nonumber \\
 & & -\mnq \Big[ \DDijqp{LLLR}\mlp \{ \Tt_1-(\msli^2-\ml^2)\Tt_0 \}
     + \DDijqp{LLRL}\ml \{ \Tt_1-(\mslpj^2-\mlp^2)\Tt_0 \} \Big] \nonumber \\
 & & -\mnp \Big[ \DDijqp{LRLL}\mlp \{ \Tt_1-(\msli^2-\ml^2)\Tt_0 \}
     + \DDijqp{RLLL}\ml \{ \Tt_1-(\mslpj^2-\mlp^2)\Tt_0 \} \Big] \Bigg],
\nonumber \\
 & & 
\end{eqnarray}
where
\begin{eqnarray}
\DDijqp{LLLL} & = & 
         C^{\neu{i}\slp{a}^*\ell *}_L C^{\neu{i}\slp{b}^{'*}\ell' *}_L
         C^{\neu{j}\slp{a}^*\ell  }_L C^{\neu{j}\slp{b}^{'*}\ell'  }_L
                    + (L \rightarrow R), \nonumber \\
\DDijqp{LLRR} & = & 
         C^{\neu{i}\slp{a}^*\ell *}_L C^{\neu{i}\slp{b}^{'*}\ell' *}_L
         C^{\neu{j}\slp{a}^*\ell  }_R C^{\neu{j}\slp{b}^{'*}\ell'  }_R
                    + (L \leftrightarrow R), \nonumber \\
\DDijqp{LRRL} & = & 
         C^{\neu{i}\slp{a}^*\ell *}_L C^{\neu{i}\slp{b}^{'*}\ell' *}_R
         C^{\neu{j}\slp{a}^*\ell  }_R C^{\neu{j}\slp{b}^{'*}\ell'  }_L
                    + (L \leftrightarrow R), \nonumber \\
\DDijqp{LRLR} & = & 
         C^{\neu{i}\slp{a}^*\ell *}_L C^{\neu{i}\slp{b}^{'*}\ell' *}_R
         C^{\neu{j}\slp{a}^*\ell  }_L C^{\neu{j}\slp{b}^{'*}\ell'  }_R
                    + (L \leftrightarrow R), \nonumber \\
\DDijqp{LLLR} & = & 
         C^{\neu{i}\slp{a}^*\ell *}_L C^{\neu{i}\slp{b}^{'*}\ell' *}_L
         C^{\neu{j}\slp{a}^*\ell  }_L C^{\neu{j}\slp{b}^{'*}\ell'  }_R
                    + (L \leftrightarrow R), \nonumber \\
\DDijqp{LLRL} & = & 
         C^{\neu{i}\slp{a}^*\ell *}_L C^{\neu{i}\slp{b}^{'*}\ell' *}_L
         C^{\neu{j}\slp{a}^*\ell  }_R C^{\neu{j}\slp{b}^{'*}\ell'  }_L
                    + (L \leftrightarrow R), \nonumber \\
\DDijqp{LRLL} & = & 
         C^{\neu{i}\slp{a}^*\ell *}_L C^{\neu{i}\slp{b}^{'*}\ell' *}_R
         C^{\neu{j}\slp{a}^*\ell  }_L C^{\neu{j}\slp{b}^{'*}\ell'  }_L
                    + (L \leftrightarrow R), \nonumber \\
\DDijqp{RLLL} & = & 
         C^{\neu{i}\slp{a}^*\ell *}_R C^{\neu{i}\slp{b}^{'*}\ell' *}_L
         C^{\neu{j}\slp{a}^*\ell  }_L C^{\neu{j}\slp{b}^{'*}\ell'  }_L
                    + (L \leftrightarrow R). \nonumber 
\end{eqnarray}

\noindent
Finally, there are slepton--neutralino annihilation channels into a
lepton and either a gauge or a Higgs boson. For each, there is an
analogous one involving the opposite electric charge.

\subsection{\boldmath ${\slp{a}\chi\rightarrow Z\ell}$}
\label{slchizel:sec}
This process involves 
the $s$--channel lepton ($\ell$) exchange, 
the $t$--channel slepton ($\slp{a}$, $a=1,2$) exchange,
and the $u$--channel neutralino ($\neu{i}$, $i=1,2,3,4$) exchange
\begin{eqnarray}
 \wt_{\slp{a}\chi\rightarrow Z\ell} = 
        \wt_{Z\ell}^{(\ell)} + \wt_{Z\ell}^{(\slp{})}
      + \wt_{Z\ell}^{(\neu{})} + \wt_{Z\ell}^{(\ell - \slp{})} 
      + \wt_{Z\ell}^{(\ell - \neu{})} + \wt_{Z\ell}^{(\slp{}-\neu{})}: 
\end{eqnarray}
%
%
$\bullet$
 \underline{lepton ($\ell$) exchange:}
\begin{eqnarray}
\wt_{Z\ell}^{(\ell)} & = & 
\frac{1}{2\mz^2(s-\ml^2)^2} \nonumber \\
 & & \times \Bigg[ 
\Big\{ C_{+11}^{aa} C_{+}^{\ell\ell Z}
       + D_{+11}^{aa} D_{+}^{\ell\ell Z} 
      + \frac{\ml^2}{s}(C_{+11}^{aa} C_{+}^{\ell\ell Z}
                        - D_{+11}^{aa} D_{+}^{\ell\ell Z}) \Big\}
\nonumber \\
 & &   \times (s-\msli^2+\mchi^2)
                 \{ (s-\ml^2)^2 + \mz^2(s+\ml^2) -2\mz^4 \} 
\nonumber \\
 & &   + 4 C_{-11}^{aa} C_{+}^{\ell\ell Z}\ml\mchi
         \{ (s-\ml^2)^2 + \mz^2(s+\ml^2) -2\mz^4 \}
\nonumber \\
 & &   -12 C_{+11}^{aa} C_{-}^{\ell\ell Z}\ml^2\mz^2 (s-\msli^2+\mchi^2)
       -12 C_{-11}^{aa} C_{-}^{\ell\ell Z}\ml\mchi\mz^2 (s+\ml^2)
\Bigg],
\end{eqnarray}
where 
\begin{eqnarray}
C_{\pm}^{\ell\ell Z} & = & (C_V^{\ell\ell Z})^2 \pm (C_A^{\ell\ell Z})^2, \\
D_{+}^{\ell\ell Z} & = & 2 C_V^{\ell\ell Z}C_A^{\ell\ell Z}; 
\end{eqnarray}
%
$\bullet$
 \underline{slepton ($\slp{b}$) exchange:}
\begin{eqnarray}
\wt_{Z\ell}^{(\slp{})} & = & 
  \frac{1}{\mz^2} \sum_{b,c=1}^2 
    C^{\slp{b}^*\slp{a}Z}C^{\slp{c}^*\slp{a}Z *} \nonumber \\
 & & \times 
\Bigg[ C_{+11}^{cb} \Big\{ 
- \Tt_3 + (2\msli^2+\mchi^2+2\mz^2+\ml^2)\Tt_2 \nonumber \\
 & & -[ \msli^4 + 2\msli^2(\mchi^2-\mz^2+\ml^2) 
                + \mz^2(2\mchi^2+\mz^2+2\ml^2)]\Tt_1 \nonumber \\
 & & +(\mchi^2+\ml^2)(\msli^2-\mz^2)^2\Tt_0 \Big\} \nonumber \\
 & & + C_{-11}^{cb}\cdot 2\ml\mchi \Big\{ 
\Tt_2 - 2(\msli^2+\mz^2)\Tt_1 +(\msli^2-\mz^2)^2 \Tt_0 \Big\} \Bigg];
\end{eqnarray}
%
$\bullet$
 \underline{neutralino ($\neu{i}$) exchange:}
\begin{eqnarray}
\wt_{Z\ell}^{(\neu{})} & = & 
  \frac{1}{\mz^2} \sum_{i,j=1}^4 \CAzq \CAzpc 
\Bigg[ 
C_{+ji}^{aa} \Big\{
\Tu_3 + (s-\msli^2-2\mchi^2-\mz^2)\Tu_2 \nonumber \\
 & & + [ -s(\mchi^2+2\mz^2) + \mchi^4 
         + \mchi^2(2\msli^2+2\mz^2-\ml^2)+2\ml^2\mz^2 ]\Tu_1 \nonumber \\
 & & -(\msli^2-\ml^2)(\mchi^2-\mz^2)(\mchi^2+2\mz^2)\Tu_0 \nonumber \\
 & & + \mchi(\mnpmn) \cdot 3\mz^2 [ \Tu_1-(\msli^2-\ml^2)\Tu_0 ] 
\nonumber \\
 & & + \mnmn [ (-s+2\mz^2+\ml^2)\Tu_1 \nonumber \\
 & & + \{s(\mchi^2+2\mz^2)-\mz^2(\msli^2+\mchi^2+2\mz^2) 
     -\ml^2(\mchi^2+\mz^2)\}\Tu_0 ] \Big\} \nonumber \\ 
 & & 
+ C_{-ji}^{aa} \cdot \ml \Big\{ 6\mchi\mz^2\Tu_1 
     + (\mnpmn) [ \Tu_2-(2\mchi^2-\mz^2)\Tu_1 \nonumber \\ 
 & & + (\mchi^2-\mz^2)(\mchi^2+2\mz^2)\Tu_0 ] 
     + 6\mchi\mnmn\mz^2\Tu_0 \Big\} \Bigg];
\end{eqnarray}
%
$\bullet$
 \underline{lepton\ ($\ell$) -- slepton\ ($\slp{b}$) interference:}
\begin{eqnarray}
\wt_{Z\ell}^{(\ell - \slp{})} & = & 
  \frac{1}{\mz^2} \Re \sum_{b=1}^2 
    \frac{C^{\slp{b}^*\slp{a}Z}}{s-\ml^2}
 \Bigg[ 
  (C_V^{\ell\ell Z*}C_{+11}^{ab}+C_A^{\ell\ell Z*}D_{+11}^{ab}) I_1^{Z\ell}
  \nonumber \\
 & & 
 - (C_V^{\ell\ell Z*}C_{+11}^{ab}-C_A^{\ell\ell Z*}D_{+11}^{ab})
                                                       \ml^2 I_2^{Z\ell}
 + 4 C_V^{\ell\ell Z*}C_{-11}^{ab}\ml\mchi I_3^{Z\ell} \Bigg],
\end{eqnarray}
where
\begin{eqnarray}
I_1^{Z\ell} & = & -s^2 -s(\msli^2+\mchi^2+\mz^2-2\mslj^2-\ml^2)
      +(\msli^2-\mchi^2)(3\mz^2-\ml^2) \nonumber \\
 & & + 2 \Big\{ s[ \mslj^4 - \mslj^2(\msli^2+\mchi^2+\mz^2)
                  + \mchi^2(\msli^2-\mz^2) ] \nonumber \\
 & & + 2\mz^2[ \mslj^2(\msli^2-\mchi^2)
                +\mchi^2(\mchi^2+\mz^2-\msli^2) ] \nonumber \\
 & & -\ml^2[(\mslj^2-\msli^2)(\msli^2-\mchi^2)
                +\mz^2(\msli^2+\mchi^2)] \Big\}\Ft, \\
I_2^{Z\ell} & = & s + 2\mslj^2 + \mchi^2 -3\msli^2 -\mz^2 -\ml^2 
           - \frac{1}{s}(\mz^2-\ml^2)(\msli^2-\mchi^2) \nonumber \\
 & & + 2 \Big\{ s(\mslj^2-\msli^2+\mz^2)+(\mslj^2-\msli^2)^2
                               -\mslj^2(\mz^2+\ml^2) \nonumber \\
 & &    -\msli^2(\mz^2-\ml^2)- \mz^2(2\mchi^2-\ml^2) \Big\}\Ft, \\
I_3^{Z\ell} & = & -s - \mz^2 + \ml^2 + \Big\{ s(\msli^2-\mslj^2-\mz^2) 
               -(\mslj^2-\mz^2)(\mz^2-\ml^2) \nonumber \\
 & &    -\msli^2(\mz^2+\ml^2) + 2 \mz^2\mchi^2 \Big\}\Ft;
\end{eqnarray}
$\bullet$
 \underline{lepton\ ($\ell$) -- neutralino\ ($\neu{i}$) interference:}
\begin{eqnarray}
\wt_{Z\ell}^{(\ell - \neu{})} & = & 
  \frac{1}{\mz^2} \Re \sum_{i=1}^4
    \frac{\CAzq}{s-\ml^2} \nonumber \\
 & & \times 
 \Bigg[ 
  (\CAzc C_{+1i}^{aa}+\CVzc D_{+1i}^{aa}) I_4^{Z\ell}
   + (\CAzc C_{-1i}^{aa}+\CVzc D_{-1i}^{aa}) \cdot 2\ml I_5^{Z\ell}
\nonumber \\
 & & -(\CAzc C_{+1i}^{aa}-\CVzc D_{+1i}^{aa})\ml^2 I_6^{Z\ell}
     -(\CAzc C_{-1i}^{aa}-\CVzc D_{-1i}^{aa})\cdot 2\ml I_7^{Z\ell} \Bigg],
\nonumber \\
 & & 
\end{eqnarray}
where
\begin{eqnarray}
I_4^{Z\ell} & = & 
  s^2 +s(2\mnq^2+3\mz^2-\msli^2-\mchi^2-\ml^2)
      -(\msli^2-\mchi^2)(3\mz^2-\ml^2) \nonumber \\
 & & + 2 \Big\{ s^2 \mnq(\mchi+\mnq) + s[ -2\mz^2(\msli^2-\ml^2) \nonumber \\
 & &   +(\mnq^2-\mchi^2)(\mnq^2-\msli^2)
       -\ml^2(\mnq+\mchi)^2 + \mz^2\mnq(\mnq+\mchi) ] \nonumber \\
 & & -(\mz^2-\ml^2)(2\mz^2+\ml^2)\mchi\mnq + 2\mz^2\msli^4 \nonumber \\
 & & -\msli^2 [2\mz^2\mchi^2+\ml^2(\mchi^2+2\mz^2)] \nonumber \\
 & & -\mnq^2(2\mz^2-\ml^2)(\msli^2-\mchi^2)
              +\ml^2\mchi^2(\mchi^2+\mz^2+\ml^2) \Big\}\Fu, \\
I_5^{Z\ell} & = & 
  \mnq(-s+2\mz^2+\ml^2)+3\mz^2\mchi
       + \Big\{ s\mnq(\mchi^2+2\mz^2-\mnq^2) \nonumber \\
 & & + \mnq [\mz^2(2\mnq^2-\mchi^2-\msli^2-2\mz^2)
             +\ml^2(\mnq^2-\mchi^2-\mz^2) ] \nonumber \\
 & &    +3\mchi\mz^2(\mnq^2-\msli^2+\ml^2) \Big\}\Fu, \\
I_6^{Z\ell} & = & 
  s-2\mnq^2-\msli^2+3\mchi^2-3\mz^2-\ml^2
     -\frac{1}{s}(\mz^2-\ml^2)(\msli^2-\mchi^2) \nonumber \\
 & & - 2 \{ (\mnq+\mchi)^2-\mz^2 \}
               \{ (\mnq-\mchi)^2+2\mz^2 \} \Fu, \\
I_7^{Z\ell} & = & 
  \mchi(s+\mz^2-\ml^2) + \Big\{ 
s[ \mchi(\mnq^2-\mchi^2+\mz^2)-3\mz^2\mnq  ] \nonumber \\
 & & + \mchi [ \mnq^2(\mz^2-\ml^2) + \mz^2(\msli^2-2\mchi^2+2\mz^2)
               + \ml^2(\mchi^2-2\mz^2) ] \nonumber \\
 & & + 3 \mnq \mz^2 (\msli^2-\mchi^2) \Big\}\Fu ;
\end{eqnarray}
$\bullet$
 \underline{slepton\ ($\slp{b}$) -- neutralino\ ($\neu{i}$) interference:}
\begin{eqnarray}
\wt_{Z\ell}^{(\slp{}-\neu{})} & = & 
  \frac{1}{\mz^2} \Re \sum_{b=1}^2 \sum_{i=1}^4
    \CZsl{b}{a} \CAzq \nonumber \\
 & & \times \Bigg[ D_{+1i}^{ba} \Big\{ \Y_3 
- [s+\msli^2-\mchi^2+\mz^2+\ml^2 \nonumber \\
 & &    +\frac{3}{s}(\mz^2-\ml^2)(\msli^2-\mchi^2) ]\Y_2 
        + I_8^{Z\ell} \Y_1 + I_9^{Z\ell} \Y_0 \Big\} \nonumber \\
 & & 
 - D_{-1i}^{ba} \ml(\mnq-\mchi) \Big\{ 
         -2\Y_2 + (\msli^2-\mchi^2)\Y_1 \nonumber \\
 & & + [s(\msli^2+\mchi^2-2\mz^2)-\msli^4+\msli^2(3\mz^2-\ml^2) \nonumber \\
 & &    - \mchi^2(\mchi^2-3\mz^2)-\ml^2(\mchi^2+2\mz^2) ]\Y_0 \nonumber \\
 & & + \frac{1}{s}(\mz^2-\ml^2)(\msli^2-\mchi^2)
                               [\Y_1-(\msli^2-\mchi^2)\Y_0] \nonumber \\
 & & - \frac{1}{2s^2}(\mz^2-\ml^2)^2(\msli^2-\mchi^2)^2 \Y_0 \Big\} \Bigg],
\end{eqnarray}
where
\begin{eqnarray}
I_8^{Z\ell} & = & 
\mchi\mnq (s+\mz^2-\ml^2)+\frac{1}{2}\msli^2(2\mchi^2-3\mz^2) \nonumber \\
 & & - \frac{1}{2}\mz^2\mchi^2 
     + \frac{1}{2}\ml^2(\msli^2-\mchi^2+2\mz^2) \nonumber \\
 & & + \frac{1}{2s}(\mz^2-\ml^2)(\msli^2-\mchi^2)
                   (\msli^2-\mchi^2+\mz^2+\ml^2) \nonumber \\
 & & + \frac{3}{4s^2}(\mz^2-\ml^2)^2(\msli^2-\mchi^2)^2, \\
I_9^{Z\ell} & = & 
\frac{1}{2}s[ \msli^2(3\mz^2+3\ml^2-2\mchi^2)
               +\mchi^2(\mz^2+\ml^2)-6\ml^2\mz^2 ] \nonumber \\
 & & - \msli^4(\mchi^2+\mz^2+2\ml^2)
     + \msli^2 \{ \mchi^4+\mz^2\mchi^2-\mz^4
                  +\ml^2(\mchi^2+4\mz^2-2\ml^2) \} \nonumber \\
 & & - \mz^2 \{ \mchi^4+\mz^2\mchi^2
                  -\ml^2(4\mchi^2-\mz^2-\ml^2) \} \nonumber \\
 & & - \frac{1}{4s}(\mz^2-\ml^2)(\msli^2-\mchi^2)
       \{ 2\msli^4 + \msli^2(8\mchi^2-3\mz^2+7\ml^2) \nonumber \\
 & &            +2\mchi^4+3\mz^2\mchi^2+2\mz^4
                  +\ml^2(\mchi^2+8\mz^2+2\ml^2) \} \nonumber \\
 & & - \frac{1}{s}\mchi\mnq(s+\msli^2-\mchi^2)
          \{ (s-\ml^2)^2-2\mz^2(s+\ml^2)+\mz^4 \} \nonumber \\
 & & - \frac{1}{4s^2}(\mz^2-\ml^2)^2(\msli^2-\mchi^2)^2
         (\msli^2-\mchi^2+\mz^2+\ml^2) \nonumber \\
 & & - \frac{1}{4s^3}(\mz^2-\ml^2)^3(\msli^2-\mchi^2)^3.
\end{eqnarray}
%

\subsection{\boldmath ${\slp{a}\chi\rightarrow \gm\ell}$}
\label{slchigmel:sec}
This process proceeds via
the $s$--channel lepton ($\ell$) exchange, 
and the $t$--channel slepton ($\slp{a}$, $a=1,2$) exchange
\begin{eqnarray}
 \wt_{\slp{a}\chi\rightarrow \gm\ell} = 
        \wt_{\gm\ell}^{(\ell)} + \wt_{\gm\ell}^{(\slp{})}
                                  + \wt_{\gm\ell}^{(\ell - \slp{})}: 
\end{eqnarray}
%
$\bullet$
 \underline{lepton ($\ell$) exchange:}
\begin{eqnarray}
\wt_{\gm\ell}^{(\ell)} & = & 
  \frac{e^2}{(s-\ml^2)^2} \Bigg[ 
   C_{+11}^{aa} (s^2-6\ml^2 s+\ml^4)
    \left\{ 1-\frac{1}{s}(\msli^2-\mchi^2) \right\} \nonumber \\
 & & - C_{-11}^{aa} \cdot 4 \ml\mchi(s+\ml^2) \Bigg];
\end{eqnarray}
%
$\bullet$
 \underline{slepton ($\slp{a}$) exchange:}
\begin{eqnarray}
\wt_{\gm\ell}^{(\slp{})} & = & 
 2e^2 \Bigg[ C_{+11}^{aa} \Big\{ 
  \Tt_2 + (\msli^2-\mchi^2-\ml^2)\Tt_1 
 - (\mchi^2+\ml^2)\msli^2\Tt_0 \Big\} \nonumber \\
 & & - C_{-11}^{aa}\cdot 2\ml\mchi \Big\{ \Tt_1+\msli^2 \Tt_0 \Big\} \Bigg];
\end{eqnarray}
%
$\bullet$
 \underline{lepton\ ($\ell$) -- slepton\ ($\slp{a}$) interference:}
\begin{eqnarray}
\wt_{\gm\ell}^{(\ell - \slp{})} & = & 
  - \frac{2e^2}{s-\ml^2} \Re 
 \Bigg[ 
  C_{+11}^{aa} \Big\{ s+2\mchi^2-2\msli^2-2\ml^2 \nonumber \\
 & & + [ s(\msli^2+\mchi^2+\ml^2)
         -2(\msli^2-\mchi^2)^2
         -\ml^2(\msli^2+\mchi^2-\ml^2) ]\Ft \Big\} \nonumber \\
 & & + C_{-11}^{aa} \cdot 2\ml\mchi 
\Big\{ 1+(s+2\msli^2-2\mchi^2+\ml^2)\Ft \Big\} \Bigg].
\end{eqnarray}

\subsection{\boldmath ${\slp{a}\chi\rightarrow W^-\nu_{\ell}}$}
\label{slchiwnu:sec}
This process proceeds via
the $s$--channel lepton ($\ell$) exchange, 
the $t$--channel sneutrino ($\sn{\ell}$) exchange,
and the $u$--channel chargino ($\cha{k}$, $k=1,2$) exchange
\begin{eqnarray}
 \wt_{\slp{a}\chi\rightarrow W\nu_{\ell}} = 
     \wt_{W\nu_{\ell}}^{(\ell)} + \wt_{W\nu_{\ell}}^{(\sn{})}
      + \wt_{W\nu_{\ell}}^{(\cha{})} + \wt_{W\nu_{\ell}}^{(\ell - \sn{})}
      + \wt_{W\nu_{\ell}}^{(\ell - \cha{})} 
      + \wt_{W\nu_{\ell}}^{(\sn{}-\cha{})}: 
\end{eqnarray}
%
$\bullet$
 \underline{lepton ($\ell$) exchange:}
\begin{eqnarray}
\wt_{W\nu_{\ell}}^{(\ell)} & = & 
\frac{(C^{\ell\nu_{\ell}W^+})^2}{\mw^2(s-\ml^2)^2} 
          (s-\mw^2)(s+2\mw^2) \nonumber \\
 & & \times \Bigg[ \Big\{ \Big| C_L^{\chi\slp{a}\ell} \Big|^2
        + \frac{\ml^2}{s} \Big| C_R^{\chi\slp{a}\ell} \Big|^2 \Big\} 
       (s-\msli^2+\mchi^2)
+ C_{-11}^{aa} \cdot 4 \ml\mchi \Bigg]; \nonumber \\
 & & 
\end{eqnarray}
%
$\bullet$
 \underline{sneutrino ($\sn{\ell}$) exchange:}
\begin{eqnarray}
\wt_{W\nu_{\ell}}^{(\sn{})} & = & 
  \frac{2}{\mw^2} 
    \left| C^{\sn{\ell}^*\slp{a}W^+}C_L^{\chi\sn{\ell}\nu_{\ell}} \right|^2
\Bigg[ -\Tt_3 + (\mchi^2+2\msli^2+2\mw^2)\Tt_2 \nonumber \\
 & & -\{ (\msli^2-\mw^2)^2+2\mchi^2(\msli^2+\mw^2) \}\Tt_1 
                + \mchi^2(\msli^2-\mw^2)^2\Tt_0 \Bigg]; \nonumber \\
 & & 
\end{eqnarray}
%
$\bullet$
 \underline{chargino ($\cha{k}$) exchange:}
\begin{eqnarray}
\wt_{W\nu_{\ell}}^{(\cha{})} & = & 
  \frac{2}{\mw^2} \sum_{k,l=1}^2 
C_L^{\chi_k^+\slp{a}\nu_{\ell} *} C_L^{\chi_l^+\slp{a}\nu_{\ell}}
\Bigg[ 
\CWLp \CWLr \Big\{
\Tu_3 + (s-\msli^2-2\mchi^2-\mw^2)\Tu_2 \nonumber \\
 & & + [ -s(\mchi^2+2\mw^2) + \mchi^2(2\msli^2+\mchi^2+2\mw^2) ]\Tu_1 
      \nonumber \\
 & & -\msli^2(\mchi^2-\mw^2)(\mchi^2+2\mw^2)\Tu_0 \Big\} \nonumber \\
 & & + \CWRp \CWRr \mcmc \Big\{ -(s-2\mw^2)\Tu_1 \nonumber \\
 & & + [ s(\mchi^2+2\mw^2)-\mw^2(\msli^2+\mchi^2+2\mw^2) ] \Tu_0 
                                                  \Big\} \nonumber \\
 & & + \mchi \Big( \mcp\CWRp\CWLr + \mcr\CWLp\CWRr \Big) 
     \cdot 3\mw^2 \Big\{ -\Tu_1+\msli^2\Tu_0 \Big\} \Bigg]; \nonumber \\
\end{eqnarray}
%
$\bullet$
 \underline{lepton\ ($\ell$) -- sneutrino\ ($\sn{\ell}$) interference:}
\begin{eqnarray}
\wt_{W\nu_{\ell}}^{(\ell - \sn{})} & = & 
  - \frac{2}{\mw^2} \Re 
    \frac{C^{\ell\nu_{\ell} W^+ *} C^{\sn{\ell}^*\slp{a}W^+}  
                    C_L^{\chi\sn{\ell}\nu_{\ell}}}{s-\ml^2} \nonumber \\
 & & \times \Bigg[ C_L^{\chi\slp{a}\ell}
    \Big\{ s^2 + s(\msli^2+\mchi^2+\mw^2-2\msn^2)
                            -3\mw^2(\msli^2-\mchi^2) \nonumber \\
 & & -2 \Big[ s \{ \msn^4-\msn^2(\msli^2+\mchi^2+\mw^2)
                             +\mchi^2(\msli^2-\mw^2)\} \nonumber \\
 & & + 2\mw^2 \{ \msn^2(\msli^2-\mchi^2)
                 + \mchi^2(\mchi^2+\mw^2-\msli^2) \} \Big]\Ft \Big\}
\nonumber \\
 & & 
 + C_R^{\chi\slp{a}\ell} \cdot 2\ml\mchi 
\Big\{ s+\mw^2 + \Big[ s(\msn^2+\mw^2-\msli^2) \nonumber \\
 & &     +\mw^2(\msli^2+\msn^2-2\mchi^2-\mw^2) \Big]\Ft \Big\} \Bigg];
\end{eqnarray}
%
$\bullet$
 \underline{lepton\ ($\ell$) -- chargino\ ($\cha{k}$) interference:}
\begin{eqnarray}
\wt_{W\nu_{\ell}}^{(\ell - \cha{})} & = & 
  \frac{2}{\mw^2} \Re \sum_{k=1}^2
    \frac{C^{\ell\nu_{\ell} W^+ *}
          C_L^{\chi_k^+\slp{a}\nu_{\ell}}}{s-\ml^2} \nonumber \\
 & & \times 
 \Bigg[ C_L^{\chi\slp{a}\ell}\CWLp 
   \Big\{ s^2+s(2\mcp^2+3\mw^2-\msli^2-\mchi^2)
                       -3\mw^2(\msli^2-\mchi^2) \nonumber \\
 & & + 2 \Big[ s^2\mcp^2 + s\{ (\mcp^2-\mchi^2)(\mcp^2-\msli^2)
                               +\mw^2(\mcp^2-2\msli^2) \} \nonumber \\
 & & + 2\mw^2(\msli^2-\mchi^2)(\msli^2-\mcp^2) \Big]\Fu \Big\} \nonumber \\
 & & 
 + C_R^{\chi\slp{a}\ell}\CWRp \cdot 2\ml\mcp 
   \Big\{ s-2\mw^2 \nonumber \\
 & & + \Big[ s(\mcp^2-\mchi^2-2\mw^2)
     +\mw^2(\msli^2+\mchi^2+2\mw^2-2\mcp^2) \Big]\Fu \Big\} \nonumber \\
 & & 
 + C_R^{\chi\slp{a}\ell}\CWLp \cdot 6\ml\mchi\mw^2 
       \Big\{ 1 + (\mcp^2-\msli^2)\Fu \Big\} \nonumber \\
 & & 
 - C_L^{\chi\slp{a}\ell}\CWRp \cdot 2\mchi\mcp (s-\mw^2)(s+2\mw^2)\Fu
     \Bigg]; 
\end{eqnarray}
%
$\bullet$
 \underline{sneutrino\ ($\sn{\ell}$) -- chargino\ ($\cha{k}$) interference:}
\begin{eqnarray} 
\wt_{W\nu_{\ell}}^{(\sn{}-\cha{})} & = & 
  \frac{2}{\mw^2} \Re \sum_{k=1}^2 
    C^{\sn{\ell}^*\slp{a}W^+ *} C_L^{\chi\sn{\ell}\nu_{\ell}}
                            C_L^{\chi_k^+\slp{a}\nu_{\ell}}
      \nonumber \\
 & & \times \Bigg[ \CWLp \Big\{ \Y_3 
- \Big[ s+\msli^2-\mchi^2+\mw^2+\frac{3}{s}\mw^2(\msli^2-\mchi^2) \Big]\Y_2
\nonumber \\
 & & + \frac{1}{2} \Big[ 
        \msli^2 (2\mchi^2-3\mw^2)-\mw^2\mchi^2
        +\frac{3}{2s^2}\mw^4(\msli^2-\mchi^2)^2 \nonumber \\
 & & + \frac{1}{s}\mw^2(\msli^2-\mchi^2)(\msli^2-\mchi^2+\mw^2)
            \Big] \Y_1 + I^{W\nu} \Y_0 \Big\} \nonumber \\
 & & 
 + \CWRp \mchi\mcp 
     \Big\{ -(s+\mw^2)\Y_1 
     + (s-\mw^2)^2 \Big[ 1+\frac{1}{s}(\msli^2-\mchi^2)\Big] \Y_0
                                                   \Big\} \Bigg],
\nonumber \\
 & & 
\end{eqnarray}
where
\begin{eqnarray}
I^{W\nu} & = & 
\frac{1}{2}s[ \msli^2(3\mw^2-2\mchi^2)
               +\mw^2\mchi^2 ] \nonumber \\
 & & - \msli^4(\mchi^2+\mw^2)
     + \msli^2(\mchi^4+\mw^2\mchi^2-\mw^4)
     - \mw^2\mchi^2(\mchi^2+\mw^2) \nonumber \\
 & & - \frac{1}{4s}\mw^2(\msli^2-\mchi^2)
       \{ 2\msli^4 + \msli^2(8\mchi^2-3\mw^2) 
                   +2\mchi^4+3\mw^2\mchi^2+2\mw^4 \} \nonumber \\
 & & - \frac{1}{4s^2}\mw^4(\msli^2-\mchi^2)^2(\msli^2-\mchi^2+\mw^2)
     - \frac{1}{4s^3}\mw^6(\msli^2-\mchi^2)^3. \nonumber \\
 & & 
\end{eqnarray}
%

\subsection{\boldmath ${\slp{a}\chi \rightarrow h\ell, H\ell}$}
\label{slchihHel:sec}
The process ${\slp{a}\chi \rightarrow h\ell}$ involves 
the $s$--channel lepton ($\ell$) exchange, 
the $t$--channel slepton ($\slp{a}$, $a=1,2$) exchange,
and the $u$--channel neutralino ($\neu{i}$, $i=1,2,3,4$) exchange
\begin{eqnarray}
 \wt_{\slp{a}\chi \rightarrow h\ell} = 
        \wt_{h\ell}^{(\ell)} + \wt_{h\ell}^{(\slp{})}
      + \wt_{h\ell}^{(\neu{})} + \wt_{h\ell}^{(\ell - \slp{})} 
      + \wt_{h\ell}^{(\ell - \neu{})} + \wt_{h\ell}^{(\slp{}-\neu{})}: 
\end{eqnarray}
%
$\bullet$
 \underline{lepton ($\ell$) exchange:}
\begin{eqnarray}
\wt_{h\ell}^{(\ell)} & = & 
\frac{1}{2} \left| \frac{C_S^{\ell\ell h}}{s-\ml^2} \right|^2
\Bigg[ 
4 C_{-11}^{aa}\ml\mchi(2s+2\ml^2-\mh^2) \nonumber \\
 & & 
+ C_{+11}^{aa} \Big\{ s^2-s(\mh^2-6\ml^2)-\ml^2(\mh^2-\ml^2) \Big\}
             \Big\{ 1-\frac{1}{s}(\msli^2-\mchi^2) \Big\} \Bigg]; 
\nonumber \\
 & & 
\end{eqnarray}
%
$\bullet$
 \underline{slepton ($\slp{b}$) exchange:}
\begin{eqnarray}
\wt_{h\ell}^{(\slp{})} & = & 
  \sum_{b,c=1}^2 C_S^{\slp{b}^*\slp{a}h} C_S^{\slp{c}^*\slp{a}h *}
  \Big[ C_{+11}^{bc} \{ -\Tt_1 + (\mchi^2+\ml^2)\Tt_0 \}
                        + 2 C_{-11}^{bc}\ml\mchi \Tt_0 \Big]; 
\nonumber \\
 & & 
\end{eqnarray}
%
$\bullet$
 \underline{neutralino ($\neu{i}$) exchange:}
\begin{eqnarray}
\wt_{h\ell}^{(\neu{})} & = & 
\sum_{i,j=1}^4 C_S^{\neu{i}\chi h} C_S^{\neu{j}\chi h *}
\Bigg[ C_{+ji}^{aa} \Big\{ 
  -(s-\mchi^2-\ml^2)\Tu_1 - (\msli^2-\ml^2)(\mchi^2-\mh^2)\Tu_0 \nonumber \\
 & & + \mchi(\mnpmn) [\Tu_1 - (\msli^2-\ml^2)\Tu_0] 
     + \mnmn [\Tu_1 + (s-\msli^2-\mh^2)\Tu_0] \Big\} \nonumber \\
 & & + C_{-ji}^{aa}\ml \Big\{ 
 2\mchi\Tu_1 +(\mnpmn)[\Tu_1 + (\mchi^2-\mh^2)\Tu_0] 
     + \mnmn \cdot 2\mchi \Tu_0 \Big\} \Bigg]; \nonumber \\
\end{eqnarray}
%
$\bullet$
 \underline{lepton\ ($\ell$) -- slepton\ ($\slp{b}$) interference:}
\begin{eqnarray}
\wt_{h\ell}^{(\ell - \slp{})} & = & 
  2 \Re \sum_{b=1}^2 
    \frac{C_S^{\ell\ell h *} C_S^{\slp{b}^*\slp{a}h}}{s-\ml^2}
 \Big[ C_{-11}^{ab}\mchi(s-\mh^2+3\ml^2)\Ft  \nonumber \\
 & &   + C_{+11}^{ab}\ml 
           \Big\{ -1 + (s+2\mchi^2+\ml^2-\msli^2-\mslj^2)\Ft \Big\} \Big];
\end{eqnarray}
%
$\bullet$
 \underline{lepton\ ($\ell$) -- neutralino\ ($\neu{i}$) interference:}
\begin{eqnarray}
\wt_{h\ell}^{(\ell - \neu{})} & = & 
  2 \Re \sum_{i=1}^4
    \frac{C_S^{\ell\ell h *} C_S^{\neu{i}\chi h}}{s-\ml^2}
 \Big[ 
 C_{+1i}^{aa}\ml \Big\{ \mnq+\mchi \nonumber \\
 & &  + [ 2\mnq s + (\mnq+\mchi)(\mnq^2+\mchi^2-2\msli^2-\mh^2)
                                   + 2\mchi\ml^2 ]\Fu \Big\} \nonumber \\
 & & + C_{-1i}^{aa} \Big\{ s+\ml^2 
    + [ s\mnq(\mnq+\mchi) - \mchi\mnq(\mh^2-3\ml^2)\nonumber \\
 & &    - \msli^2\mh^2 + \ml^2(\mnq^2+2\mchi^2-\mh^2) ]\Fu \Big\} \Big];
\nonumber \\
 & & 
\end{eqnarray}
%
$\bullet$
 \underline{slepton\ ($\slp{b}$) -- neutralino\ ($\neu{i}$) interference:}
\begin{eqnarray}
\wt_{h\ell}^{(\slp{}-\neu{})} & = & 
  \Re \sum_{b=1}^2 \sum_{i=1}^4
   C_S^{\slp{b}^*\slp{a}h *} C_S^{\neu{i}\chi h} \nonumber \\
 & & \times \Bigg[ 
C_{+1i}^{ba} \Big\{ -(\mnq+\mchi)\Y_1 \nonumber \\
 & & + \Big[
 -\mnq(s-\mh^2+\ml^2)\{1-\frac{1}{s}(\msli^2-\mchi^2) \} \nonumber \\
 & & + \mchi \{ s+\msli^2-\mh^2-\mchi^2-3\ml^2
        -\frac{1}{s} (\mh^2-\ml^2)(\msli^2-\mchi^2) \} \Big] \Y_0 \Big\}
 \nonumber \\
 & & + C_{-1i}^{ba} \ml \Big\{ \Y_1 
 + \Big[ s-\msli^2+\mh^2-3\mchi^2-\ml^2-4\mchi\mnq \nonumber \\
 & &    -\frac{1}{s} (\mh^2-\ml^2)(\msli^2-\mchi^2) \Big]\Y_0 \Big\} \Bigg].
\nonumber \\
 & & 
\end{eqnarray}
The contribution from ${\slp{a}\chi \rightarrow H\ell}$ can be
obtained by a simple substitution $h\rightarrow H$. 

\subsection{\boldmath ${\slp{a}\chi \rightarrow A\ell}$}
\label{slchiael:sec}
This process involves 
the $s$--channel lepton ($\ell$) exchange, 
the $t$--channel slepton ($\slp{a}$, $a=1,2$) exchange,
and the $u$--channel neutralino ($\neu{i}$, $i=1,2,3,4$) exchange
\begin{eqnarray}
 \wt_{\slp{a}\chi \rightarrow A\ell} = 
        \wt_{A\ell}^{(\ell)} + \wt_{A\ell}^{(\slp{})}
      + \wt_{A\ell}^{(\neu{})} + \wt_{A\ell}^{(\ell - \slp{})} 
      + \wt_{A\ell}^{(\ell - \neu{})} + \wt_{A\ell}^{(\slp{}-\neu{})}: 
\end{eqnarray}
%
$\bullet$
 \underline{lepton ($\ell$) exchange:}
\begin{eqnarray}
\wt_{A\ell}^{(\ell)} & = & 
\frac{1}{2} \left| \frac{C_P^{\ell\ell A}}{s-\ml^2} \right|^2
\Bigg[ 
- 4 C_{-11}^{aa}\ml\mchi\ma^2 \nonumber \\
 & & 
+ C_{+11}^{aa} \Big\{ (s-\ml^2)^2-\ma^2(s+\ml^2) \Big\}
             \Big\{ 1-\frac{1}{s}(\msli^2-\mchi^2) \Big\} \Bigg];
\end{eqnarray}
%
$\bullet$
 \underline{slepton ($\slp{b}$) exchange:}
\begin{eqnarray}
\wt_{A\ell}^{(\slp{})} & = & 
  \sum_{b,c=1}^2 C_P^{\slp{b}^*\slp{a}A} C_P^{\slp{c}^*\slp{a}A *}
  \Big[ C_{+11}^{bc} \{ -\Tt_1 + (\mchi^2+\ml^2)\Tt_0 \}
                        + 2 C_{-11}^{bc}\ml\mchi \Tt_0 \Big];
\nonumber \\
 & & 
\end{eqnarray}
%
$\bullet$
 \underline{neutralino ($\neu{i}$) exchange:}
\begin{eqnarray}
\wt_{A\ell}^{(\neu{})} & = & 
\sum_{i,j=1}^4 C_P^{\neu{i}\chi A} C_P^{\neu{j}\chi A *}
\Bigg[ C_{+ji}^{aa} \Big\{ 
  -(s-\mchi^2-\ml^2)\Tu_1 - (\msli^2-\ml^2)(\mchi^2-\ma^2)\Tu_0 \nonumber \\
 & & - \mchi(\mnpmn) [\Tu_1 - (\msli^2-\ml^2)\Tu_0] 
     + \mnmn [\Tu_1 + (s-\msli^2-\ma^2)\Tu_0] \Big\} \nonumber \\
 & & - C_{-ji}^{aa}\ml \Big\{ 
 2\mchi\Tu_1 -(\mnpmn)[\Tu_1 + (\mchi^2-\ma^2)\Tu_0] 
     + \mnmn \cdot 2\mchi \Tu_0 \Big\} \Bigg]; \nonumber \\
 & & 
\end{eqnarray}
%
$\bullet$
 \underline{lepton\ ($\ell$) -- slepton\ ($\slp{b}$) interference:}
\begin{eqnarray}
\wt_{A\ell}^{(\ell - \slp{})} & = & 
  2 \Re \sum_{b=1}^2 
    \frac{C_P^{\ell\ell A *} C_P^{\slp{b}^*\slp{a}A}}{s-\ml^2}
 \Big[ D_{-11}^{ab}\mchi(s-\ma^2-\ml^2)\Ft  \nonumber \\
 & &   + D_{+11}^{ab}\ml 
           \Big\{ -1 - (s+\mslj^2-\msli^2-\ml^2)\Ft \Big\} \Big]; 
\end{eqnarray}
%
$\bullet$
 \underline{lepton\ ($\ell$) -- neutralino\ ($\neu{i}$) interference:}
\begin{eqnarray}
\wt_{A\ell}^{(\ell - \neu{})} & = & 
  2 \Re \sum_{i=1}^4
    \frac{C_P^{\ell\ell A *} C_P^{\neu{i}\chi A}}{s-\ml^2}
 \Big[ 
 C_{+1i}^{aa}\ml \Big\{ \mnq-\mchi \nonumber \\
 & &  + [ (\mnq-\mchi)(\mnq^2-\mchi^2)-(\mnq+\mchi)\ma^2 ]\Fu \Big\}
\nonumber \\
 & & + C_{-1i}^{aa} \Big\{ -s+\ml^2 
    + [ -s\mnq(\mnq-\mchi) - \mchi\mnq(\ma^2+\ml^2)\nonumber \\
 & &    + \msli^2\ma^2 + \ml^2(\mnq^2-\ma^2) ]\Fu \Big\} \Big];
\nonumber \\
 & & 
\end{eqnarray}
%
$\bullet$
 \underline{slepton\ ($\slp{b}$) -- neutralino\ ($\neu{i}$) interference:}
\begin{eqnarray}
\wt_{A\ell}^{(\slp{}-\neu{})} & = & 
  \Re \sum_{b=1}^2 \sum_{i=1}^4
   C_P^{\slp{b}^*\slp{a}A *} C_P^{\neu{i}\chi A} \nonumber \\
 & & \times \Bigg[ 
D_{+1i}^{ba} \Big\{ (\mnq-\mchi)\Y_1 \nonumber \\
 & & + \Big[
 -\mnq(s-\ma^2+\ml^2)\{1-\frac{1}{s}(\msli^2-\mchi^2) \} \nonumber \\
 & & - \mchi \{ s+\msli^2-\ma^2-\mchi^2-3\ml^2
        -\frac{1}{s} (\ma^2-\ml^2)(\msli^2-\mchi^2) \} \Big] \Y_0 \Big\}
 \nonumber \\
 & & - D_{-1i}^{ba} \ml \Big\{ \Y_1 
 + \Big[ s-\msli^2+\ma^2-3\mchi^2-\ml^2+4\mchi\mnq \nonumber \\
 & &    -\frac{1}{s}(\ma^2-\ml^2)(\msli^2-\mchi^2) \Big]\Y_0 \Big\} \Bigg].
\nonumber \\
 & & 
\end{eqnarray}

\subsection{\boldmath ${\slp{a}\chi\rightarrow H^-\nu_{\ell}}$}
\label{slchihnu:sec}
This process proceeds via
the $s$--channel lepton ($\ell$) exchange, 
the $t$--channel sneutrino ($\sn{\ell}$) exchange,
and the $u$--channel chargino ($\cha{k}$, $k=1,2$) exchange
\begin{eqnarray}
 \wt_{\slp{a}\chi \rightarrow H^-\nu_{\ell}} = 
        \wt_{H^-\nu_{\ell}}^{(\ell)} + \wt_{H^-\nu_{\ell}}^{(\sn{})}
      + \wt_{H^-\nu_{\ell}}^{(\cha{})} + \wt_{H^-\nu_{\ell}}^{(\ell - \sn{})}
      + \wt_{H^-\nu_{\ell}}^{(\ell - \cha{})} 
      + \wt_{H^-\nu_{\ell}}^{(\sn{}-\cha{})}: 
\end{eqnarray}
%
$\bullet$
 \underline{lepton ($\ell$) exchange:}
\begin{eqnarray}
\wt_{H^-\nu_{\ell}}^{(\ell)} & = & 
\left| \frac{C_S^{\nu_{\ell}\ell H^+}}{s-\ml^2} \right|^2 (s-\mch^2)
\Bigg[ 
4 C_{-11}^{aa}\ml\mchi \nonumber \\
 & & 
+ \Big( s \Big|C_R^{\chi\slp{a}\ell} \Big|^2 
       + \ml^2 \Big| C_L^{\chi\slp{a}\ell} \Big|^2 \Big)
             \Big\{ 1-\frac{1}{s}(\msli^2-\mchi^2) \Big\} \Bigg];
\nonumber \\
 & & 
\end{eqnarray}
%
$\bullet$
 \underline{sneutrino ($\sn{\ell}$) exchange:}
\begin{eqnarray}
\wt_{H^-\nu_{\ell}}^{(\sn{})} & = & 
 2 \Big| C_S^{\sn{\ell}^*\slp{a}H^+} C_L^{\chi\sn{\ell}\nu_{\ell}} \Big|^2 
    \Big[ -\Tt_1 + \mchi^2\Tt_0 \Big]; 
\end{eqnarray}
%
$\bullet$
 \underline{chargino ($\cha{k}$) exchange:}
\begin{eqnarray}
\wt_{H^-\nu_{\ell}}^{(\cha{})} & = & 
2 \sum_{k,l=1}^2 C_L^{\chi_k^+\slp{a}\nu_{\ell} *}
                 C_L^{\chi_l^+\slp{a}\nu_{\ell}}  
 \nonumber \\
 & &  \times
\Bigg[ C_R^{\chi_k^+\chi H^- *} C_R^{\chi_l^+\chi H^-} 
 \Big\{ -(s-\mchi^2)\Tu_1 + \msli^2(\mch^2-\mchi^2)\Tu_0 \Big\} \nonumber \\
 & & 
-\mchi \Big( \mcp C_L^{\chi_k^+\chi H^- *} C_R^{\chi_l^+\chi H^-} 
            + \mcr C_R^{\chi_k^+\chi H^- *} C_L^{\chi_l^+\chi H^-} \Big)
\Big\{ -\Tu_1 + \msli^2\Tu_0 \Big\} \nonumber \\
 & & + C_L^{\chi_k^+\chi H^- *} C_L^{\chi_l^+\chi H^-}
        \mcmc \Big\{ \Tu_1 + (s-\mch^2-\msli^2)\Tu_0 \Big\} \Bigg];
\end{eqnarray}
%
$\bullet$
 \underline{lepton\ ($\ell$) -- sneutrino\ ($\sn{\ell}$) interference:}
\begin{eqnarray}
\wt_{H^-\nu_{\ell}}^{(\ell - \sn{})} & = & 
  4 \Re \frac{C_S^{\sn{\ell}\slp{a}H^+} C_L^{\chi\sn{\ell}\nu_{\ell} *} 
                         C_S^{\nu_{\ell}\ell H^+ *}}{s-\ml^2}\nonumber \\
 & & \times 
 \Big[ C_L^{\chi\slp{a}\ell} \ml \{ -1-(\msn^2-\mchi^2)\Ft \}
     + C_R^{\chi\slp{a}\ell} \mchi (s-\mch^2)\Ft \Big]; \nonumber \\
 & &
\end{eqnarray}
%
$\bullet$
 \underline{lepton\ ($\ell$) -- chargino\ ($\cha{k}$) interference:}
\begin{eqnarray}
\wt_{H^-\nu_{\ell}}^{(\ell - \cha{})} & = & 
  4 \Re \sum_{k=1}^2
    \frac{C_S^{\nu_{\ell}\ell H^+ *} 
          C_L^{\chi_k^+\slp{a}\nu_{\ell}}}{s-\ml^2}
 \Bigg[ 
       C_R^{\chi\slp{a}\ell }C_R^{\chi_k^+\chi H^- *}
    \Big\{ s + (s\mcp^2-\mch^2\msli^2)\Fu \Big\} \nonumber \\
 & & + C_L^{\chi\slp{a}\ell }C_L^{\chi_k^+\chi H^- *}\ml\mcp
    \Big\{ 1 + (s+\mcp^2-\msli^2-\mch^2)\Fu \Big\} \nonumber \\
 & & + C_R^{\chi\slp{a}\ell }C_L^{\chi_k^+\chi H^- *}\mchi\mcp
                 (s-\mch^2)\Fu  \nonumber \\
 & & + C_L^{\chi\slp{a}\ell }C_R^{\chi_k^+\chi H^- *}\ml\mchi
    \Big\{ 1 + (\mcp^2-\msli^2)\Fu \Big\} \Bigg];
\nonumber \\
 & & 
\end{eqnarray}
%
$\bullet$
 \underline{sneutrino\ ($\sn{}$) -- chargino\ ($\cha{k}$) interference:}
\begin{eqnarray}
\wt_{H^-\nu_{\ell}}^{(\sn{}-\cha{})} & = & 
  -2 \Re \sum_{k=1}^2 
   C_S^{\sn{\ell}^*\slp{a}H^+ *} C_L^{\chi\sn{\ell}\nu_{\ell}} 
                             C_L^{\chi_k^+\slp{a}\nu_{\ell}} \nonumber \\
 & & \times \Bigg[ 
 -\Big(\mcp C_L^{\chi_k^+\chi H^- *} 
       + \mchi C_R^{\chi_k^+\chi H^- *} \Big)\Y_1 \nonumber \\
 & & + \Big[ \mcp  C_L^{\chi_k^+\chi H^- *} 
              \Big\{ 1-\frac{1}{s}(\msli^2-\mchi^2) \Big\}\nonumber \\
 & &        -\mchi C_R^{\chi_k^+\chi H^- *} 
              \Big\{ 1+\frac{1}{s}(\msli^2-\mchi^2) \Big\} \Big]
                                                (s-\mch^2)\Y_0 \Bigg]. 
\nonumber \\
 & & 
\end{eqnarray}
%

\begin{figure}[h!]
\vspace*{-0.1in}
\hspace*{-.12in}
\begin{minipage}{8in}
\epsfig{file=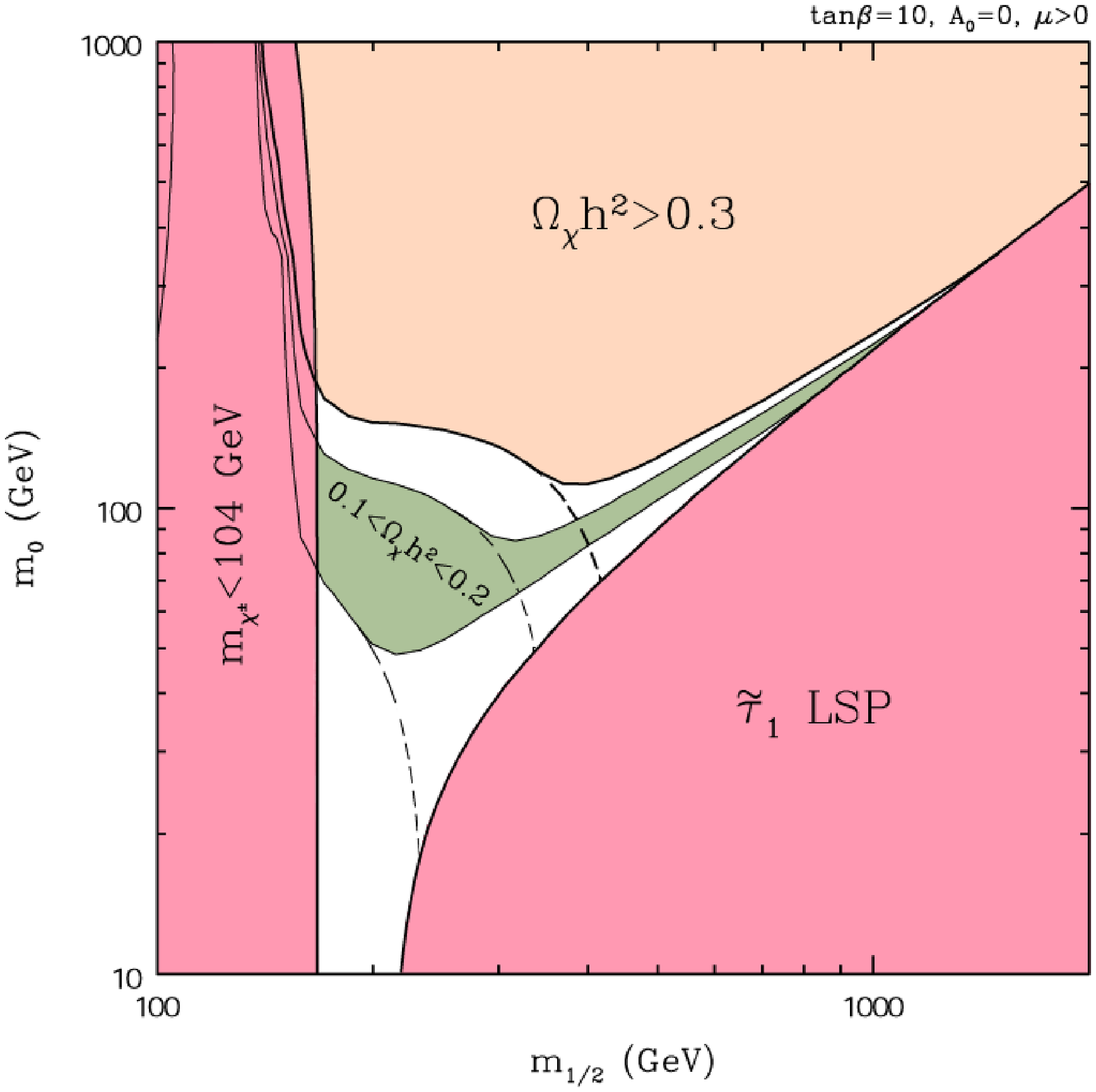,height=3.15in} 
\hspace*{-0.03in}
\epsfig{file=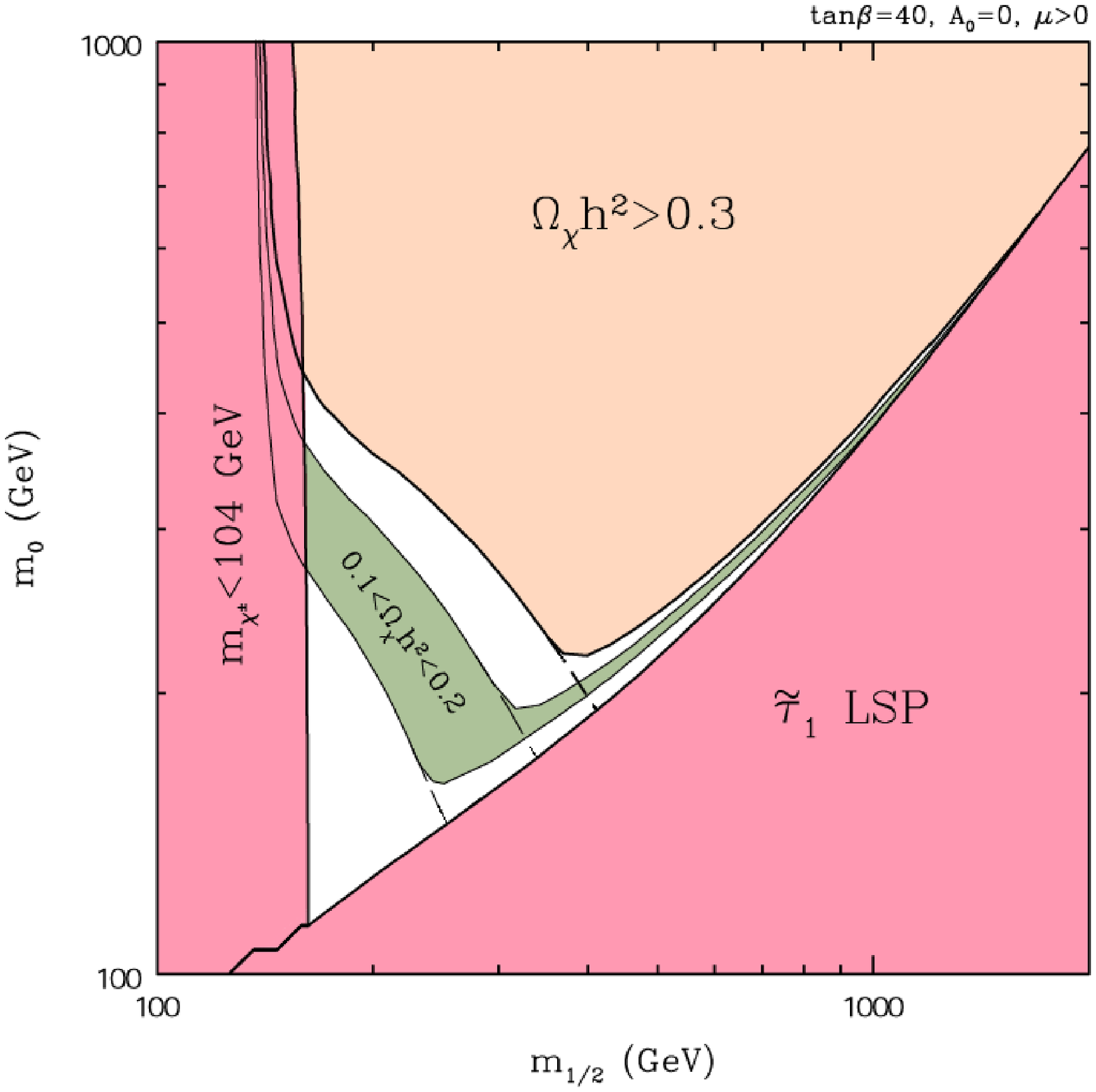,height=3.15in} 
\end{minipage}
\caption{\label{fig:oh2cont} 
Contours of $\abundchi$ in the plane 
($\mhalf,\mzero$) for $\tanb=10$ (left window) and
$\tanb=40$ (right window), and for
$\azero=0$, $\mu>0$, $\mtpole=175\gev$ and
$\mbmbsmmsbar=4.20\gev$. The red regions bands are
excluded by chargino searches at LEP and corresponds to
the lighter stau being the LSP. The light orange regions of
$\abundchi>0.3$ are excluded by cosmology while the narrow green bands
correspond to the expected range $0.1<\abundchi<0.2$. Also shown are
the semi--oval contours of $\abundchi$ in the absence of coannihilation.
}
\end{figure}

\section{Numerical Analysis}\label{numanalysis:sec}

In this Section we present some numerical examples to illustrate the
effect of the neutralino--slepton coannihilation. We will work in the
framework of the CMSSM where the effect considered in this paper is
particularly important.  In computing the neutralino WIMP relic
abundance we will employ the exact expressions for the cross--sections
of neutralino pair--annihilation derived in~\cite{nrr2} and the
neutralino--slepton coannihilation ones listed above. To generate mass
spectra we will use the package SUSPECT~(v.2.05)~\cite{suspectweb},
which includes full one--loop radiative corrections to
sfermion masses as well as to the effective potential.

We begin by presenting in Fig.~\ref{fig:oh2cont} contours of the relic
abundance $\abundchi$ in the plane $(\mhalf,\mzero)$ for $\tanb=10$
(left window) and $40$ (right window), and for $\azero=0$ and
$\mu>0$. The solid (dashed) curves correspond to including
(neglecting) the coannihilation effect. The light orange region of
$\abundchi>0.3$ is inconsistent with the age of the Universe while the
green band corresponding to $0.1<\abundchi<0.2$ is favored by direct
measurements of the dark matter component in the Universe. For the
sake of clarity, we only denote (in red) the regions of the plane
where the lighter stau $\stauone$ is the LSP (and in some part of it
one of the Higgs boson mass--square is negative), as well as those
excluded by the LEP limit on the lightest chargino but no other
experimental bounds. In particular, we do not indicate the regions
inconsistent with the lightest Higgs boson mass bound from LEP, nor
with $\br( B\ra X_s \gamma )$. (Their effect has been presented
in~\cite{rrn1} with updates in~\cite{lrmoriond02}.)  We only note that
these bounds, if taken at face value, exclude the favored regions of
$0.1<\abundchi<0.2$ (green bands) of $\mhalf\lsim320\gev$ ($\tanb=10$)
and $\mhalf\lsim550\gev$ ($\tanb=40$). In reality, theoretical
uncertainties may considerably weaken these bounds, but in any case,
it is clear that the remaining region is allowed mostly only by the
coannihilation effect.

Note that the narrow band opened up by
coannihilation eventually ends at the boundary of equal neutralino and
stau masses. This is in qualitative agreement with the results
obtained in~\cite{efo98,efos00,efgos01} using the usual partial wave
expansion but not with the more recent analysis~\cite{bbb02} where
the cosmologically allowed region appears to lie basically parallel
to the boundary even at very large $\mhalf$. We will come back to
discussing the upper limit on $\mchi$ below.

\begin{figure}[t!]
\vspace*{-0.1in}
\hspace*{-.12in}
\begin{minipage}{8in}
\epsfig{file=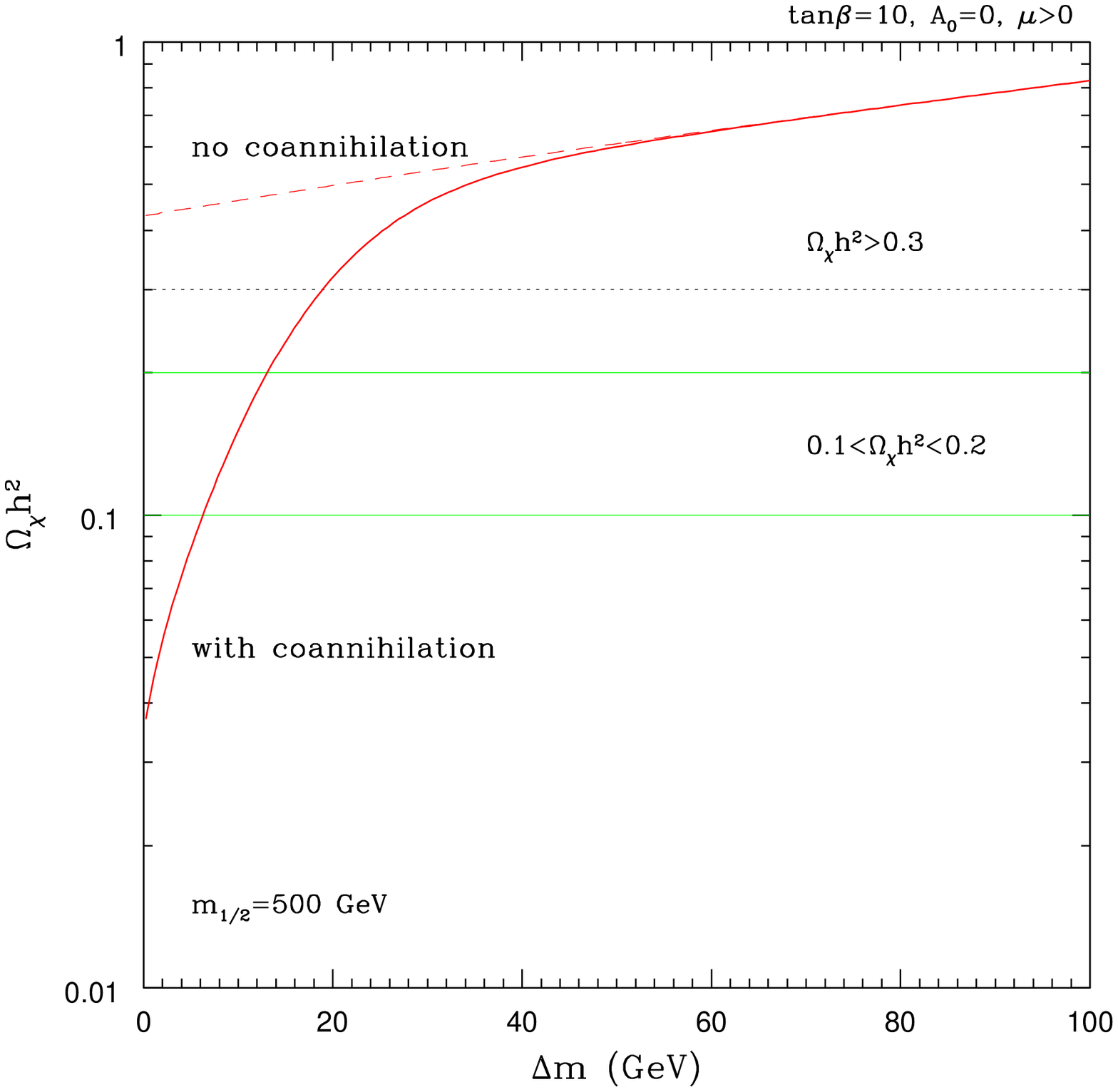,height=3.15in} 
\hspace*{-0.03in}
\epsfig{file=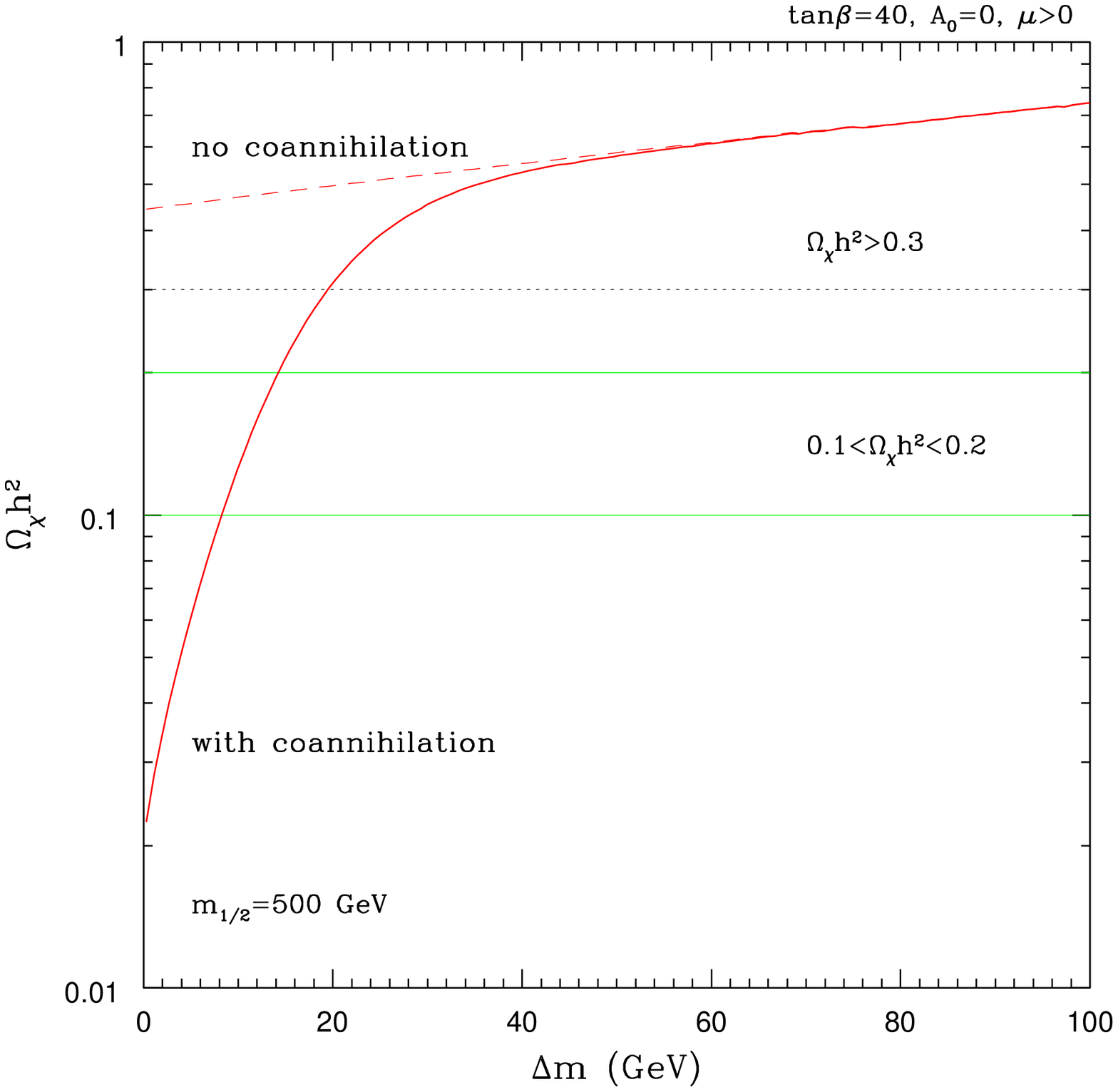,height=3.15in} 
\end{minipage}
\caption{\label{fig:oh2delm} The relic abundance $\abundchi$ with
(solid) and without (dash) coannihilation 
\vs\ $\Delta m = \mchi -
m_{\stauone}$ for $\tanb=10$ (left window) and $40$ (right window) at
a fixed value of $\mhalf=500\gev$. Also marked are the cosmologically
excluded ($\abundchi>0.3$) and favored ($0.1<\abundchi<0.2$) regions. 
}
\end{figure}

The effect of coannihilation becomes dramatic when the two
coannihilating particles are nearly degenerate in mass, but can be
significant even for the mass difference of some $20-40\gev$. This is
illustrated in Fig.~\ref{fig:oh2delm} where the relic abundance
$\abundchi$ is plotted as a function of $\Delta m = \mchi -
m_{\stauone}$ for $\tanb=10$ (left window) and $40$ (right window) at
a fixed value of $\mhalf=500\gev$. Also marked are the cosmologically
excluded ($\abundchi>0.3$) and favored ($0.1<\abundchi<0.2$) regions.

\begin{figure}[t!]
\vspace*{-0.1in}
\hspace*{-.12in}
\begin{minipage}{8in}
\epsfig{file=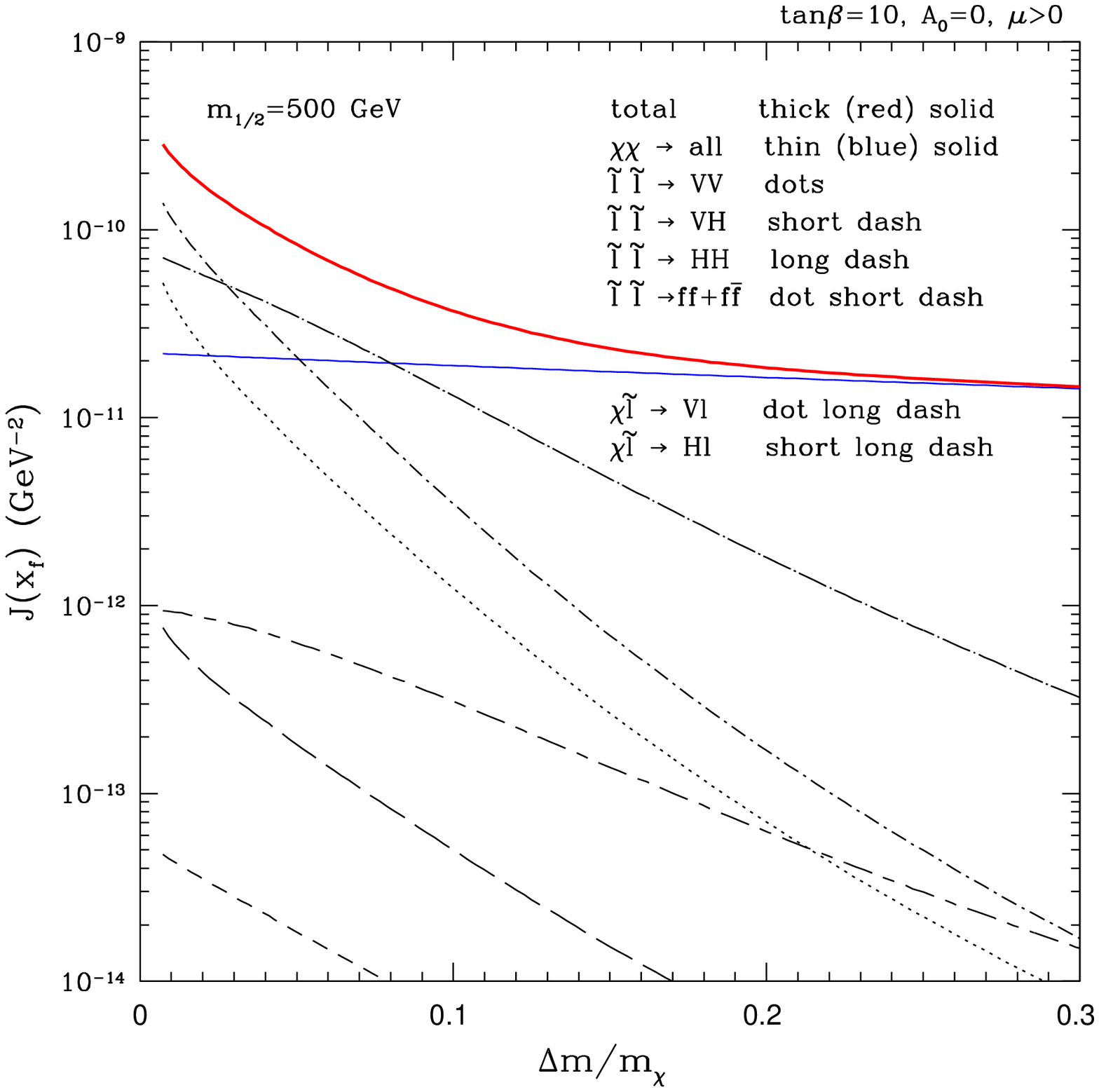,height=3.15in} 
\hspace*{-0.03in}
\epsfig{file=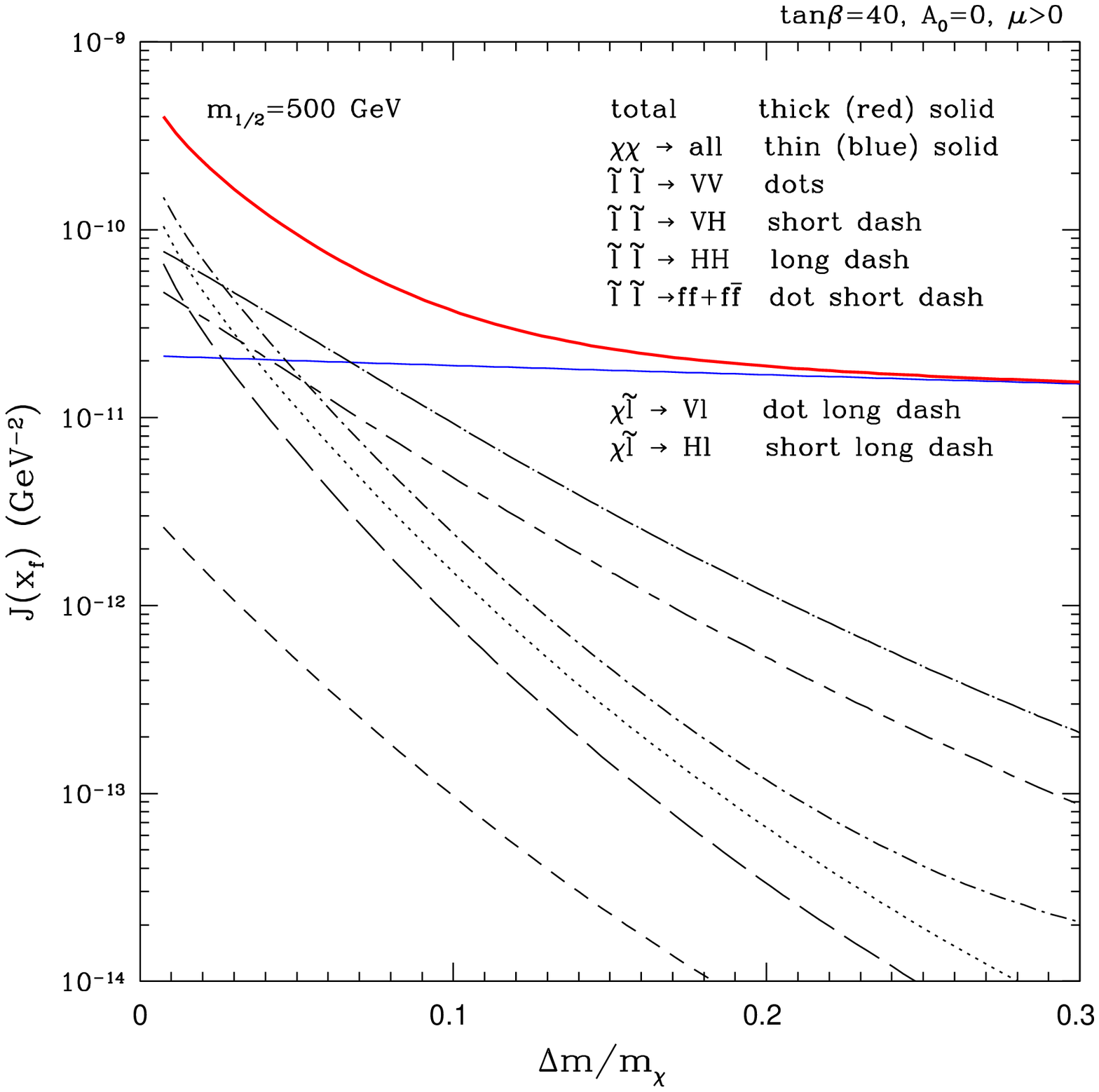,height=3.15in} 
\end{minipage}
\caption{\label{fig:jxf} Total and partial contributions to $\jxf$
from the various classes of processes listed in
Table~{\protect\ref{tableone}} as a function of $\Delta m/\mchi = (\mchi -
m_{\stauone})/\mchi$ 
for the two choices of $\tanb=10,40$ of
Fig.~{\protect\ref{fig:oh2cont}}.
}
\end{figure}

In Fig.~\ref{fig:jxf} we present the total as well as the individual
contributions to the quantity $\jxf\equiv \int_0^{x_f}dx \langle\sigma
v_{\rm M\o l}\rangle(x)$ from the various classes of processes listed in
Table~\ref{tableone}. (Roughly, $\abundchi\sim 1/\jxf$ -- see,
\eg,~\cite{nrr2}.) This is done for a slice of constant 
$\mhalf=500\gev$ as a function of $\Delta m/\mchi = (\mchi -
m_{\stauone})/\mchi$ 
for the two choices of $\tanb=10,40$ of
Fig.~\ref{fig:oh2cont}. At a smaller mass difference it is the
slepton--slepton annihilation into lepton pairs (mostly
$\stauone\stauone\rightarrow \tau\tau$) that appears to be
dominant but it drops quickly and
at larger $\Delta m/\mchi$ it is overtaken by the neutralino--slepton
coannihilation channel (mostly the $\gamma\tau$
final state). Here our conclusions qualitatively agree with~\cite{efos00}. 

\begin{figure}[t!]
\vspace*{-0.1in}
\hspace*{-.12in}
\begin{minipage}{8in}
\epsfig{file=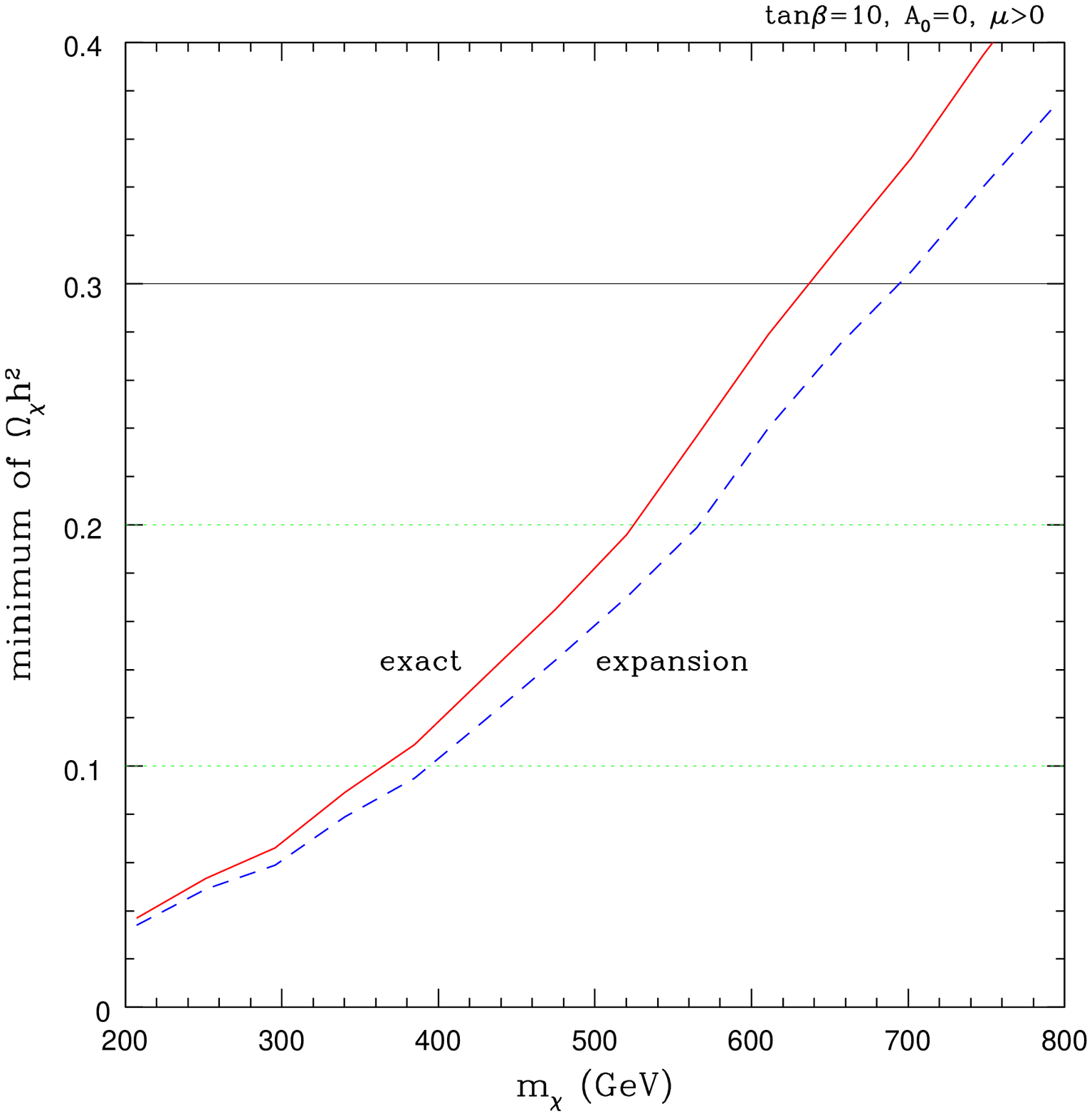,height=3.15in} 
\hspace*{-0.03in}
\epsfig{file=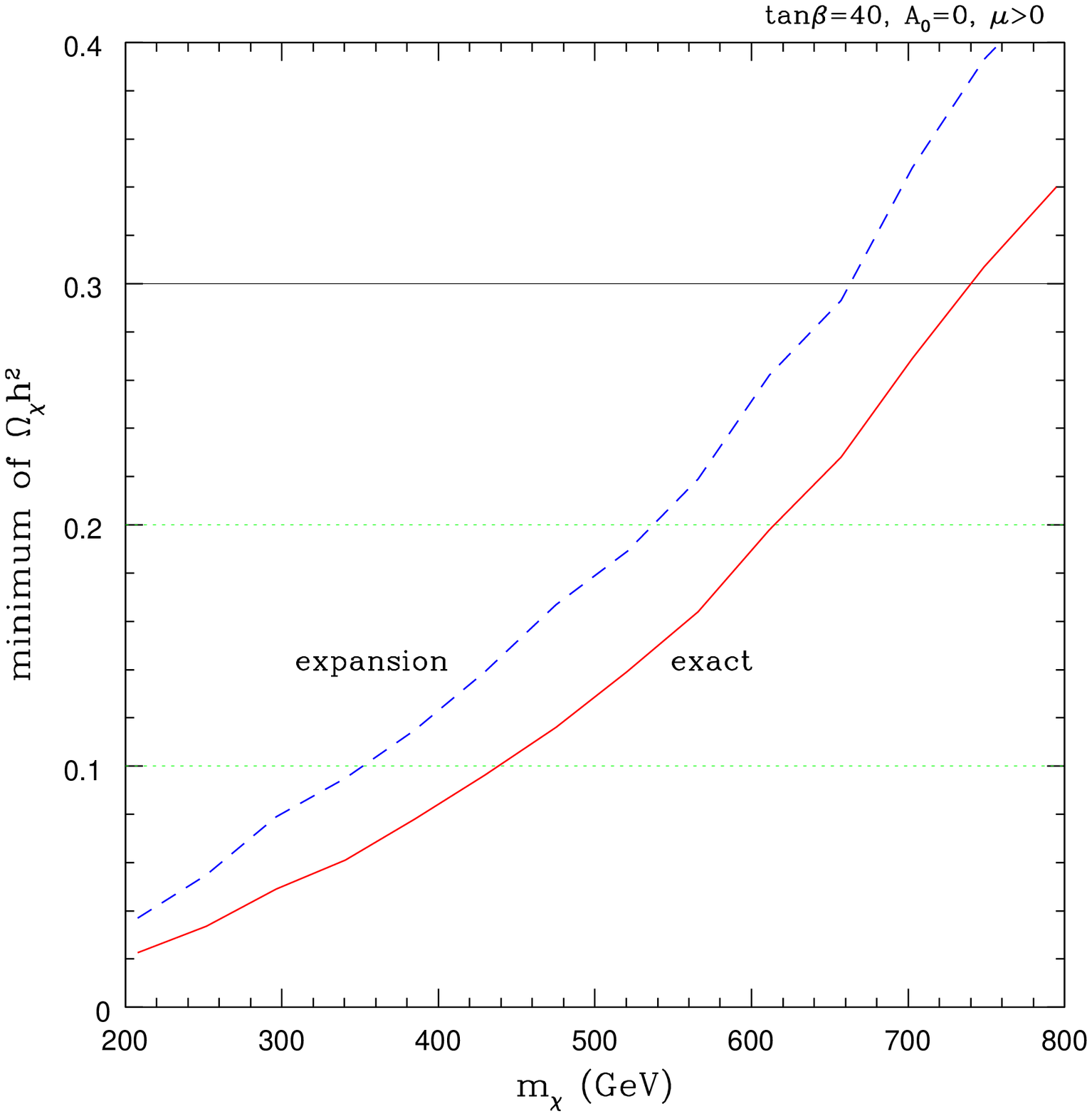,height=3.15in} 
\end{minipage}
\caption{\label{fig:oh2min} The minimum of $\abundchi$ (\ie,
$\abundchi$ along the line $\mchi=m_{\stauone}$) as a function of
$\mchi$ for the two representative choices of
Fig.~{\protect\ref{fig:oh2cont}}. The solid (red) curve correspods
to the exact neutralino--slepton coannihilation cross sections, while
the dashed (blue) ones to the approximate ones of~{\protect\cite{efos00}}. 
}
\end{figure}

As mentioned in the Introduction,~\cite{efos00} contains the only
set of analytic expresssions for the neutralino--slepton
coannihilation that is at present available in the literature. As
noted by the authors, the expressions given there are approximate as
they were derived using partial wave expansion. They also become less
reliable at large values of $\tanb\gsim20$~\cite{efgos01} because the
effects of $\ytau$ and of the $\stauone-\stautwo$ mixing, and in some
channels of the mass of the $\tau$, were neglected. We have made an
attempt at improving the expressions of~\cite{efos00} by
including in the propagators the widths of the gauge and Higgs bosons
and the neutralinos; otherwise they become singular.

While keeping these points in mind, it is nevertheless interesting to
compare them with the exact expressions derived here. We do
this numerically in Fig~\ref{fig:oh2min} for the cases displayed in
Fig.~\ref{fig:oh2cont}. Along the boundary of $\mchi=m_{\stauone}$
(and in fact in most of the $(\mhalf,\mzero)$ plane) $\mchi\simeq0.44
\mhalf-2.8\sin 2\beta$~\cite{bkao98}. This allows us to 
plot the minimum value of
$\abundchi$ along the boundary as a function of $\mchi$.  
It is clear that $\abundchi$ increases with
$\mchi$ and at some point becomes inconsistent with
$\abundchi<0.3$. The solid (red) curve correspond to the exact results
while the dash (blue) ones were obtained by using the approximate
expressions of~\cite{efos00}. For $\tanb=10$ the upper limit from
$\abundchi<0.3$ on $\mchi$ is strengthened from $\sim700\gev$ down to
some $640\gev$. In contrast, at large $\tanb=40$ (where, we repeat, the
approximate expressions are not really applicable~\cite{efos00}), an
upper bound becomes considerably weaker ($\mchi\lsim 765\gev$) than
what one would get by naively applying the expressions
of~\cite{efos00}. The ranges of $\mchi$ corresponding to the favored 
range $0.1<\abundchi<0.2$ also shift accordingly.

%
\section{Summary}\label{summary:sec}
The accuracy of determining the abundance of the dark matter in the
Universe is continuously improving. This requires theoretical
computations of the neutralino relic abundance to be performed with at
least the same, if not better, level of precision, if one wants to
reliably compare theoretical predictions with observations.

In this paper we have derived a full set of exact, analytic
expressions for the neutralino--slepton coannihilation cross sections into
all tree--level two--body final states. While these formulae are
applicable in the framework of the general MSSM, they are of
particular importance in the context of the Constrained MSSM. In this
framework, which is often considered a ``reference'' SUSY model and is
thus of much interest to the community, much of the allowed regions
are a result of the neutralino--slepton coannihilation. Our results
should help in allowing one to determine these regions more precisely.

%
\bigskip

\acknowledgments
T.N. is grateful to Lancaster University for kind hospitality
extended during his visit. R.RdA is supported by the EC
``Supersymmetry and the Early Universe'' grant (RTN contract number:
HPRN--CT--2000--00152) of which he and L.R. are members.
%
\newpage
\appendix
\section{Lagrangian Terms and Couplings}\label{appxa}

In this Appendix, we define the couplings 
which appear in the main text and which have not been defined
in~\cite{nrr2}. We follow the same conventions and notation as
in that paper.  

\vspace*{0.2cm}\noindent
\underline{\bf slepton -- slepton -- Higgs}
\begin{eqnarray}
  {\cal L} & = & 
      \Big( C^{\slp{b}^*\slp{a}h} h 
          + C^{\slp{b}^*\slp{a}H} H 
          + C^{\slp{b}^*\slp{a}A} A \Big)\slp{b}^*\slp{a} 
   + \Big[ C^{\sn{\ell}^*\slp{a}H^+} \sn{\ell}^*\slp{a}H^+ + {\rm h.c.} \Big],
\end{eqnarray}
where
\begin{eqnarray}
C^{\slp{b}^*\slp{a}r} & = & 
 C_{LL}^r (\vsl)_{1b}^*(\vsl)_{1a} + C_{RR}^r (\vsl)_{2b}^*(\vsl)_{2a}
 \nonumber \\
 & & 
+ C_{LR}^r (\vsl)_{1b}^*(\vsl)_{2a} + C_{RL}^r (\vsl)_{2b}^*(\vsl)_{1a}
 \ \ \ (r=h,H,A), \\
C^{\sn{\ell}^*\slp{a}H^+} & = &
 C_{LL}^{H^+}(\vsl)_{1a} + C_{LR}^{H^+}(\vsl)_{2a},
\end{eqnarray}
with
\begin{eqnarray}
C_{LL}^h & = & -\frac{g\mz}{\cw}\left(\frac{1}{2}-\swsq \right)\sinapb 
               + \frac{g\ml^2}{\mw\cosb}\sina, \\
C_{RR}^h & = & -\frac{g\mz}{\cw}\swsq\sinapb 
               + \frac{g\ml^2}{\mw\cosb}\sina, \\ 
C_{LR}^h & = & \frac{g\ml}{2\mw\cosb}\left(A_{\ell}\sina+\mu \cosa \right), \\
C_{RL}^h & = & C_{LR}^h, \\ 
C_{LL}^H & = & \frac{g\mz}{\cw}\left(\frac{1}{2}-\swsq \right)\cosapb 
               - \frac{g\ml^2}{\mw\cosb}\cosa, \\
C_{RR}^H & = & \frac{g\mz}{\cw}\swsq\cosapb 
               - \frac{g\ml^2}{\mw\cosb}\cosa, \\
C_{LR}^H & = & -\frac{g\ml}{2\mw\cosb}\Big(A_{\ell}\cosa-\mu \sina \Big), \\
C_{RL}^H & = & C_{LR}^H, \\
C_{LL}^A & = & 0, \\
C_{RR}^A & = & 0, \\
C_{LR}^A & = & - \frac{ig\ml}{2\mw}\Big(A_{\ell}\tanb+\mu \Big), \\
C_{RL}^A & = & - C_{LR}^A, \\
C_{LL}^{H^+} & = & -\frac{g\mw}{\sqrt{2}}
       \left[ \sin 2\beta - \frac{\ml^2}{\mw^2} \tanb \right], \\
C_{LR}^{H^+} & = & \frac{g\ml}{\sqrt{2}\mw}\Big( A_{\ell}\tanb+\mu \Big).
\end{eqnarray}
Note that, for $a=b$, the pseudoscalar coupling
$C^{\slp{b}^*\slp{a}A}$ vanishes in the absence of CP violating 
phases: $C^{\slp{a}^*\slp{a}A}$ $=0$. 

\noindent
\underline{\bf slepton -- slepton -- gauge}
\begin{eqnarray}
  {\cal L} & = & 
      \Big[ iC^{\sn{\ell}^*\slp{a}W^+} (\sn{\ell}^*\ \delmux \slp{a})W_\mu^+ 
                                + {\rm h.c.} \Big] 
     + i(C^{\slp{b}^*\slp{a}Z}Z_\mu)
                       (\slp{b}^* \delmux \slp{a}), \nonumber \\
 & & 
\end{eqnarray}
where
\begin{eqnarray}
C^{\sn{\ell}^*\slp{a}W^+} & = & 
       -\frac{g}{\sqrt{2}}(\vsl)_{1a}, \\
C^{\slp{b}^*\slp{a}Z} & = & -\frac{g}{\cw}\left[ 
\left(-\frac{1}{2}+\swsq \right)(\vsl)_{1b}^*(\vsl)_{1a}
                      + \swsq(\vsl)_{2b}^*(\vsl)_{2a} \right]. 
\end{eqnarray}

\noindent
\underline{\bf slepton -- slepton -- Higgs -- Higgs}
\begin{eqnarray}
  {\cal L} & = & 
      \Bigg[ \frac{1}{2} C^{\slp{b}^*\slp{a}hh} h^2
           + \frac{1}{2} C^{\slp{b}^*\slp{a}HH} H^2
                       + C^{\slp{b}^*\slp{a}hH} hH \nonumber \\
 & &       + \frac{1}{2} C^{\slp{b}^*\slp{a}AA} A^2
           + C^{\slp{b}^*\slp{a}H^+H^-} H^+H^- \Bigg]\slp{b}^*\slp{a},
\end{eqnarray}
where
\begin{eqnarray}
C^{\slp{b}^*\slp{a}X} & = & 
  C_{LL}^X (\vsl)_{1b}^*(\vsl)_{1a} + C_{RR}^X (\vsl)_{2b}^*(\vsl)_{2a}
\nonumber \\
 & &  \hspace{4cm} (X=hh,hH,AA,H^+H^-),
\end{eqnarray}
and 
\begin{eqnarray}
C_{LL}^X & = & \frac{g^2}{2} \left[ 
                  C_1^X \left( \frac{-1+2\swsq}{2\cos^2\theta_W} \right)
                  -\frac{\ml^2}{\mw^2}C_2^X \right], \\
C_{RR}^X & = & \frac{g^2}{2} \left[ 
                  - \tan^2\theta_W C_1^X
                  -\frac{\ml^2}{\mw^2}C_2^X \right], \\
C_{LL}^{H^+H^-} & = & \frac{g^2\cos 2\beta}{4 \cos^2\theta_W}, \\
C_{LR}^{H^+H^-} & = & -\frac{g^2}{2} \left[
\cos 2\beta \tan^2\theta_W
       + \frac{\ml^2}{\mw^2}\tan^2\beta \right],
\end{eqnarray}
with $C_1^X$ and $C_2^X$ ($X=hh, HH, hH, AA$) in the following
\begin{eqnarray}
C_1^{hh} & = & \cos 2\alpha, \\
C_2^{hh} & = & \frac{\sin^2\alpha}{\cos^2\beta}, \\
C_1^{HH} & = & -\cos 2\alpha, \\
C_2^{HH} & = & \frac{\cos^2\alpha}{\cos^2\beta}, \\
C_1^{hH} & = & \sin 2\alpha, \\
C_2^{hH} & = & -\frac{\sin 2\alpha}{2\cos^2\beta}, \\
C_1^{AA} & = & \cos 2\beta, \\
C_2^{AA} & = & \tan^2\beta.
\end{eqnarray}

\noindent
\underline{\bf slepton -- slepton -- gauge -- gauge}
\begin{eqnarray}
  {\cal L} & = & 
      \Bigg[ \frac{1}{2} C^{\slp{b}^*\slp{a}ZZ} Z_\mu Z^\mu
             + C^{\slp{b}^*\slp{a}Z\gm} Z_\mu A^\mu 
    + C^{\slp{b}^*\slp{a}WW} W^+_\mu W^{-\mu} \Bigg]\slp{b}^*\slp{a},
\end{eqnarray}
where
\begin{eqnarray}
C^{\slp{b}^*\slp{a}ZZ} & = & 
\frac{2g^2}{\cos^2\theta_W}\left[
\left(-\frac{1}{2}+\swsq \right)^2(\vsl)_{1b}^*(\vsl)_{1a}
                      + \sin^4\theta_W(\vsl)_{2b}^*(\vsl)_{2a} \right], \\
C^{\slp{b}^*\slp{a}Z\gm} & = & 
 - 2g^2\tw\left[ 
\left(-\frac{1}{2}+\swsq \right)(\vsl)_{1b}^*(\vsl)_{1a}
                      + \swsq(\vsl)_{2b}^*(\vsl)_{2a} \right], \nonumber \\
 & & \\
C^{\slp{b}^*\slp{a}WW} & = & \frac{1}{2}g^2(\vsl)_{1b}^*(\vsl)_{1a}.
\end{eqnarray}


Some more couplings including the neutrinos: 
\begin{eqnarray}
  {\cal L} & = & 
C_L^{\chi_k^+\slp{a}\nu_{\ell}} 
     \slp{a} \bar{\nu}_{\ell} (1+\gamma_5)\chi_k^+
+ C_L^{\ell\nu_{\ell} H^-} 
     \bar{\ell}(1-\gamma_5)\nu_{\ell} H^+ \nonumber \\
 & & \hspace{3cm}
+ C_L^{\ell\nu_{\ell} W} 
     \bar{\nu}_{\ell} \gamma^{\mu} (1-\gamma_5) \ell W_{\mu}^+
+ {\rm h.c.} 
\end{eqnarray}
where 
\begin{eqnarray}
C_L^{\chi_k^+\slp{a}\nu_{\ell}} & = & -\frac{1}{2}g 
\left[ U_{k1}(\vsl)_{1a} 
       - \frac{\ml}{\sqrt{2}\mw\cosb}U_{k2}(\vsl)_{2a} \right], \\
C_L^{\ell\nu_{\ell} H^-} & = & \frac{g\ml \tanb}{2\sqrt{2}\mw}, \\
C_L^{\ell\nu_{\ell} W} & = & -\frac{g}{2\sqrt{2}}.
\end{eqnarray}

\noindent
\underline{\bf neutralino -- lepton -- slepton}
\begin{eqnarray}
   {\cal L}& =&
\sum_{\ell=e,\mu,\tau} \sum_{i=1}^4  
\sum_{a=1}^2 \left\{
\overline{\neu{i}} \left( 
   C^{\neu{i}\slp{a}^*\ell}_L \frac{1-\gamma_5}{2}
 + C^{\neu{i}\slp{a}^*\ell}_R \frac{1+\gamma_5}{2} \right) 
                             \ell \, \slp{a}^* + {\rm h.c.} \right\}, 
\label{eq:n-f-sf}
\end{eqnarray}
where
\begin{eqnarray}
\label{eq:lambdal}
C^{\neu{i}\slp{a}^*\ell}_L & = &
- \frac{g}{\sqrt{2}} \left[
     (\widetilde{V}_{\ell})_{a1}^* \ell^{\chi^0_i \ell L}
   + (\widetilde{V}_{\ell})_{a2}^* \ell^{\chi^0_i \ell R} \right], \\
\label{eq:lambdar}
C^{\neu{i}\slp{a}^*\ell}_R & = &
- \frac{g}{\sqrt{2}} \left[
     (\widetilde{V}_{\ell})_{a1}^* r^{\chi^0_i \ell L}
   + (\widetilde{V}_{\ell})_{a2}^* r^{\chi^0_i \ell R} \right]. 
\end{eqnarray}
The couplings $\ell^{\chi^0_i \ell L}$, $\ell^{\chi^0_i \ell R}$,
$r^{\chi^0_i \ell L}$ and $r^{\chi^0_i \ell R}$ are defined in~\cite{nrr2}. 
Note that $\ell$ $=$ $e$, $\mu$, $\tau$ represents a charged lepton. 
We neglect generation mixing in the lepton sector. 
The charged slepton mass eigenstates $\slp{a}$ ($a=1,2$) are related 
to the slepton gauge eigenstates $\slp{L}$ and $\slp{R}$ via
\begin{eqnarray}
  \slp{a} & = & (\widetilde{V}_{\ell})_{a1}\slp{L}
              + (\widetilde{V}_{\ell})_{a2}\slp{R}, 
\end{eqnarray}
where $\widetilde{V}_{\ell}$ denotes a $2 \times 2$ matrix 
which diagonalizes the charged slepton mass matrix: 
$\widetilde{V}_{\ell}$ ${\cal M}^2_{\widetilde{\ell}}$ 
$\widetilde{V}_{\ell}^\dagger$ $=$ 
${\rm diag}(m_{\slp{1}}^2,m_{\slp{2}}^2)$.
We also use $C^{\neu{i}\slp{a}^*\ell}_S$ and $C^{\neu{i}\slp{a}^*\ell}_P$
defined by 
\begin{eqnarray}
C^{\neu{i}\slp{a}^*\ell}_S & = & 
   \frac{1}{2}\left(C^{\neu{i}\slp{a}^*\ell}_L
                    +C^{\neu{i}\slp{a}^*\ell}_R \right), \\
C^{\neu{i}\slp{a}^*\ell}_P & = & 
   \frac{1}{2}\left(C^{\neu{i}\slp{a}^*\ell}_L
                    -C^{\neu{i}\slp{a}^*\ell}_R \right). 
\end{eqnarray}


\section{Auxiliary functions}

Here we give definitions for the auxiliary functions used in the text. 
The functions $\Ft$ and $\Fu$ are given by
\begin{eqnarray}
\Ft(s,x_1,x_2,y_1,y_2,z) & = & \frac{1}{2\,F}\,
   \ln\left|\frac{D_0-D_1+F-z}{D_0-D_1-F-z}\right|, \\
\Fu(s,x_1,x_2,y_1,y_2,z) & = & \frac{1}{2\,F}\,
   \ln\left|\frac{D_0+D_1+F-z}{D_0+D_1-F-z}\right|, 
\end{eqnarray}
where 
\begin{eqnarray}
D_0 & = & -\frac{s}{2} + \frac{x_1+x_2+y_1+y_2}{2}, \\
D_1 & = & \frac{(x_1-x_2)(y_1-y_2)}{2s}, \\
F & = & \frac{1}{2}
    \sqrt{s-(\sqrt{x_1}+\sqrt{x_2})^2}
    \sqrt{1-\frac{(\sqrt{x_1}-\sqrt{x_2})^2}{s}} \nonumber \\
 & & \times \sqrt{s-(\sqrt{y_1}+\sqrt{y_2})^2}
 \sqrt{1-\frac{(\sqrt{y_1}-\sqrt{y_2})^2}{s}}. 
\end{eqnarray}
For $y_1$ $=$ $y_2$, $D_1$ vanishes so that the two functions 
$\Ft$ and $\Fu$ reduce to the same function which is denoted by $\F$. 

The functions $\Tt_i$ and $\Tu_i$ are obtained
from $\T_i$ in~\cite{nrr2} by the following replacements 
\begin{eqnarray}
\Tt_i(s,x_1,x_2,y_1,y_2,z_1,z_2) & = & 
      \T_i(D \rightarrow D_0-D_1,
           \F \rightarrow \Ft), \\
\Tu_i(s,x_1,x_2,y_1,y_2,z_1,z_2) & = & 
      \T_i(D \rightarrow D_0+D_1,
           \F \rightarrow \Fu). 
\end{eqnarray}
The functions $\T_i$ in the right-hand side represent 
those defined in~\cite{nrr2}, 
where $F$ in~\cite{nrr2} should be replaced with $F$ in this appendix. 
For $y_1$ $=$ $y_2$, the two functions $\Tt_i$ and $\Tu_i$ 
reduce to the same function which is denoted by $\T_i$. 
The expressions for the functions $\Y_i$ ($i=0,1,2,3,4$) 
are given by
\begin{eqnarray}
\Y_0 &=& 
\frac{1}{z_1+z_2-2 D_0}[\Ft(s,x_1,x_2,y_1,y_2,z_1)
                        +\Fu(s,x_1,x_2,y_1,y_2,z_2)], \\
\Y_1 &=& \frac{2}{z_1+z_2-2 D_0}
        [(z_1-D_0+D_1)\Ft(s,x_1,x_2,y_1,y_2,z_1) \nonumber \\
 & & \hspace{4.5cm} - (z_2-D_0-D_1)\Fu(s,x_1,x_2,y_1,y_2,z_2)], 
                                                    \\ 
\Y_2 &=& 
1 +\frac{1}{z_1+z_2-2 D_0}
[(z_1+D_1)(z_1-2D_0+D_1)\Ft(s,x_1,x_2,y_1,y_2,z_1) \nonumber \\
& & \hspace{3.5cm} + (z_2-D_1)(z_2-2D_0-D_1)\Fu(s,x_1,x_2,y_1,y_2,z_2)],
                                                     \\
\Y_3 &=& 
2(z_1-z_2+2D_1) \nonumber \\
& & +\frac{2}{z_1+z_2-2 D_0}
      [(z_1+D_1)(z_1-D_0+D_1)(z_1-2D_0+D_1)\Ft(s,x_1,x_2,y_1,y_2,z_1)
        \nonumber \\
& & \hspace{3.0cm} -(z_2-D_1)(z_2-D_0-D_1)(z_2-2D_0-D_1)
                                \Fu(s,x_1,x_2,y_1,y_2,z_2)], \nonumber \\
& &                                                    \\  
\Y_4 &=& 
-D_0^2+3D_1^2-D_0(z_1+z_2)+3D_1(z_1-z_2)
+(z^2_1+z^2_2-z_1\,z_2)+\frac{1}{3} F^2 \nonumber \\
& & \hspace{1.5cm}  + \,\frac{1}{z_1+z_2-2 D_0}
    [(z_1+D_1)^2(z_1-2D_0+D_1)^2 \Ft(s,x_1,x_2,y_1,y_2,z_1) \nonumber \\
& &   \hspace{4.7cm} + \,(z_2-D_1)^2(z_2-2D_0-D_1)^2
                                      \Fu(s,x_1,x_2,y_1,y_2,z_2)], \nonumber \\
& &                                                      \\  \nonumber
\end{eqnarray}
where $\Y_i$ $=$ $\Y_i(s,x_1,x_2,y_1,y_2,z_1,z_2)$. 


\newpage

\end{document}